\documentclass[a4paper,fleqn]{cas-sc}
\usepackage{natbib}

\usepackage{orcidlink} 

\usepackage[T1]{fontenc}
\usepackage[utf8]{inputenc}
\usepackage[english]{babel}

\usepackage{graphicx}

\usepackage{amsmath,amssymb}
\usepackage[version=4]{mhchem}

\usepackage{subcaption}

\usepackage{tikz}
\usetikzlibrary{shapes.geometric, arrows.meta, patterns, calc}

\usepackage{lineno}

\usepackage{csquotes}
\usepackage{hyperref}

\usepackage{xcolor}
\usepackage{listings}
\definecolor{codegreen}{rgb}{0,0.6,0}
\definecolor{codegray}{rgb}{0.5,0.5,0.5}
\definecolor{codepurple}{rgb}{0.58,0,0.82}
\definecolor{backcolour}{rgb}{0.97,0.97,0.98}
\definecolor{codeblue}{rgb}{0.1,0.1,0.7}

\begin{document}

\title[mode=title]{Europa smooth plains formation: a SPH cryovolcanic model}
\shorttitle{Cryovolcanic modeling of smooth plains of Europa}
\author[1]{Bastien Bodin}[
    orcid=0000-0003-3602-2589,
    linkedin=bastien-bodin,
]
\cormark[1]
\ead{bastien.bodin@proton.me}

\author[2]{Daniel Cordier}[orcid=0000-0003-4515-6271]

\author[3]{Ashley Gerard Davies}[orcid=0000-0003-1747-8142]

\affiliation[1]{
  organization={Laboratoire Environnements et Atmosphères Terrestres et Planétaires,
Université de Reims Champagne-Ardenne},
  city={Reims},
  country={France}
}
\affiliation[2]{
  organization={Laboratoire de Plan\'etologie et G\'eodynamique, CNRS UMR 6112, Universit\'e de Nantes},
  postcode={44322},
  city={Nantes},
  country={France}
}
\affiliation[3]{
  organization={Jet Propulsion Laboratory, California Institute of Technology},
  addressline={4800 Oak Grove Drive},
  city={Pasadena},
  state={CA},
  postcode={91109},
  country={USA}
}

\cortext[1]{Corresponding author}

\shortauthors{Bodin et al.}
\begin{abstract}
    Cryovolcanism is a fundamental geodynamical process that has been observed on several icy moons.
    Europa stands out as a primary target for investigating potential cryovolcanic activity and its role in shaping icy moon surfaces.
    This work characterizes the physical state of ascending cryomagma
    and evaluates the relevance of terrestrial analogues to explain Europa's
    surface features.
    After investigating cryomagma depressurization, we developed CryoPy,
    a numerical model based on Smoothed-Particle Hydrodynamics (SPH) for
    computational fluid dynamics and heat transfer. We simulated flow
    propagation and surface cooling using this model.
    Our simulations show that snow is a relevant physical analogue for
    cryolava, as the resulting flows closely match the smooth plains
    observed by Galileo in both extent and morphology. Additionally,
    we established the temporal thermal signatures of these events.
    Such active cryovolcanic events are
    within the detection limits of, primarily, the E-THEMIS infrared imager, 
    providing a predictive framework for the Europa Clipper mission.
\end{abstract}


\begin{highlights}
    \item Europa's smooth plains are interpreted as cryolava 
    flows.
    \item We propose a snow-like rheology for cryolava erupted 
    via depressurization.
    \item These cryolava flows are modeled using Smoothed 
    Particle Hydrodynamics (SPH).
    \item Our simulations successfully reproduce the observed 
    spatial extent of the plains.
    \item Flow thermal signatures are estimated for detectability 
    by Clipper's E-THEMIS.
\end{highlights}


\maketitle


\section{Introduction}
Europa, first observed by \cite{galileo1610sidereus}, is recognized
as a prominent member of the ocean worlds class of celestial bodies,
with outstanding exobiological interest. Although the moons of Jupiter
have been observed by multiple spacecraft (Pioneer 10 and 11,  Voyager 1 and 2, Galileo, Cassini, New Horizons, and Juno), the greatest
scientific advances concerning Europa over the past half century
were made possible by Voyager~2 (1979)  \citep{smith1979jupiter} and Galileo (1996-2003)
\citep{johnson1992space}. Subsequent observations by the Hubble Space Telescope provided evidence of ongoing plume activity \citep{roth2014vapor}.  Recently, the Juno mission obtained images and near-infrared spectra of Europa \citep{becker2023SRUeuropa, hansen2024juno, filacchione2019europa}. Spacecraft imagery reveal an ice-covered world with a young surface, as evidenced by a relative lack of primary impact craters \citep{moore2001IcarusEurGEM, leonard2023Europageomap}. Some
surface features may have their origins in cryovolcanism.  As summarized in \cite{pappalardo2009europa} and  \cite{daubar2024planned}, cryovolcanic activity leading to the resurfacing of Europa might manifest in the form of surface flows, domes, melt-through features, and geyser-like plumes. Low-albedo mantlings, that may have been ballistically emplaced, surround lineaments and lenticulae \citep{Fagents2000IcarusEurVolc}. Chaos regions may be the surface manifestation of internal cryomagmatic processes, such as diapirism \citep[e.g.,][]{wilson1997JGRflowsEuropa, pappalardo1998geological, prockter2005macula}.  Galileo image and spectral data in particular revealed features suggestive of effusive cryovolcanic processes, including low-albedo plains and extrusive domes \citep[e.g.,][]{fagents2003considerations, kattenhorn2014NatGeo, quick2017cryovolcanic, quick2022Icarus}, and intrusions \citep{figueredo2004resurfacing, sotin2002europa, manga2017formation, chivers2021JGRE}. Geyser-like plumes of cryoclastic material, like those seen on Enceladus \citep{dougherty2006SciEncPlume, hansen2006enceladus, porco2006cassini, spahn2006cassini, waite2006SciEncPlume}, may also be a resurfacing mechanism \citep[e.g.,][]{Fagents2000IcarusEurVolc, quick2013constraints, quick2020characterizing}. The search for such features and active processes is a primary objective of the Europa Clipper mission \citep{pappalardo2024SSR}.

Cryovolcanism is important for two reasons. Firstly,
it represents a complex combination of physical processes by which
materials and energy are exchanged between the interior and the
surface of an icy body. This exchange is of great importance from an
astrobiological perspective, as the interior of Europa may harbor
prebiotic or even biological activity \citep{vance2023investigating}.
Secondly, cryovolcanism likely plays a major geophysical role in the
thermo-physical evolution of Europa and the formation and evolution of surface features \citep{daubar2024planned}.

In this work, we focus on a scenario of \enquote{smooth plains} formation. As
their name suggests, Europa's smooth plains are geological features
that appear much smoother than the surrounding terrain. Several
hypotheses have been proposed regarding their formation: (1) the
flow of brine onto the surface, possibly associated with a cryovolcanic
vent; (2) a local enhancement of sublimation due to a subsurface heat
source; and (3) a cryoclastic deposit \citep{daubar2024planned}. In
this paper, we adopt a scenario consistent with the first hypothesis,
that is, a cryolava flow fed by material erupting from a vent.

\begin{figure}
    \centering
    \includegraphics[width=0.99\linewidth]{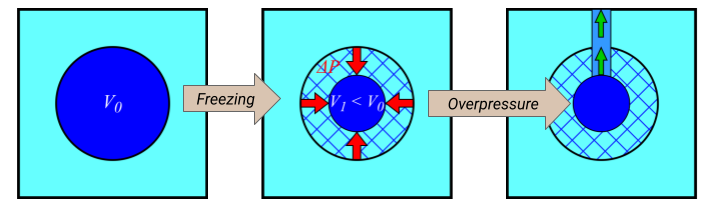}
    \caption
    [Illustration of pressure-driven cryomagma reservoirs]
    {Illustration of the model imagined by \citet{fagents2003considerations} and developed by
    \citet{lesage2020cryomagma,lesage2022simulation}. The liquid within
    the reservoir, composed of pure or saline water, partially freezes,
    leading to a significant overpressure. When hitting a critical
    threshold, the ice crust begins to fracture, allowing cryomagma
    to erupt at speeds up to 20 m/s at the edge of the reservoir.}
    \label{fig:fagents2003}
\end{figure}

Such cryovolcanic processes, particularly those involving eruptions
from subsurface reservoirs, have been extensively modeled, for
instance, by \citet{fagents2003considerations} and further developed by
\citet{lesage2020cryomagma,lesage2022simulation}.
These models propose that cryovolcanic eruptions on
Europa are driven by the
pressurization of subsurface cryomagmatic reservoirs
(Figure~\ref{fig:fagents2003}).

This pressurization occurs
as the liquid within the reservoir, composed of pure or saline water (brine),
begins to freeze and expand, leading to pressurization of the remaining liquid.
When this overpressure exceeds a critical threshold (5--35 MPa, depending on
depth and ice strength), it causes the overlying ice crust to fracture,
allowing cryomagma to exploit any fissure or conduit that forms and erupt at speeds up to 20 m/s \citep{lesage2020cryomagma}. Later work by
\citet{lesage2022simulation} incorporated the visco-elastic
relaxation of the reservoir walls, showing its impact on eruption
behavior. We consider this model as a starting point for our
investigations.

A study by \cite{morrison2023reevaluation} investigated
the hydrodynamic behavior of water containing sodium chloride under
Europa's surface conditions, assuming the cryolava reached the surface in a liquid state, and
underwent progressive solidification, thus increasing its viscosity. This work
represented a first coupling of hydrodynamics and thermodynamics for
cryolava flows.



Phase and rheological conditions within the conduit are complex. To better understand the eruption process, we have incorporated additional complexity into our model by considering variation in pressure, temperature, and composition during ascent. We also incorporate phase change within the ascending liquid due to depressurization and cooling, which affects rheology and flow dynamics. 

To investigate the smooth plain formation, we developed a
Smoothed-Particle Hydrodynamics (SPH) model that incorporates the complexities described above. SPH is a Lagrangian
numerical method originally introduced in astrophysics to model non-spherical polytropes
\citep{gingold1977smoothed}. Since then, SPH has been applied to a
wide variety of problems in hydrodynamics, impact physics, heat
transport, and radiative transfer \citep{zhang2022smoothed}. This
approach is particularly suitable for systems with a free surface,
which is why it has been successfully employed for numerical
simulations of terrestrial lava flow emplacement
\citep{prakash2011three,bilotta2016gpusph}. The mesh-free nature
and ability to handle large deformations and complex fluid-solid
interactions make SPH ideal for simulating the dynamic and evolving
morphology of cryolava flows on Europa's surface, especially when
considering potential phase transitions and non-Newtonian rheologies
that arise from cooling and depressurization. SPH also allows for the
direct incorporation of thermal processes, crucial for understanding
the solidification and detectability of cryovolcanic deposits.

Consequently, we have adapted the SPH framework to the problem being investigated, and the
implementation of the models has been enabled with the help of the
open-source, object-oriented Python package PySPH
\citep{ramachandran2021pysph}.

This paper is organized as follows: the fracturing of Europa's icy
crust and the pressure drop during cryolava ascent are first
discussed; the behavior of liquid water undergoing depressurization
and the evolving cryolava rheology are then examined; and the
dynamics of lava flowing on the surface of Europa are studied,
along with the evolving thermal emission, as might be detected from a spacecraft. A section dedicated to conclusions
and perspectives closes the article.

\section{The Ascent of Cryolava on Europa: From Fracturing to Eruption}
\label{sec:ascent_cryolava}

The ascent of cryolava from subsurface reservoirs to Europa's surface
is governed by a cascade of physical processes, including the
fracturing of the ice shell, pressure drops during ascent, and
phase transitions near the surface. These processes determine whether
an eruption can occur and the final state of the erupted material.
This section discusses the fracturing of the ice crust, as well as the behavior of water as it rises toward the surface.

\subsection{Fracturing of Europa's Ice Shell}
\label{ssec:fracturing}

\begin{figure}
    \centering
    \includegraphics[width=0.5\linewidth]{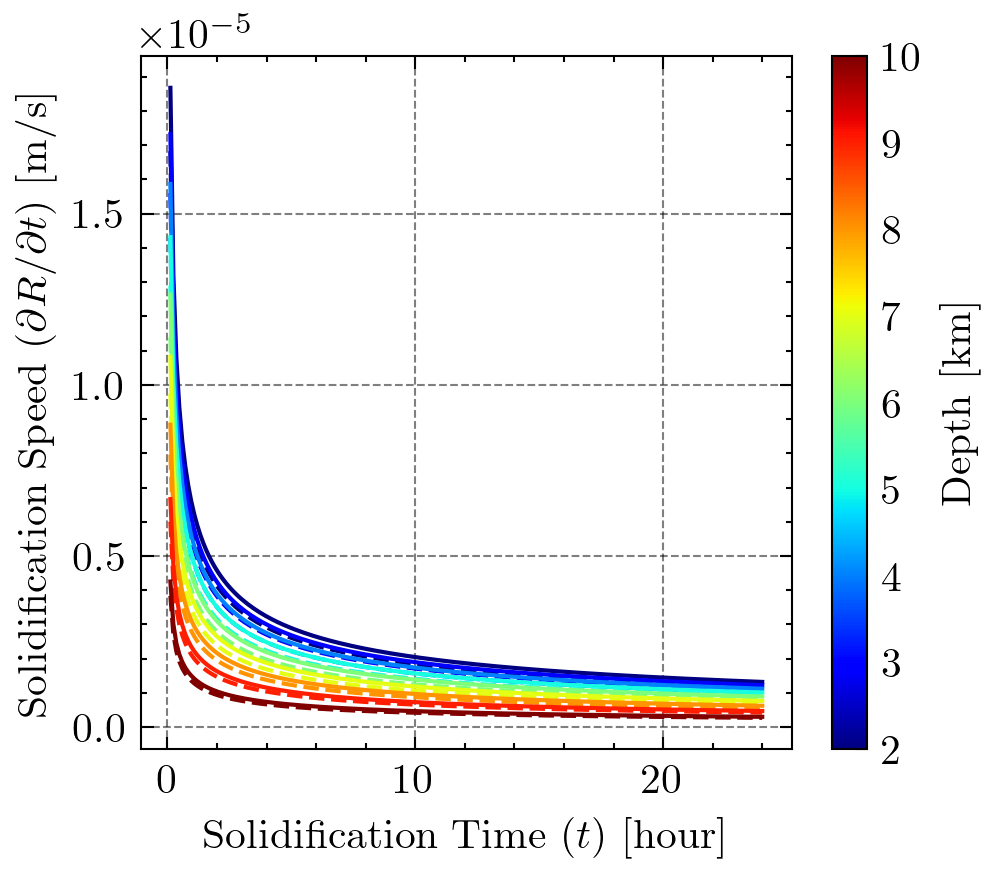}
    \caption{Solidification speed $\partial R/\partial t$ as a function of time for various reservoir depths $H$ (represented by the color scale). The results are derived from the analytical solution of the Stefan problem \citep{Carslaw1959}. Solid lines represent pure water cryolava, while dashed lines corresponds to a "briny" cryomagma composition (e.g., \ce{H_2O-MgSO_4-Na_2SO_4} mixture)}
    \label{fig:stefan_solidification_speed}
\end{figure}

Eruptions are initiated when the pressure in a subsurface reservoir of liquid water
exceeds the tensile strength of the overlying ice shell, leading to
fracture formation \citep{lesage2020cryomagma, lesage2022simulation}.
The propagation speed of these fractures must exceed the growth rate of
the ice layer solidifying on the reservoir walls, a process that can
be modeled using the Stefan problem \citep{Carslaw1959}. Ice growth
rates, derived from this model as a function of reservoir depth (\emph{cf.} Figure \ref{fig:stefan_solidification_speed}), are relatively
slow ($10^{-6}$--$10^{-5}~ \text{m/s}$ after several hours), while fracture
propagation speeds in ice range from 20 m/s to 400 m/s
\citep{parsons1987preliminary, dempsey1999scale, vogt2008speed}.
This disparity ensures that fractures can propagate from the reservoir
to the surface before being sealed by ice regrowth.

The energy required to create a fracture of depth $H$ and perimeter $p$
is estimated using the surface energy of ice,
$\gamma_{\text{ice}} \approx 80$ mJ m$^{-2}$
\citep{boinovich2014experimental}:
\begin{equation}
    E_{\text{frac}} \approx \gamma_{\text{ice}} H p
\end{equation}
For a typical reservoir at $H = 10$ km depth and $p = 200$ m,
$E_{\text{frac}} \approx 1.6 \times 10^5$ J. However, the elastic
energy stored in the compressed reservoir,
$E_{\text{comp}} = \frac{1}{2} \chi_{H_2O,\text{(l)}} V_0 (\Delta P)^2$, is
significantly larger (\textit{i.e.} we found $E_{\text{comp}} =10^{11} -10^{15}$~J),
due to the high overpressure ($\Delta P \approx 257$ bar) and large
reservoir volumes ($10^6$--$10^{10}$ m$^3$). This energy imbalance not
only supports the plausibility of fracturing but also suggests that
the resulting conduits may form complex, potentially fractal networks
rather than idealized cylindrical pathways \citep{guerra2012understanding}.

\subsection{Pressure Drops in the Volcanic Conduit}
\label{ssec:pressure_dynamics}

\citet{lesage2020cryomagma,lesage2022simulation} developed a cryovolcanic eruption model based on a high-pressure liquid reservoir. However, these authors mainly focused on the efficiency of the proposed mechanism. The average ascent speeds of the liquid ($U$) they estimated are very high, \emph{i.e.}, several tens of meters per second at the beginning of the eruption. These values clearly demonstrate the great efficiency of the described mechanism, but it is difficult to imagine the formation of Europa’s smooth plains with such a violent event, as such high ascent speeds imply short‑lived, energetic eruptions more consistent with jets or localized flows rather than the slow, widespread emplacement of low‑viscosity cryolava required to form extensive smooth plains \citep{lesage2020cryomagma,fagents2003considerations}.

In their work, \cite{lesage2020cryomagma} took into account the friction of the fluid with the conduit walls in turbulent flow using the shear stress $\tau_w$, given by:

\begin{equation}
    \tau_w = f_\tau \frac{1}{2} \rho U^2
    \label{eq:shear_stress_lesage}
\end{equation}

where the friction factor $f_\tau$ in turbulent flow is set to 0.01 \citep{bird2007transport}. This term $\tau_w$ represents an approach to include the pressure loss experienced by the fluid during its ascent.

In ideal cases where friction is negligible, when a fluid is transported through a conduit, Bernoulli's principle \citep{bernoulli_hydrodynamica_1738} applies. Otherwise, in situations where friction is no longer negligible, Bernoulli's principle must be corrected by introducing these pressure losses. This well known principle is based on the conservation of energy, and this pressure loss term represents an energy dissipation leading to a loss of pressure. We discuss here the different possible sources of pressure losses in Europa's volcanic conduits.

To account for these losses, a corrective term must be added to Bernoulli's principle, with the total pressure loss $\Delta P_\text{PL}$ (in Pa) along a conduit given by:
\begin{equation}
    \Delta P_\text{PL} = \rho g \Delta H_\text{PL}
    \label{eq:pressureloss}
\end{equation}

where $\rho$ is the density (in kg m$^{-3}$), $g$ is the acceleration due to gravity (in m s$^{-2}$), and $\Delta H_\text{PL}$ is the equivalent head loss of the liquid (in m). These pressure losses are often divided into two categories: regular pressure losses and singular pressure losses. In flows occurring in conduits without obstacles, and for moderate values of $\Delta H_\text{PL}$, regular pressure losses can be represented by the Darcy-Weisbach equation:

\begin{equation}
    \Delta H_{\text{PL, regul}} = f \frac{L}{D_h} \frac{\bar{v}^2}{2g}
\end{equation}

where $f$ is the loss coefficient, $L$ is the length of the conduit (in m), $\bar{v}$ is the average fluid velocity (in m s$^{-1}$), and $D_h$ is the hydraulic diameter (in m), itself defined by:

\begin{equation}
    D_h = \frac{4A}{p}
\end{equation}

with $A$ being the cross-sectional area (in m$^2$) and $p$ the wetted perimeter (in m). In laminar flows, $f$ follows:

\begin{equation}
    f = \frac{64}{\text{Re}}
    \label{eq:darcy}
\end{equation}

where Re is the Reynolds number. In turbulent flow, the Colebrook equation \citep{colebrook1939} applies:

\begin{equation}
    \frac{1}{\sqrt{f}} = -2 \log_{10} \left( \frac{2.51}{\text{Re} \sqrt{f}} + \frac{\epsilon}{D_h} \frac{1}{3.7} \right)
    \label{eq:colebrook}
\end{equation}

\begin{figure}
    \centering
    \includegraphics[width=0.7\linewidth]{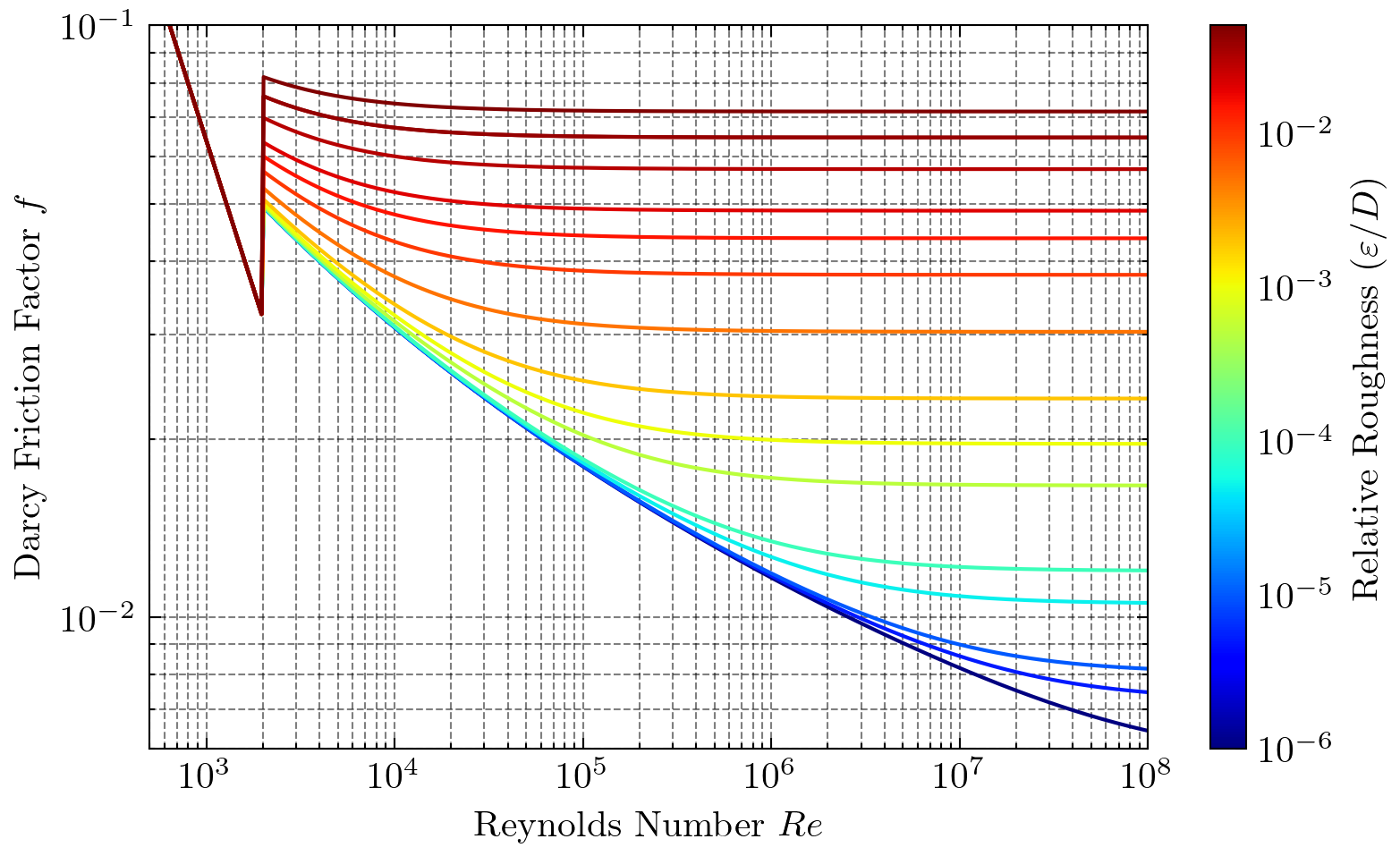}
    \caption[Moody Diagram]{Darcy friction factor $f$ as a function of Reynolds number $Re$ for various relative roughness values $\varepsilon/D$. In the laminar regime ($Re < 2{,}000$), $f$ is given analytically by~\eqref{eq:darcy} and is independent of wall roughness. The sharp discontinuity near $Re \approx 2{,}000$ marks the laminar-to-turbulent transition, beyond which $f$ is obtained by numerically solving the Colebrook equation~\eqref{eq:colebrook}.}
    \label{fig:moodydiagram}
\end{figure}

Here, the ratio $\epsilon/D$ represents the relative roughness of the conduit, \emph{i.e.}, the ratio between the roughness $\epsilon$ (in m) representing the average height of asperities on the conduit surface, and its diameter. Since this equation has no analytical solution, it is solved numerically. In Figure~\ref{fig:moodydiagram}, the solutions of equations \eqref{eq:darcy} and \eqref{eq:colebrook} are represented as a function of the Reynolds number $Re$ for different roughness values. This figure, called the Moody diagram, highlights greater pressure losses for rough surfaces compared to smooth surfaces.

From \eqref{eq:pressureloss} and \eqref{eq:darcy}, it is possible to directly relate the regular pressure loss to its average volumetric kinetic energy:

\begin{equation}
    \Delta P_\text{PL} = \left(f \frac{L}{D_h}\right) \frac{1}{2} \rho \bar{v}^2
    \label{eq:regularpressuredrop}
\end{equation}

\begin{figure}
    \centering
    \includegraphics[width=0.95\linewidth]{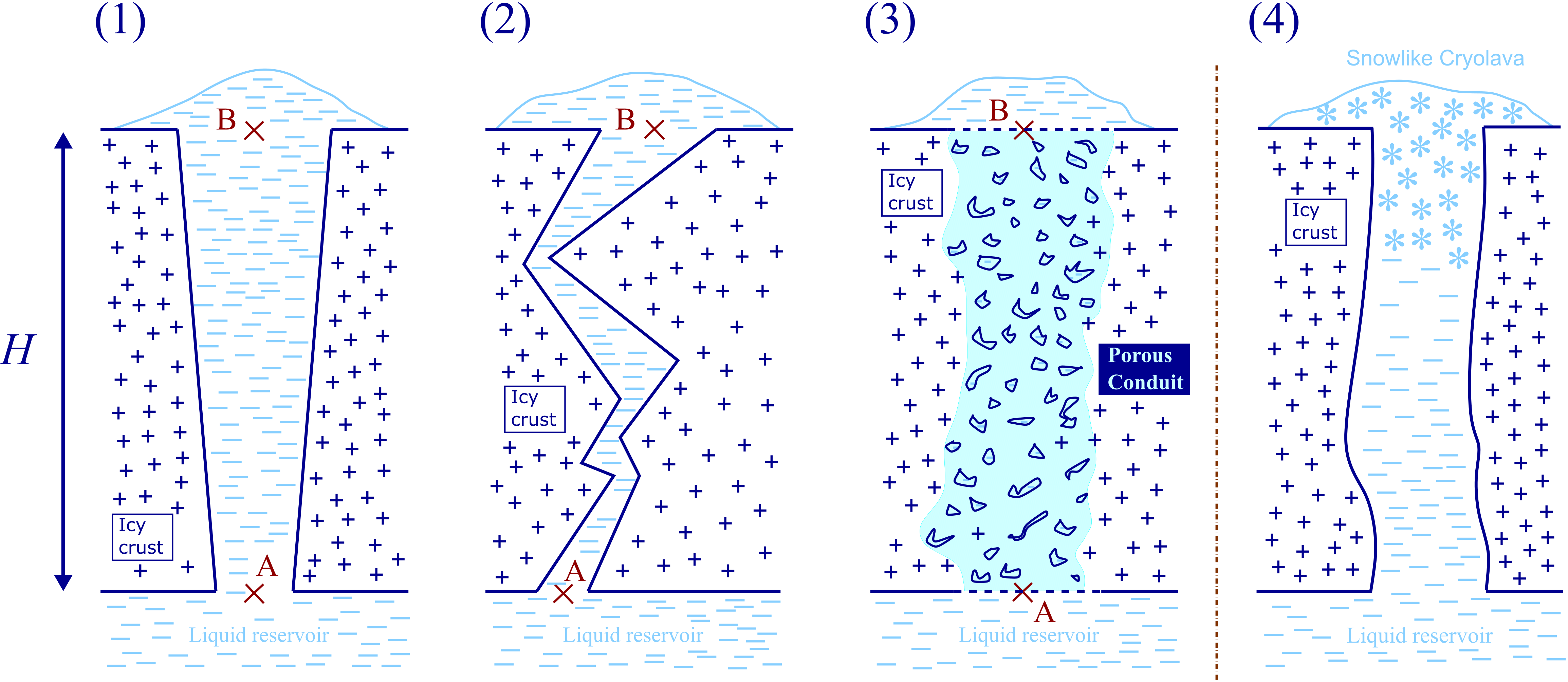}
    \caption[Vertical cross-sectional view of three fracture geometries through Europa's crust]{Three possible fracture geometries through Europa's crust, presented in a vertical cross-sectional view. The ice is represented in dark blue with crosses, while the cryolava is indicated in light blue hatched lines. These fracture systems establish a physical connection between a pressurized liquid water reservoir beneath the moon's surface and the crust itself, in accordance with the scenario proposed by \cite{lesage2020cryomagma}. Each geometry corresponds to a specific hydrodynamic regime: case (1) shows a simple conduit without obstacles, case (2) presents a more complex geometry where the fluid encounters obstacles on its way to the surface, and case (3) represents porous fractures allowing the fluid to seep to the surface. In all cases, to facilitate reasoning, reference points are indicated: "A" at the top of the reservoir beneath the surface and "B" at the conduit outlet. The horizontal sections of the conduit, at A and B, are denoted as $S_A$ and $S_B$, respectively. The total depth $H$ considered by \cite{lesage2020cryomagma} ranges between 1 km and 10 km. Panel (4) outlines the scenario adopted in this work: the depressurization experienced by the liquid water, as it approaches the surface, rapidly triggers partial vaporization into a mixture of water vapor and ice grains, similar to snow. In panel (4), the conduit is generic and can represent any combination of the cases illustrated in the previous panels (1 to 3).}
    \label{fig:pressuredrop}
\end{figure}

This scenario is represented by Figure~\ref{fig:pressuredrop}-(1). This expression has a form similar to \eqref{eq:shear_stress_lesage}, used by \cite{lesage2020cryomagma}. Using their values, i.e., $A=200$ m$^2$ and $p=100$ m, and considering a relative roughness $\epsilon/D = 0.06$ (cf. Figure~\ref{fig:moodydiagram}), the obtained $fL/D_h$ terms are 4 to 5 orders of magnitude greater than the friction coefficient $f_\tau$ considered by \cite{lesage2020cryomagma}. This represents a drastic reduction in eruption speed, which could become less than 1 m/s (\emph{cf.} \cite{lesage2020cryomagma}, Eq. (25)).

When the conduit between the reservoir and the surface has only a few bends as obstacles, as illustrated by case (2) in Figure~\ref{fig:pressuredrop}, the regular pressure loss must be accompanied by a singular loss \citep{Ito1960,CROOKS1997123}. This type of pressure loss can be expressed as a fraction of the volumetric kinetic energy:

\begin{equation}
    \Delta P_{\text{PL, singul}} = \left( \sum_{i=1}^{N_{\text{sing}}} \zeta_i \right) \frac{1}{2} \rho \bar{v}^2
    \label{eq:singpressure}
\end{equation}

where each $\zeta_i$ is an empirical parameter characterizing an obstacle encountered by the fluid. The number $N_\text{sing}$ of obstacles encountered during the ascent is unknown; however, since $\zeta_i$ parameter of an individual obstacle can have a value that can be several tens of percent, or even greater than unity, the total contribution of the singular pressure drop is likely much greater than that corresponding to the coefficient $f_\tau$ used by \cite{lesage2020cryomagma}.

A purely geometric effect can influence the effusion speed at point B (see Figure~\ref{eq:pressureloss}), i.e., the ratio of the horizontal cross-sectional areas $S_B/S_A$. We can show that the speed $v_B$ at the chimney can be written, in the domain of large values $S_B/S_A$:

\begin{equation}
    v_B = \frac{1}{\rho \frac{S_B}{S_A}} \left[ \frac{16 \eta H}{D_h^2} + \sqrt{\left( \frac{16 \eta H}{D_h^2} \right)^2 - 2 \rho \Delta P_c} \right]
\end{equation}

For simplicity, the pressure drop was assumed to be regular, with laminar flow. Adopting the notation of \cite{lesage2020cryomagma}, $\eta$ is the viscosity of the fluid, $H$ is the height difference between $A$ and $B$, and $\Delta P_c$ is the critical overpressure at $B$. This expression shows that $v_B$ tends to $0$ as the ratio $S_B/S_A$ becomes arbitrarily large. This shows another argument in favor of a strong modulation of the effusion speed $v_B$. Obviously, a ratio $S_B/S_A \ll 1$ has the opposite effect by accelerating the fluid at the conduit outlet.

An extreme case is when the liquid passes through a porous medium to reach the surface (see Figure~3, case 3). In such a situation, the total liquid flow rate $Q$ (m$^3$ s$^{-1}$) at the surface follows the well-known Darcy's law \citep{darcy1856fontaines}:

\begin{equation}
    Q = \frac{kA}{\eta L} \Delta P
\end{equation}

where $k$ is the permeability of the medium (in m$^2$), $A$ is the cross-sectional area of the conduit (in m$^2$), $\eta$ is the dynamic viscosity (in Pa s), $L$ is the length of the conduit (in m), and $\Delta P$ represents the pressure difference (in Pa) between the top and bottom for a vertical medium. The hydraulic permeability is related to the hydraulic conductivity $K$ (in m s$^{-1}$), which characterizes the porous medium, with $K$ defined by:

\begin{equation}
    K = \frac{k \rho g}{\eta}
\end{equation}

where $\rho$ is the density of the liquid (in kg m$^{-3}$) and $g$ is the local acceleration due to gravity (in m s$^{-2}$). For example, the hydraulic conductivity $K$ of fractured rocks or metamorphic rocks ranges over a wide range. Indeed, in the literature, we find $8 \times 10^{-9} \leq K \leq 3 \times 10^{-4}$ m s$^{-1}$ \citep{menzies2018past}. Considering Earth's gravity, this leads to $8.2 \times 10^{-16} \leq k \leq 3.1 \times 10^{-11}$ m$^2$. Using typical values from \citet{lesage2020cryomagma,lesage2022simulation}, \emph{i.e.}, $\Delta P = \Delta P_c = 2.57 \times 10^7$ Pa and $A = 1 \times 100$ m$^2$, this results in an extremely low eruption speed between $2.1 \times 10^{-9}$ m s$^{-1}$ and $8 \times 10^{-5}$ m s$^{-1}$. These values may seem unrealistic, but when considering a high value of hydraulic conductivity $K_{\text{max}} = 2 \times 10^{-2}$ m s$^{-1}$, it follows that $Q_{\text{max}} \approx 0.52$ m$^3$ s$^{-1}$, corresponding to a maximum speed $v_{\text{max}} = Q/A \approx 5.2$ mm s$^{-1}$.

In conclusion, the speed of the fluid exiting the volcanic chimney, driven by a mechanism such as that proposed by \citet{lesage2020cryomagma}, could range from zero to several tens of meters per second. However, this discussion adopted the implicit assumption of a liquid state throughout the ascent. This is a strong assumption, and several physical processes could occur: phase transitions through the formation of water vapor or ice grains, ice deposition on the conduit walls, etc. The subject of the next section is dedicated to these processes.

\subsection{Phase Transitions Near the Surface}
\label{ssec:phase_transitions}

As cryolava approaches Europa's surface, it encounters a near-vacuum
environment, with surface pressures as low as 1--10~µPa
\citep{hall1995detection, roth2021stable}. These conditions are far
below the triple point of water (611.657 Pa;
\citet{guildner1976vapor,murphy2005review}), causing rapid depressurization and phase
transitions.
This phenomenon drives rapid transformation of the liquid water.
Since the fluid's ascent velocity is much lower than the speed of
sound in the liquid (~1402.5 m/s at 0°C; \citet{chen1977speed}),
the entire volume of liquid water undergoes near-instantaneous
depressurization, triggering explosive nucleation of water vapor
bubbles and ice grains. This process is analogous to terrestrial
phenomena such as Loss of Coolant Accidents (LOCA) in nuclear reactors
\citep{bartak1990study,elias1993flashing} and industrial pressure
shift freezing (PSF) \citep{li2022influence}, where sudden
pressure drops induce phase transitions. The final properties of
the erupted material depend on the interplay between
thermodynamics and flow dynamics during ascent.

The evaporation flux of water molecules, $n_{\text{H}_2\text{O}}$,
is governed by the Knudsen equation \citep{eames_etal_1997}:
\begin{equation}
    n_{\text{H}_2\text{O}} = \frac{P_s(T_{\text{H}_2\text{O}})}
    {\sqrt{2 \pi m_{\text{H}_2\text{O}} k_B T_{\text{H}_2\text{O}}}}
\end{equation}
with $m_{\text{H}_2\text{O}}$ is the mass of one single water molecule, $k_B$ the Boltzmann constant, $P_s$ is the saturation vapor pressure at the temperature $T_{\text{H}_2\text{O}}$.
Therefore, the related latent flux can be expressed as:
\begin{equation}
    \Phi_{\text{evap,H}_2\text{O}} = \epsilon_{\text{evap}} \frac{n_{\text{H}_2\text{O}}}{N_A} \Delta H_{\text{vap,H}_2\text{O}}
    \label{eq:evap_h2o}
\end{equation}

where $\Delta H_{\text{vap,H}_2\text{O}} = 48.86 \, \text{kJ/mol}$ is
the enthalpy of vaporization, $N_A$ is Avogadro's number, and
$\epsilon_{\text{evap}}$ is a correction factor for non-ideal
effects such as collisions between molecules (typically
$\epsilon_{\text{evap}} \approx 10^{-2}$, \citet{eames_etal_1997}).
At the temperature $T_{\text{H}_2\text{O}} = 273.15 \, \text{K}$,
\eqref{eq:evap_h2o} gives:
\begin{equation}
    \Phi_{\text{evap,H}_2\text{O}} \approx 18.64 \, \text{kW/m}^2
\end{equation}

Subsequently, after evaporation caused by sudden depressurization, the resulting water vapor largely escapes into space.  The evaporation energy loss cools the adjacent liquid water.  A portion of this water freezes into ice crystals that are then deposited onto Europa's surface, along with the liquid portion and some entrained vapor.  The resulting clastogenic cryolava is therefore only partially solid. 
The mass fraction of water converted into ice,
$x_{\text{Ih-ice}}$, can be estimated from energy conservation, such that
\begin{equation}
    x_{\text{Ih-ice}} = \frac{\Delta H_{\text{vap,H}_2\text{O}}}
    {\Delta H_{\text{sol,H}_2\text{O}} + c_p \Delta T +
    \Delta H_{\text{vap,H}_2\text{O}}}
    \label{eq:mass_fraction}
\end{equation}
For pure water, this yields $x_{\text{Ih-ice}} \sim 0.82$, resulting
in a mixture of vapor and ice grains analogous to snow
(Figure~\ref{fig:pressuredrop}, panel 4).


{Within the framework of the scenario explored here, depressurized water undergoes two phase transitions: vaporization, which in turn promotes the formation of ice crystals through highly efficient cooling of the remaining liquid. If vaporization does not occur, the water remains in a 
metastable liquid state. However, such metastability is considered unlikely, as multiple 
context-dependent factors are expected to perturb the system sufficiently to rapidly drive it 
out of this state. In particular, the presence of shock waves, dissolved gases, various impurities 
and ice debris can trigger phase transitions \citep{debenedetti_1997a}. During a mechanical rupture of the ice 
shell, the generation of shock waves and debris is especially plausible. The presence of dissolved 
gases is also likely, as they are observed in the plumes of Enceladus' geysers 
\citep{mitchell_etal_2024}, and endogenous CO$_2$ has been detected at Europa's surface 
\citep{villanueva_etal_2023}.

The degassing of dissolved gases induced by depressurization leads to the formation
of microbubbles that can act as nucleation sites for water. The formation of these
microbubbles of minor species, as well as those of water vapor, follows classical
nucleation theory \citep{brennen_1995}. The formation of a microbubble
in the absence of contact with a solid phase is called homogeneous nucleation.
In this case, the nucleation rate, {\it i.e.}, the number of bubbles $J_{\rm hom}$ formed
per unit volume and per unit time, can be written as

\begin{equation}
   J_{\rm hom} \propto \exp\left(-\frac{\Delta G^*_{\rm hom}}{k_B T}\right),
\end{equation}

with $\Delta G^*_{\rm hom}$ denoting the energy barrier that must be overcome to
reach the critical radius beyond which the bubble grows indefinitely. This barrier
is given by

\begin{equation}\label{Ebarr}
  \Delta G^*_{\rm hom} = \frac{16\pi \sigma^3}{3 (\Delta P)^2},
\end{equation}

$\sigma$ being the surface tension of liquid water and $\Delta P$ the
vapor--liquid pressure difference. In the present case, this difference is
essentially equal to the vapor pressure of water, as the liquid pressure is
nearly zero. An estimate using numerical values at 20$^{\circ}$C yields
$\Delta G^*_{\rm hom}/k_B T \simeq 3\times 10^{11}$, implying a virtually negligible
homogeneous nucleation rate $J_{\rm hom}$.\\

At locations where the liquid is in contact with a solid, here liquid water in
contact with ice (walls of the volcanic conduit or debris of various sizes), the
energy barrier given by Eq.~\eqref{Ebarr} is modified. Heterogeneous nucleation can be
modeled by introducing a factor $f(\theta_c)$ that depends on the contact angle
$\theta_c$ between the liquid and the solid, leading to

\begin{equation}
    \Delta G^*_{\rm het} = f(\theta_c)\,\Delta G^*_{\rm hom}.
\end{equation}
Following \citep[][Eq.~4.13, p.~79]{turnbull_1950,kashchiev_2000},
\begin{equation}
    f(\theta_c) = \frac{(2 + \cos \theta_c)\,(1 - \cos\theta_c)^2}{4}.
\end{equation}
For the ice--liquid water system, the contact angle $\theta_c$ is quasi-zero
\citep{ketcham_Hobbs_1969,knight_1971,cordier_etal_2024}, and for small contact angles
\begin{equation}
    f(\theta_c) \simeq \frac{3}{16}\,\theta_c^4.
\end{equation}
This demonstrates that $\Delta G^*_{\rm het}$ rapidly approaches zero for small
contact angles, potentially leading to a heterogenous nucleation rate $J_{\rm het} \gg 1$. 
The exact value depends on the number of available nucleation sites, which may be very 
large on conduit walls as well as in the form of debris of various sizes produced during 
ice-shell fracturing at the time of the eruption.\\
  Therefore, owing to the fracturing of the ice shell and the resulting abundance of ice debris, heterogeneous nucleation 
of bubbles of a minor species, or directly of water vapor, is strongly promoted. This process 
leads to widespread vaporization throughout the entire liquid volume, with the consequence of 
substantial ice crystal formation. As long as a significant overpressure persists within the 
reservoir, all these materials are transported toward the surface.
}


Considering the processes discussed above, we have adopted a
scenario in which the active fluid (\enquote{cryolava}), is
composed predominantly of ice crystals, with minor amounts of
residual water vapor occupying the inter-granular spaces and
progressively escaping into space. There is no liquid water. Although we have presented
several arguments supporting the formation of such a material,
the complexity of the physical processes leading to the development
of this system lies beyond the scope of the present study
\citep{broz_etal_2025}.

Understanding the thermal and rheological
properties of this \enquote{cryolava snow} is required for interpreting both past geological
activity and potential future observation.  These properties 
are discussed in the following section.

\section{Properties of Cryolava Snow: A Key to
Understanding Europa's Icy Volcanism}


\subsection{Cryolava Snow: Composition and Structure}

As stated above, cryolava snow on Europa is expected to be composed
primarily of fine ice grains and water vapor, formed
during the rapid depressurization of the cryomagmatic chamber while
reaching the surface.
This process, described in
Section~\ref{ssec:pressure_dynamics}, suggests that less than 20\% of the
original cryomagma mass vaporizes (\emph{cf.} Eq \ref{eq:mass_fraction}). Although this represents a limited mass 
fraction, the resulting vapor occupies a significantly larger volume than the 
solid phase at 273~K (where a gas phase density would be
$\sim1-2~\text{kg/m}^3$). This process might lead to the formation of a highly
porous material with a significantly reduced bulk density. While the exact 
density remains uncertain, values ranging from 500 to 800~kg/m$^3$ appear 
physically plausible. However, due to the scarcity of rheological data for
such materials, we adopt a density range of 400 to 600~kg/m$^3$ to remain 
consistent with available terrestrial snow experimental data 
\citep{stockli2000characteristics}, for which rheological properties are
well-documented.
The resulting porosity of this material, ranging from $\sim 40$\% to
$\sim 60$\%, plays a critical role in its thermal and
mechanical behavior, influencing how it spreads
across Europa's surface and retains heat.

\subsection{Thermal Properties of Cryolava Snow}

Snow, whether of terrestrial atmospheric origin or considered
in a cryovolcanic context, is under the form of hexagonal crystalline
ice (\ce{Ih}). In the water phase diagram, the Ih phase is the only
stable solid phase for pressures below $200~\text{MPa}$ and temperatures
ranging from $72~\text{K}$ to the melting point \citep{petrenko1999physics}.
Although other phases such as cubic ice \ce{Ic} or amorphous phases
may appear through condensation at low temperatures
($T<130~\text{K}$), they are metastable and undergo an irreversible
transition to the \ce{Ih} phase when heated, even slightly \citep{hobbs2010ice}.
During this study, we assimilate cryolava snow as a porous medium
made of ice \ce{Ih} grains. As we aim to model the cooling of cryolava flows, we therefore focus on the thermal properties of this porous ice medium.

\subsubsection{Thermal Behavior of Ice Ih}

The specific heat capacity of ice Ih (\(c_{\text{p,Ih}}\))
is strongly dependent on temperature, and several
empirical models have been proposed to describe this
relationship. These models typically express
\(c_{\text{p,Ih}}\) as a linear function of temperature:

\begin{equation}
    c_{\text{p,Ih}} = A + B \times T,
\end{equation}

where \(A\) (\(\text{J}\,\text{kg}^{-1}\,\text{K}^{-1}\))
and \(B\) (\(\text{J}\,\text{kg}^{-1}\,\text{K}^{-2}\))
are experimentally determined coefficients.

\cite{anderson_1976} first characterized this
relationship for terrestrial snow, with coefficients
\(A = 1.67\) and \(B = 0.133\) (converted from the original
values in \(\text{cal}\,\text{g}^{-1}\,\text{K}^{-1}\)),
though the model's validity is restricted to temperatures
\(T \geq 193\,\text{K}\). \cite{yen_1981} later introduced
a piecewise model for snow and water ice, expressed in
\(\text{J}\,\text{mol}^{-1}\,\text{K}^{-1}\), which
includes discontinuities -- most notably at
\(T = 150\,\text{K}\) -- while remaining within experimental
uncertainty bounds:

\begin{equation}
    c_{\text{p,Ih}} =
    \begin{cases}
        -0.8994 + 0.1710\,T & \text{for }15 < T < 95\,\text{K}, \\
        2.2841 + 0.1350\,T & \text{for }95 < T < 150\,\text{K}, \\
        2.7442 + 0.1282\,T & \text{otherwise}.
    \end{cases}
\end{equation}

Conversion to \(\text{J}\,\text{kg}^{-1}\,\text{K}^{-1}\) is
achieved using the molar mass of water
(\(M_{\text{H}_2\text{O}} = 18.015\,\text{g}\,\text{mol}^{-1}\))
\citep{prohaska_2022}. \cite{fukusako_1990} further
refined the coefficients for extended temperature ranges,
providing a model in \(\text{kJ}\,\text{kg}^{-1}\,\text{K}^{-1}\):

\begin{equation}
    c_{\text{p,Ih}} =
    \begin{cases}
        0.185 + 0.689 \times 10^{-2}\,T & \text{for }90 < T < 273.15\,\text{K}, \\
        0.895 + 10^{-2}\,T & \text{for }40 < T < 90\,\text{K}.
    \end{cases}
\end{equation}

\begin{figure}
    \centering
    \begin{subfigure}[t]{0.45\linewidth}
        \centering
        \includegraphics[height=5.8cm]{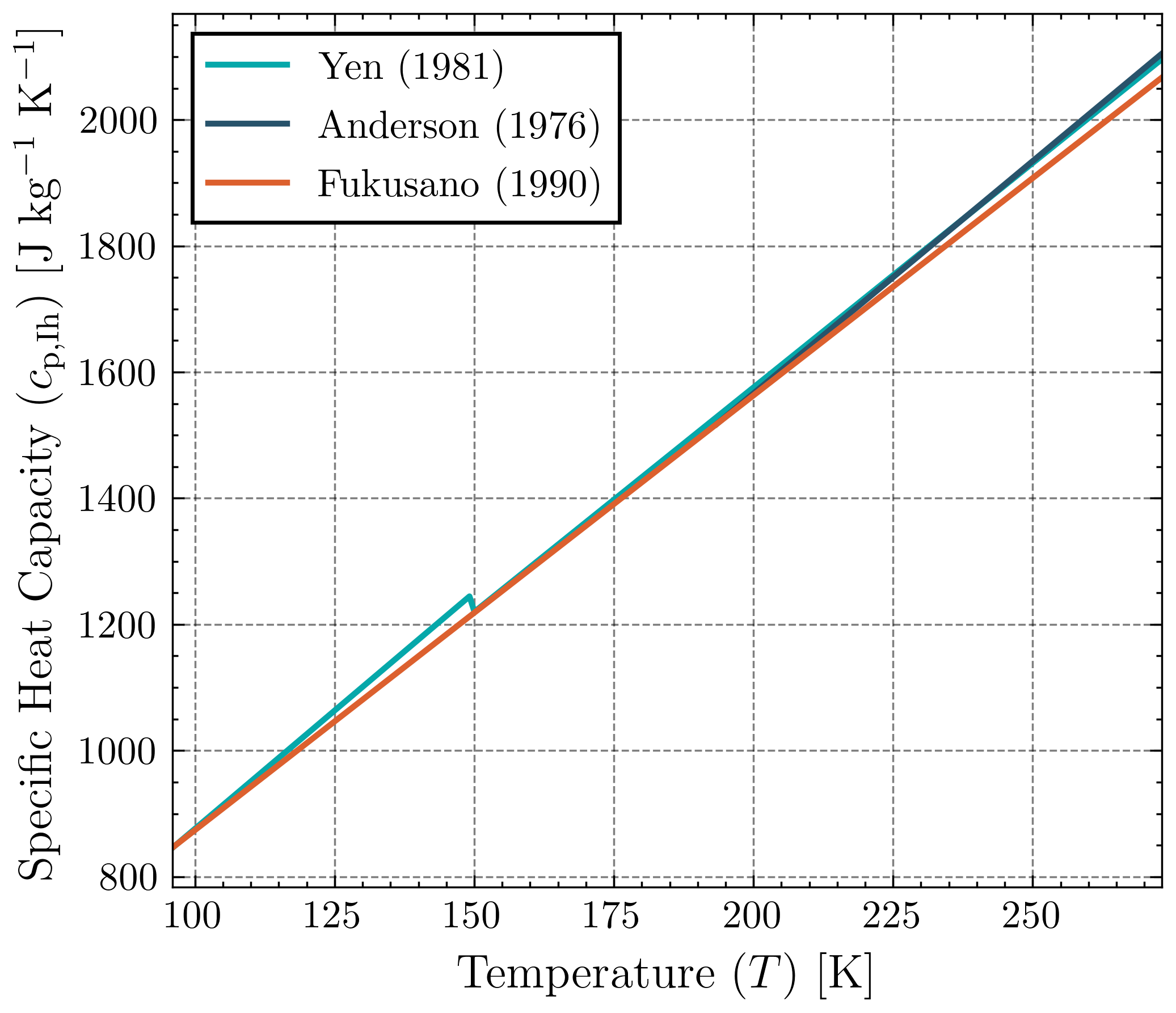}
        \caption{The variability of the specific heat capacity of Ice Ih with temperature. The difference between the different models being
            less than 3\%, the model of \cite{fukusako_1990} is adopted.}
        \label{fig:SpecificHeatCapacity}
    \end{subfigure}\hfil
    \begin{subfigure}[t]{0.45\linewidth}
        \centering
        \includegraphics[height=5.8cm]{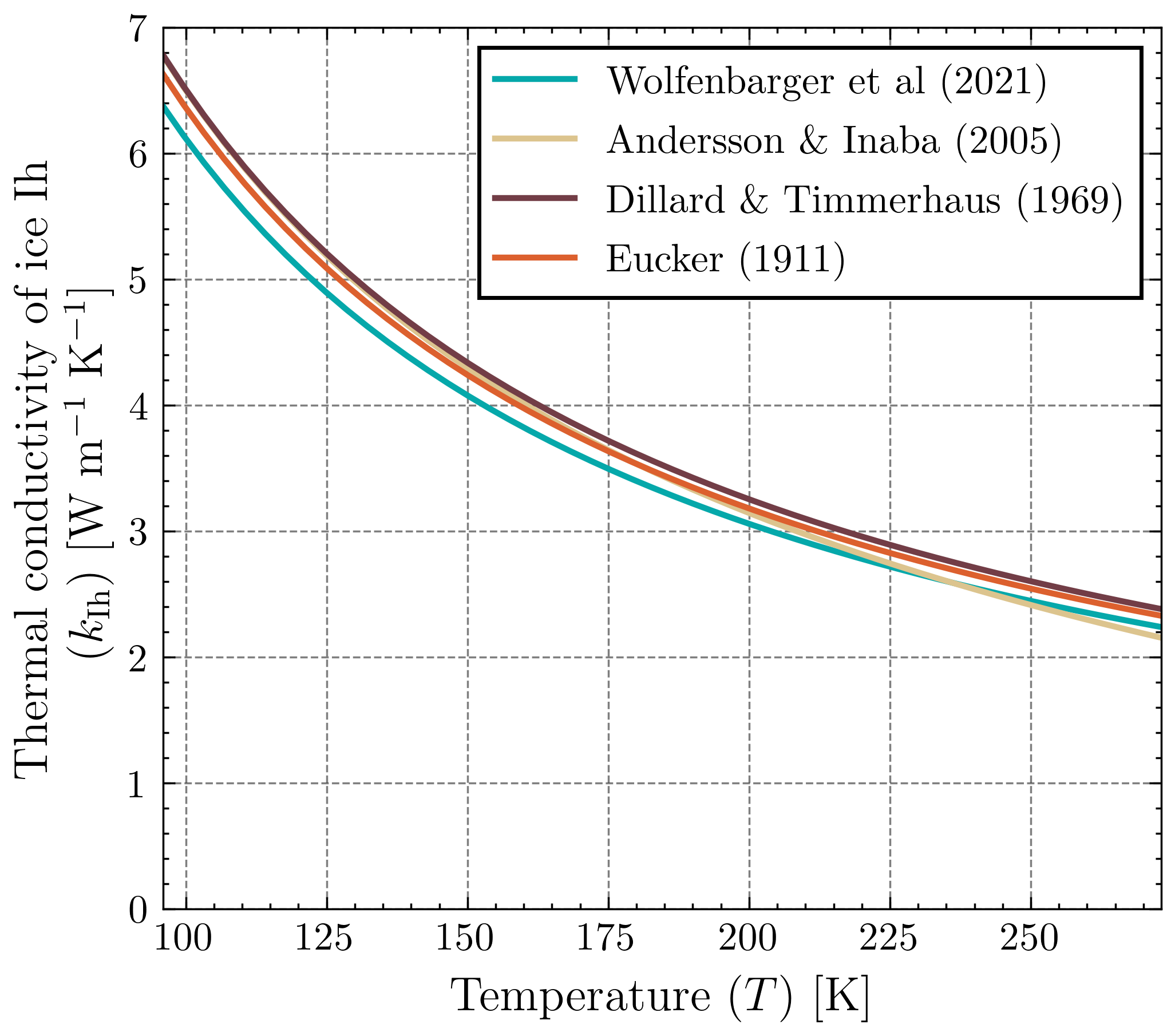}
        \caption{The variability of the thermal conductivity of Ice Ih with temperature. The relative deviation being less than 6\%, we adopt the model of \citep{wolfenbarger_etal_2021}}
        \label{fig:thermal_cond_ih}
    \end{subfigure}
    \caption{Thermal Properties of Ice Ih cryolava under Europa surface conditions ($96 - 273\text{ K}$).}
    \label{fig:thermal_props_ih}
\end{figure}

These models are compared in Fig.~\ref{fig:SpecificHeatCapacity}
over the temperature range relevant to Europa's surface conditions.

The thermal conductivity of ice Ih (\(k_{\text{Ih}}\)) was first
expressed as an inverse temperature dependence:

\begin{equation}
    k_{\text{Ih}}(T) = k_0 \times \frac{T_0}{T},
    \label{eq:k_eucken}
\end{equation}

with \(k_0 = 2.33\,\text{W}\,\text{m}^{-1}\,\text{K}^{-1}\) at
\(T_0 = 273\,\text{K}\) for \(T > 20\,\text{K}\)
\citep{eucken1911uber,Dean1963}. This reference value was
later updated to \(k_0 = 2.38\,\text{W}\,\text{m}^{-1}\,\text{K}^{-1}\)
for \(T > 80\,\text{K}\) \citep{DillardTimmerhaus1969}.

A polynomial extension was introduced by
\cite{andersson2005thermal}:

\begin{equation}
    k_{\text{Ih}}(T) = \frac{632}{T} + 0.38 + 1.97 \times 10^{-3}T,
\end{equation}

extending validity to \(T > 40\,\text{K}\). A comprehensive
meta-analysis by \citet{wolfenbarger_etal_2021}
further refined \(k_0\) to \(2.24\,\text{W}\,\text{m}^{-1}\,\text{K}^{-1}\).
These models, shown in Fig.~\ref{fig:thermal_cond_ih},
exhibit a maximum relative deviation of 6\%.

Therefore, we adopt these representations of $c_\text{p,Ih}$ and $k_\text{Ih}$ for modeling the thermal behavior of ice Ih.

\subsubsection{Effective Thermal Properties of Porous Snow}

Cryolava snow might not only be pure ice but a porous
mixture of ice and vapor (at least during the early stage of
the ascent). To account for this, we treat it as a
porous medium where the effective thermal properties
depend on both the ice and the vapor it contains. The
specific heat capacity of cryolava snow is
nearly identical to that of pure ice, as the
major part is carried out by the mass fraction of
ice~\citep{yen_1981,arenson2021physical}.

For thermal conductivity, we employ a porous model
that incorporates both density and temperature
dependencies. This model is based on the work of
\citet{Landauer1952} and accounts for the
interaction between the solid ice matrix and the vapor
within the pores:

\begin{equation}
    k_{\text{snow}} = \tfrac{1}{4}\Big[A + \sqrt{A^2 + 8k_{\text{ice}}k_{\text{vap}}}\;\Big],
\end{equation}
with \( A = k_{\text{vap}}(3\phi - 1) + k_{\text{ice}}(2 - 3\phi) \),
 $\phi$ represents the porosity, $k_{\text{vap}}$
is the thermal conductivity of water vapor, and
$k_{\text{ice}}$ is the thermal conductivity of ice. We note that we retrieve $k_\text{vap}$ and $k_\text{ice}$ when we replace $\phi$ by $1$ and $0$ respectively.

This approach provides a more realistic description
of heat transfer in cryolava snow compared to
traditional terrestrial snow models, which typically
only consider density dependence and neglect
temperature effects~\citep{yen_1981,sturm1997thermal,riche2013thermal}.
Our model shows that at Europa's surface
temperatures (90-100 K), cryolava snow with densities
of 400-600 kg/m$^3$ has a thermal conductivity of
0.1-0.5 W m$^{-1}$ K$^{-1}$, significantly lower than
that of pure ice (2-3 W m$^{-1}$ K$^{-1}$).

\begin{figure}
    \centering
    \begin{subfigure}{0.48\textwidth}
    \centering
    \includegraphics[width=0.9\linewidth]{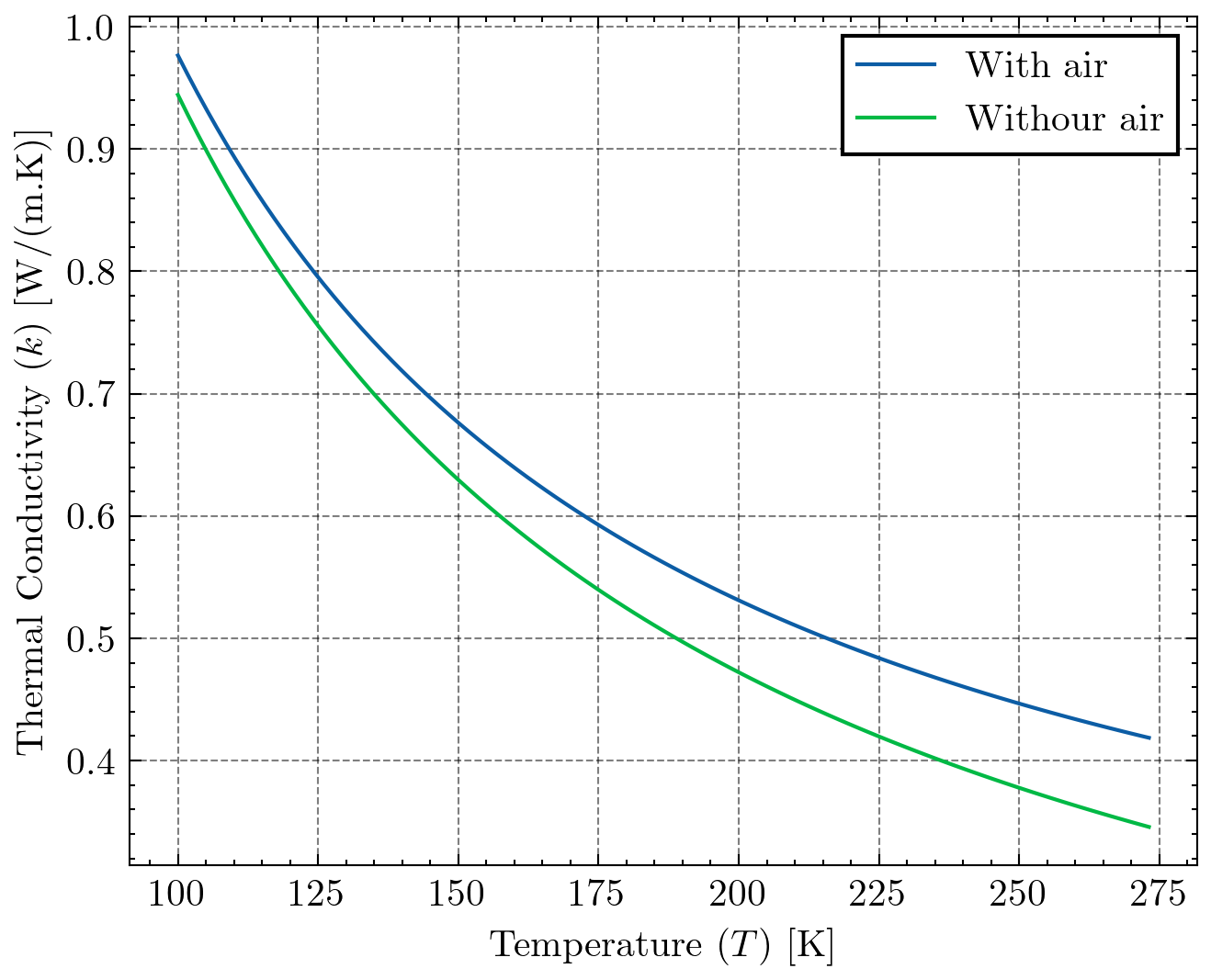}
    \caption{Snow with a bulk density of $400~\text{kg/m}^3$}
    \label{subfig:snow_400_thermalcond_comparison_air}
    \end{subfigure}\hfill
    \begin{subfigure}{0.48\textwidth}
    \centering
    \includegraphics[width=0.9\linewidth]{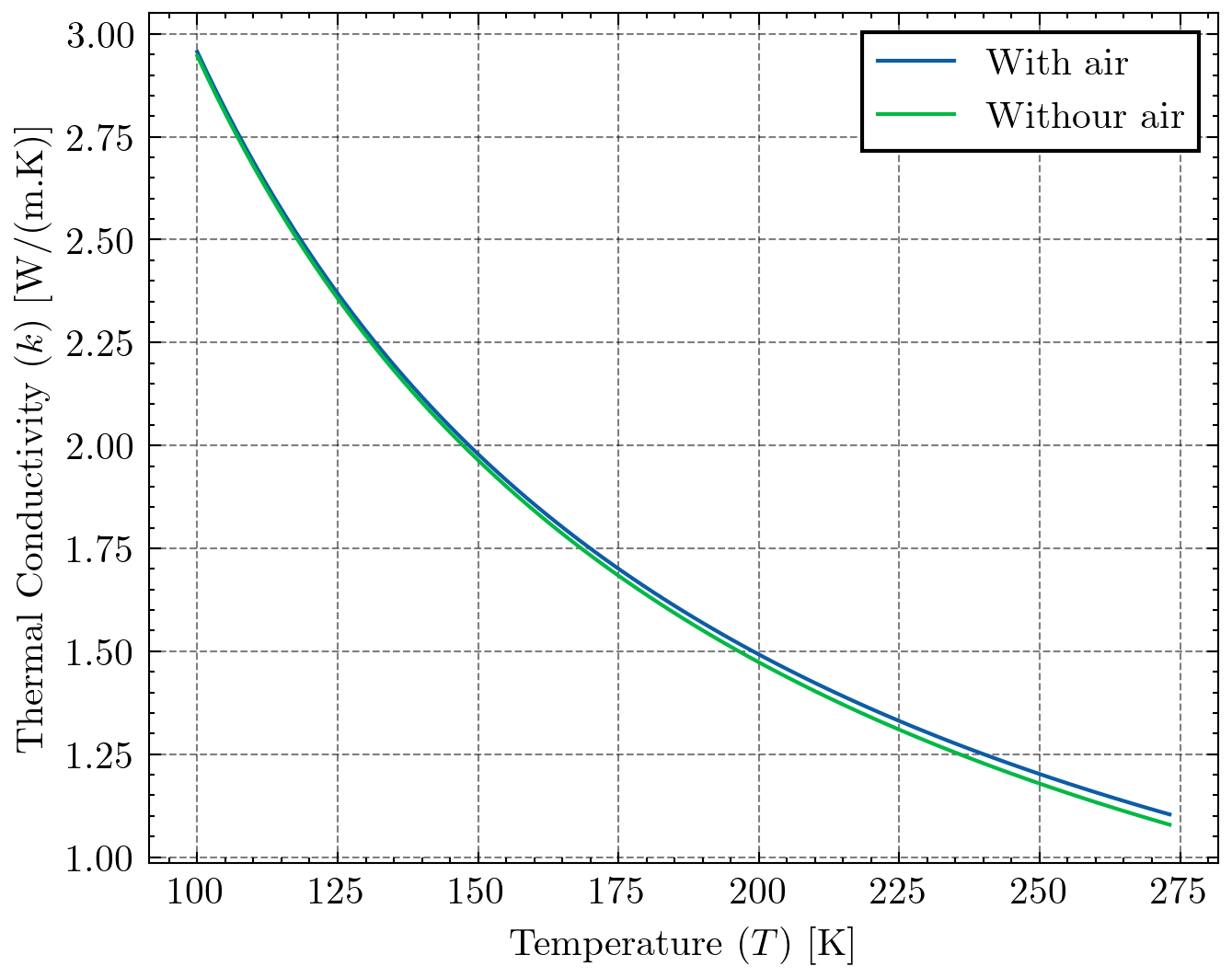}
    \caption{Snow with a bulk density of $600~\text{kg/m}^3$}
    \label{subfig:snow_600_thermalcond_comparison_air}
    \end{subfigure}
    \caption[Comparison of the thermal conductivities of snow with and without air for the densities concerned]{Comparison of the thermal conductivities of snow with the presence of air and without considering it in the apparent thermal conductivity of snow. We observe a variability similar to that of water vapor.}
    \label{fig:thermalcond_air_noair}
\end{figure}

At the melting temperature, the porous thermal model we propose shows an average difference
of 36\% compared to established snow models \citep{yen_1981,sturm1997thermal,riche2013thermal}.
Given that these reference models themselves exhibit an internal variability of 40\%, our results
demonstrate satisfactory conformity with field observations.

The impact of the thermal conductivity of air on the apparent
thermal conductivity of snow is shown in Figure~\ref{fig:thermalcond_air_noair}. 
Thermal conductivity undergoes a negligible variation for snow with $\rho=600$ kg/m$^3$ (2.25\% at most),
and a difference that can reach 17.5\% at 0\textdegree C for snow with
$\rho=400~\text{kg/m}^3$, which naturally has a higher porosity. These
differences are, moreover, similar to those observed by \citet{valovirta2004water},
who reported a variation of approximately 18\% associated with relative humidity values reaching up to 97\%.

Therefore, this model can be applied with confidence for modeling the thermal behavior of snow under Europa's surface conditions.

\subsection{Rheological Behavior of Cryolava Snow}

\subsubsection{Modeling Cryolava Snow as a Fluid}

To model the flow of cryolava snow, we adopt a
fluid-based approach as a first step. While
granular models could be used to describe individual
grain interactions, a fluid description provides a
practical framework for simulating large-scale
dynamics. This approach is justified by the
fluidization of cryolava snow, where water
vapor generated during eruption reduces
intergranular friction, allowing the mixture to flow
more like a viscous fluid. This effect is enhanced by
Europa's low gravity, where vapor drag can more easily
counteract the weight of the ice grains.

Therefore, as we assume the motion of cryolava snow to be modelled as
a fluid, it can be described by the Navier-Stokes equations:
\begin{align}
    \rho \frac{\mathrm{D} \vec{v}}{\mathrm{D} t} &= -\vec{\nabla} P + \rho \vec{g} + \vec{\nabla} \cdot \bar{\bar \tau} \label{eq:momentum} \\
    \frac{\mathrm{D} \rho}{\mathrm{D} t} &= -\rho \vec{\nabla} \cdot \vec{v} \label{eq:continuity}
\end{align}
with $\rho$ being the mass density ($\text{kg m}^{-3}$), 
$\frac{\mathrm{D}}{\mathrm{D} t} = \frac{\partial}{\partial t} 
+ \vec{v} \cdot \vec{\nabla}$ the material derivative ($\text{s}^{-1}$), 
$\vec{v}$ the flow velocity ($\text{m s}^{-1}$), $P$ the pressure 
($\text{Pa}$), $\vec{g}$ the gravitational acceleration 
($\text{m s}^{-2}$), $\bar{\bar \tau}$ the viscous stress 
tensor ($\text{Pa}$), and $\vec{\nabla}$ the nabla 
operator ($\text{m}^{-1}$).

For generalised fluids, the viscous stress tensor can be expressed as $\bar{\bar \tau} = 2\mu_\text{app} \bar{\bar{\dot{\varepsilon}}}$, with $\mu_\text{app}$ being the apparent viscosity
($\text{Pa s}$). Here, $\bar{\bar{\dot{\varepsilon}}} = \frac{1}{2}
\left[ \vec{\nabla} \vec{v} + (\vec{\nabla} \vec{v})^T \right]$ represents 
the strain rate tensor ($\text{s}^{-1}$) and $\dot{\varepsilon} 
= \sqrt{2 \bar{\bar{\dot{\varepsilon}}} : \bar{\bar{\dot{\varepsilon}}}}$ 
is the equivalent shear rate ($\text{s}^{-1}$).

To make our simulations, we use an SPH‑based numerical model (as described above; CryoPy\footnote{\url{https://github.com/bastien-bodin/CryoPy}}\footnote{\url{https://gitlab.univ-nantes.fr/icy-moons/src/cryopy}}) that includes thermal processes critical for modeling the cooling and solidification of cryolava flows.

\subsubsection{Rheological Models}
\label{sssec:rheologies}

\begin{figure}
    \centering
    \includegraphics[width=0.45\linewidth]{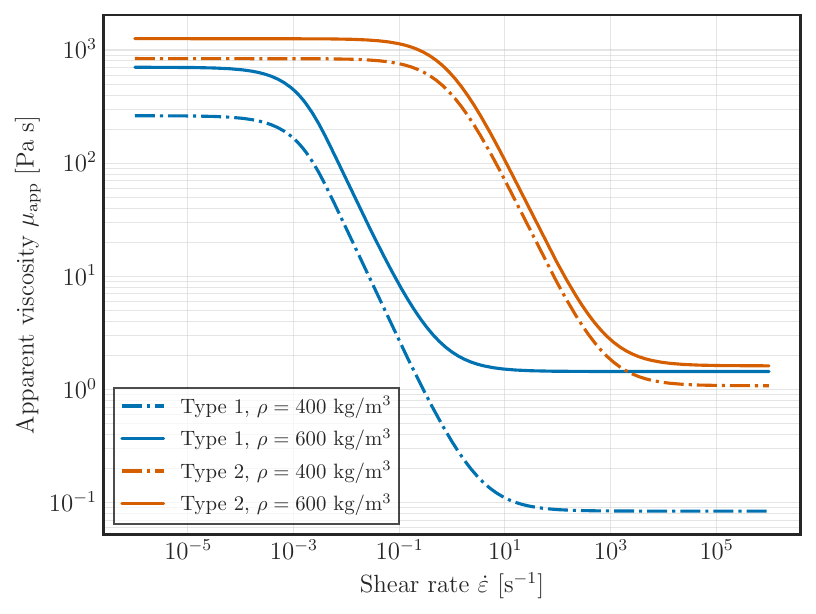}
    \caption{The rheologies considered in this work: the first type
    (called \enquote{Type 1}, in blue) based on laboratory measurements
    carried out on fluidized snow \citep{nishimura1996viscosity};
    the second type (labeled \enquote{Type 2},
    in orange) comes from works on snow sliding experiments
    along a 34~m long inclined (32°) plane~\citep{kern2004rheology}.}
    \label{fig:rheologies}
\end{figure}

We test two rheological models to describe the
behavior of cryolava snow -- the first one
based on laboratory experiments of fluidized snow (Type 1),
and the second derived from field data of avalanches of 
compact snow (Type 2). 
These two models are briefly detailed hereafter and illustrated in Figure~\ref{fig:rheologies}.

The Type 1 rheology for fluidized snow is given by
the Bingham-Papanastasiou Model~\citep{papanastasiou1987flows}:

\begin{equation}
    \mu_{\text{app}}(\dot{\varepsilon}) = \eta_0 +
    \frac{\tau_0}{\dot{\varepsilon}} (1 -
    \exp[-m\dot{\varepsilon}]),
\end{equation}

with $\eta_0$ (in Pa s) and $\tau_0$ (in Pa) depending on the 
total density $\rho$. Based on the snow fluidization experiments 
conducted by \cite{nishimura1996viscosity}, these parameters 
are defined by the following empirical relations:
\begin{equation}
    \begin{aligned}
        \eta_0(\rho) &= 2.88 \times 10^{-4} \exp(1.42 \times 10^{-2} \rho) \\
        \tau_0(\rho) &= 36.72 \times 10^{-3} \exp(4.92 \times 10^{-3} \rho)
    \end{aligned}
    \label{eq:Nishimura_laws}
\end{equation}
The parameter $m$, set to 1,000 based on the work of 
\cite{bilotta2016gpusph} on terrestrial lava flows, prevents the 
mathematical singularity present in the original Bingham model 
when $\dot \varepsilon \rightarrow 0$. Indeed, a first-order 
Taylor expansion shows that:
\begin{equation}
    \lim_{\dot \varepsilon \rightarrow 0} \mu_{app}(\dot \varepsilon) =
    \eta_0 + m \tau_0
\end{equation}

The Type 2 rheology for compact snow is given by a
simplified Cross
Model~\citep{kern2004rheology}:

\begin{equation}
    \mu_{\text{app}}(\dot{\varepsilon}) =
    \frac{\mu_0 + \mu_1 k_c \dot{\varepsilon}}
    {1 + k_c \dot{\varepsilon}},
\end{equation}

where the parameters $\mu_0$ and $\mu_1$ (in Pa s) depend on the 
bulk density of the snow $\rho$. These are defined linearly from 
the kinematic viscosity such that:
\begin{equation}
    \begin{aligned}
        \mu_0(\rho) &= \nu_0 \rho \quad \text{with} \quad \nu_0 = 2.1 
        \text{ m}^2 \text{ s}^{-1} \\
        \mu_1(\rho) &= \nu_1 \rho \quad \text{with} \quad \nu_1 = 
        2.7 \times 10^{-3} \text{ m}^2 \text{ s}^{-1}
    \end{aligned}
    \label{eq:Cross_params}
\end{equation}
The parameter $k_c = 1.1\text{ s}$ represents a time constant 
allowing the determination of the yield stress to initiate a 
viscosity regime transition.

The mobility index \citep{kargel1991rheological} is used to compare
driving and resistive forces:
\begin{equation}
    I = \log_{10}\left(\frac{\rho g}{\mu}\right) \label{eq:mobility_index}
\end{equation}

\begin{figure}
    \centering
    \includegraphics[width=0.45\linewidth]{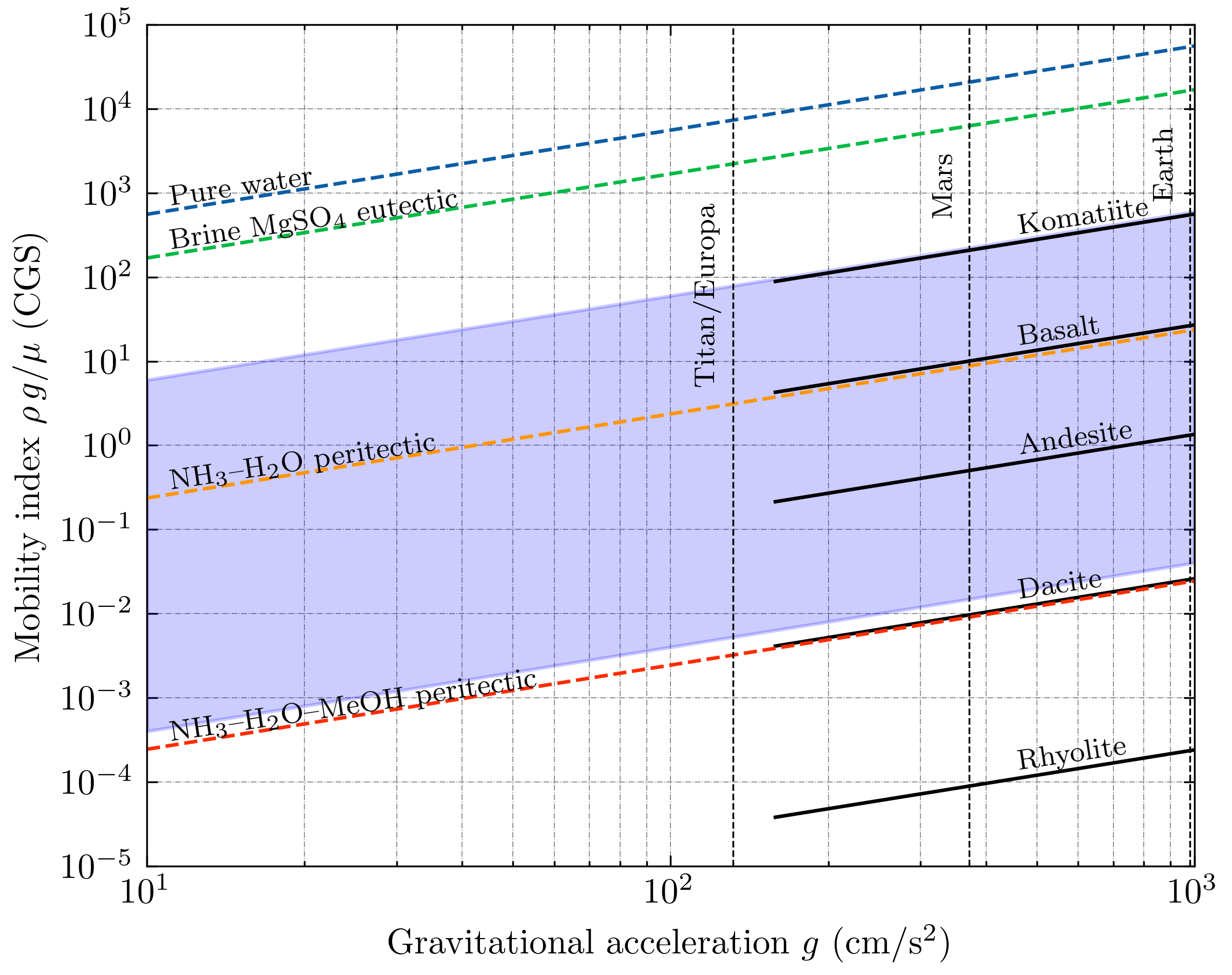}
    \caption{Mobility index of cryolava snow compared to terrestrial lavas. Vertical lines indicate gravitational accelerations for various celestial bodies. Data from \citep{kargel1994cryovolcanism}.}
    \label{fig:mobility_index}
\end{figure}

Figure~\ref{fig:mobility_index} exhibits mobility comparable to that of terrestrial komatiites and dacites, depending on the shear conditions. This
suggests that cryolava snow can be very mobile under shear, but behaves like
a solid at rest. The relaxation times suggest Type 2 flows could continue
for geological timescales, consistent with Europa's surface age estimates
of 30-70 million years \citep{zahnle2003cratering}; in practice, cooling, solidification or topography would eventually halt the motion.

On Europa, the low gravity and vapor production during eruption are expected to promote fluidization of the snowpack, motivating the use of a fluidized snow rheology (Type 1) alongside a compact snow model (Type 2).

\section{Results: Dynamics, Thermal Evolution, and Detectability of Cryolava on Europa}
\label{sec:results}

\subsection{Methodological Framework}
\begin{figure}
    \centering
    \includegraphics[width=0.45\linewidth]{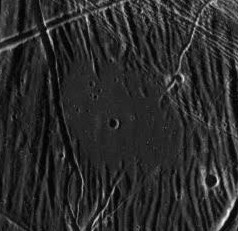}
    \caption{Smooth plain on Europa identified as a potential
    cryovolcano candidate. This image (C0374685452R) has been taken by Galileo's
    Solid-State Imaging instrument at a 30 m/px resolution. Credits: NASA/JPL/PDS-OPUS.}
    \label{fig:5452r_hydro}
\end{figure}

Smooth plains on Europa may be indicators of cryovolcanic activity \citep{head1998cryovolcanism}, and the present modelling focuses on one such feature imaged by Galileo SSI
(Figure~\ref{fig:5452r_hydro}), showing a $\sim 4\text{ km}$ diameter feature with an
estimated volume of $5 \times 10^7\text{ m}^3$ \citep{lesage2021constraints}.
We employ the \textit{CryoPy} SPH model to simulate cryolava flows,
considering two primary energy injection scenarios and two rheological
models (Bingham-Papanastasiou for fluidized snow and simplified Cross
for compact snow). Simulations were performed on the ROMEO 2018
supercomputer, using 28 cores.

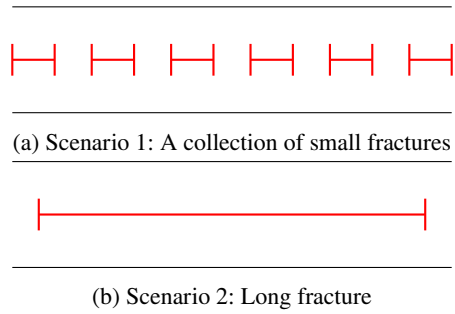
\begin{figure}
    \centering
    \begin{subfigure}{0.95\linewidth}
        \centering
        \begin{tikzpicture}[scale=0.7]
            \draw[black, thin] (0,0) -- (8.3,0);
            \draw[black, thin] (0,2) -- (8.3,2);
            \draw[red, thick] (0,1) -- (0.8,1);
            \draw[red, thick] (0,0.7) -- (0,1.3);
            \draw[red, thick] (0.8,0.7) -- (0.8,1.3);
            \draw[red, thick] (1.5,1) -- (2.3,1);
            \draw[red, thick] (1.5,0.7) -- (1.5,1.3);
            \draw[red, thick] (2.3,0.7) -- (2.3,1.3);
            \draw[red, thick] (3,1) -- (3.8,1);
            \draw[red, thick] (3,0.7) -- (3,1.3);
            \draw[red, thick] (3.8,0.7) -- (3.8,1.3);
            \draw[red, thick] (4.5,1) -- (5.3,1);
            \draw[red, thick] (4.5,0.7) -- (4.5,1.3);
            \draw[red, thick] (5.3,0.7) -- (5.3,1.3);
            \draw[red, thick] (6,1) -- (6.8,1);
            \draw[red, thick] (6,0.7) -- (6,1.3);
            \draw[red, thick] (6.8,0.7) -- (6.8,1.3);
            \draw[red, thick] (7.5,1) -- (8.3,1);
            \draw[red, thick] (7.5,0.7) -- (7.5,1.3);
            \draw[red, thick] (8.3,0.7) -- (8.3,1.3);
        \end{tikzpicture}
        \caption{Scenario 1: A collection of small fractures}
    \end{subfigure}
    \begin{subfigure}{0.95\linewidth}
        \centering
        \begin{tikzpicture}[scale=0.7]
            \draw[black, thin] (0,0) -- (8.3,0);
            \draw[black, thin] (0,2) -- (8.3,2);
            \draw[red, thick] (0.5,1) -- (7.8,1);
            \draw[red, thick] (0.5,0.7) -- (0.5,1.3);
            \draw[red, thick] (7.8,0.7) -- (7.8,1.3);
        \end{tikzpicture}
        \caption{Scenario 2: Long fracture}
    \end{subfigure}
    \caption{Scenarios of different fissure geometries considered
    for cryolava propagation.}
    \label{fig:scenario_fissures}
\end{figure}

Four fissure geometries were considered for cryolava propagation
(Figure~\ref{fig:scenario_fissures}), with a cross-sectional area
of $\mathcal{V^*}_\text{tot} = 16 \times 10^3~\text{m}^3/\text{m}$
\citep{lesage2021constraints}. As the simulations are conducted in
a 2D vertical plane, this volume is expressed per unit length in
the direction perpendicular to that plane, hence the m$^3$/m unit.

\subsection{Hydrodynamic Simulation Scenarios}
\label{subsec:hydro_scenarios}

To investigate the dynamics of cryolava flows on Europa, we designed
two primary simulation scenarios that capture distinct energy injection
mechanisms, as the final flow extent is governed by the dissipation
of the initial energy (whether potential or kinetic in nature). These
scenarios were implemented using the \textit{CryoPy}
SPH framework, optimized for Europa's environmental conditions
(gravity = 1.315 g, surface temperature = 96 K). Each scenario was
tested for four initial heights ($H_0$ = 4, 16, 32, 64 m) or maximum erupting
velocities ($v_\text{max} = 2.77, 5.54, 7.83, 11.1\text{ m/s}$), corresponding to the same amount of mechanical energy, and two
densities ($\rho$ = 400, 600 kg/m³), representing the range of cryolava
conditions inferred from Galileo's observations \citep{lesage2021constraints}.

\subsubsection{Scenario A: Gravitational Collapse}
\label{subsubsec:scenario_sa}

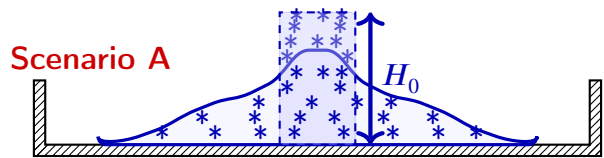
\begin{figure}
    \centering
    \begin{tikzpicture}[x=1cm,y=1cm,scale=0.5]
        \def\Xoff{1.5}
        \def\L{15}
        \def\Hwall{2.0}
        \def\Hbase{0.3}
        \def\Hpeak{2.5}
        \def\ybase{\Hbase}
        \def\Hrect{3.5}
        \def\xrect{5}
        \def\rectw{2}
        \draw[thick] (0,0) -- (\L,0) -- (\L,\Hwall) -- (\L-0.3,\Hwall) -- (\L-0.3,0.3) -- (0.3,0.3) -- (0.3,\Hwall) -- (0,\Hwall) -- cycle;
        \begin{scope}
            \clip (0,0) rectangle (\L,\Hwall);
            \fill[pattern=north east lines] (0,0) rectangle (0.3,\Hwall);
            \fill[pattern=north east lines] (\L-0.3,0) rectangle (\L,\Hwall);
            \fill[pattern=north east lines] (0,0) rectangle (\L,0.3);
        \end{scope}
        \draw[very thick,blue!70!black,rounded corners=3pt,fill=blue!10,fill opacity=0.3]
            plot[smooth,tension=0.8] coordinates {
                (0.2+\Xoff, \ybase)
                (1.5+\Xoff, \ybase+0.3)
                (3+\Xoff, \ybase+1.0)
                (4.5+\Xoff, \ybase+1.5)
                (6+\Xoff, \ybase+\Hpeak)
                (7.5+\Xoff, \ybase+1.5)
                (9+\Xoff, \ybase+1.0)
                (10.5+\Xoff, \ybase+0.3)
                (11.8+\Xoff, \ybase)
            } -- (\L - \Xoff - 0.2,\ybase) -- cycle;
        \foreach \i in {0.7,2.3}
            \foreach \j in {2.2}
                \node[blue!70!black] at (\i+\Xoff*4,\ybase+\j) {\large *};
        \foreach \i in {2.3,3.0,3.5}
            \foreach \j in {2.6}
                \node[blue!70!black] at (\i+\Xoff*3.0,\ybase+\j) {\large *};
        \foreach \i in {0.9,2.1}
            \foreach \j in {3.0}
                \node[blue!70!black] at (\i+\Xoff*4,\ybase+\j) {\large *};
        \foreach \i in {2.3,3.0,3.5}
            \foreach \j in {3.3}
                \node[blue!70!black] at (\i+\Xoff*3.1,\ybase+\j) {\large *};
        \draw[blue!50!white,fill=blue!20,opacity=0.4,rounded corners=2pt]
            (\xrect+\Xoff,\ybase) rectangle ++(\rectw,\Hrect);
        \draw[blue!70!black,dashed,thick]
            (\xrect+\Xoff,\ybase) rectangle ++(\rectw,\Hrect);
        \foreach \i in {0.7,2.1,3.3,4.2,5.5,6.8,8.0,9.}
            \foreach \j in {0.2}
                \node[blue!70!black] at (\i+\Xoff*1.8,\ybase+\j) {\large *};
        \foreach \i in {0.7,2.1,3.3,4.2,5.5,6.8}
            \foreach \j in {0.6}
                \node[blue!70!black] at (\i+\Xoff*2.5,\ybase+\j) {\large *};
        \foreach \i in {0.7,2.1,3.3,4.2}
            \foreach \j in {1.0}
                \node[blue!70!black] at (\i+\Xoff*3.5,\ybase+\j) {\large *};
        \foreach \i in {0.7,2.1}
            \foreach \j in {1.4}
                \node[blue!70!black] at (\i+\Xoff*4,\ybase+\j) {\large *};
        \foreach \i in {2.3,3.0,3.5}
            \foreach \j in {1.8}
                \node[blue!70!black] at (\i+\Xoff*3.1,\ybase+\j) {\large *};
        \draw[<->,ultra thick,blue!70!black]
            (\xrect+\rectw+0.4+\Xoff,\ybase) -- ++(0,\Hrect)
            node[midway,right] {\large $H_0$};
        \node[red!80!black,font=\bfseries\large] at (1.5,\Hwall+0.6) {Scenario A};
    \end{tikzpicture}
    \caption{Scenario A: Gravitational collapse of a cryolava column. The initial height $H_0$ represents the vertical extent of the cryolava column before collapse. The blue shaded area with asterisks (*) illustrates the spread of the collapsing cryolava material during the simulation.}
    \label{fig:scenario_a}
\end{figure}

In Scenario SA, the cryolava column collapses under Europa's
gravity, converting potential energy into kinetic energy. The
initial potential energy is given by:
\begin{equation}
    E_0 = \frac{\rho\, g\, \mathcal{V^*}\, H_0^2}{2} \label{eq:potential_energy}
\end{equation}
where $\rho$ is the cryolava density, $\mathcal{V^*}$ is the cross-sectional area ($16 \times 10^3$ m²/m), and $H_0$ is the initial height. This scenario is analogous to the sudden collapse and lateral movement of a lava column, where material is released laterally following the collapse.

Numerical simulations were conducted using the following parameters:
initial fluid column heights, \( H_0 \), ranged from 4, 16, 32,
to 64~\text{m}. Two bulk densities, \( \rho \), were considered:
400~\text{kg/m$^3$} and 600~\text{kg/m$^3$}. The time step was
adaptively adjusted according to the Courant-Friedrichs-Lewy (CFL)
stability condition to ensure numerical
convergence~\cite{courant1928partiellen}. Each simulation employed
approximately \( 10^5 \) SPH particles for domain discretization.
Dynamic boundary conditions \citep{cabrera2007boundary} were
implemented to model fluid-structure interactions.

\begin{table}
    \centering
    \begin{tabular}{c c c}
        \hline
        $H_0$ (m) & $\rho$ (kg/m³) & $E_0$ (MJ/m) \\
        \hline
        \multirow{2}{*}{4} & 400 & 16.8 \\
        \cline{2-3}
        & 600 & 25.3  \\
        \hline
        \multirow{2}{*}{16} & 400 & 67.3  \\
        \cline{2-3}
        & 600 & 101  \\
        \hline
        \multirow{2}{*}{32} & 400 & 135  \\
        \cline{2-3}
        & 600 & 203  \\
        \hline
        \multirow{2}{*}{64} & 400 & 269  \\
        \cline{2-3}
        & 600 & 404 \\
        \hline
    \end{tabular}
    \caption[Scenario A: Initial Potential Energy]{Initial potential energy $E_0$ per unit length for each combination of initial height $H_0$ and density $\rho$ considered in Scenario A.}
    \label{tab:scenario_sa_results}
\end{table}

\begin{figure*}
    \centering
    \includegraphics[width=0.95\linewidth]{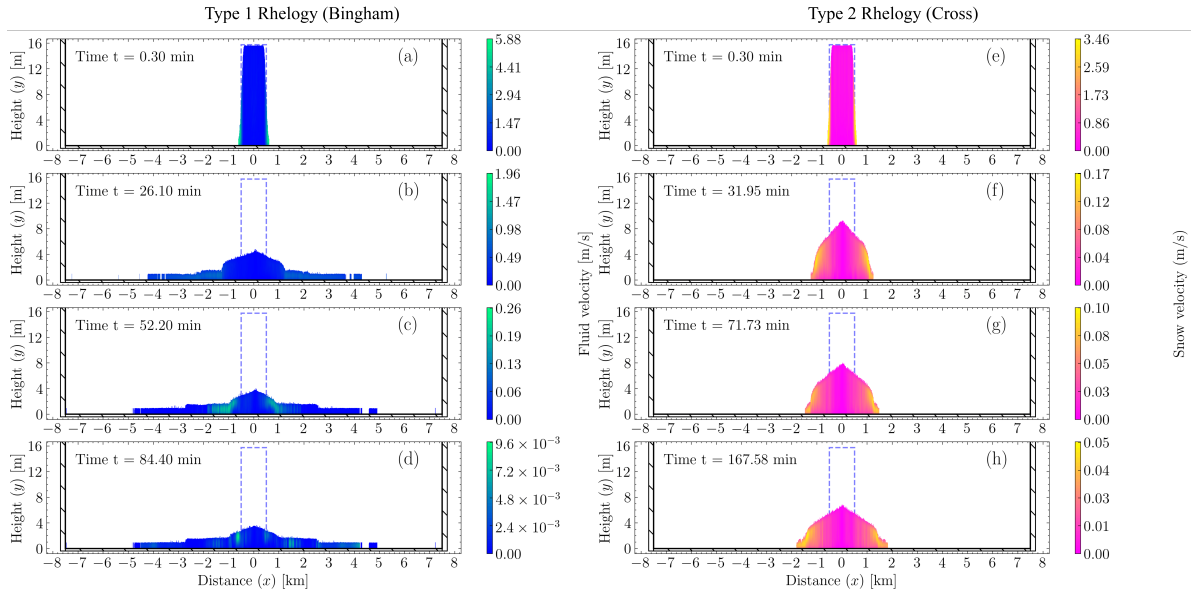}
    \caption[Scenario A: Snapshots]{Scenario A simulations for Type 1 (a--d, winter color palette) and Type 2 (e--h, summer color palette) rheologies with $\rho = 600$ kg/m³ and $H_0 = 16$ m.}
    \label{fig:results_scenario_sa}
\end{figure*}

Figure~\ref{fig:results_scenario_sa} illustrates the evolution of cryolava flows for both rheological models with an initial height of 16 m and a density of 600 kg/m³. The velocity of the extent ($v_\text{ext}$) follows a power-law relationship:

\begin{equation}
    v_\text{ext} = \alpha_\text{ext} t^{n_\text{ext}}
    \label{eq:extension_velocity}
\end{equation}

\begin{figure*}
    \centering
    \includegraphics[width=0.95\linewidth]{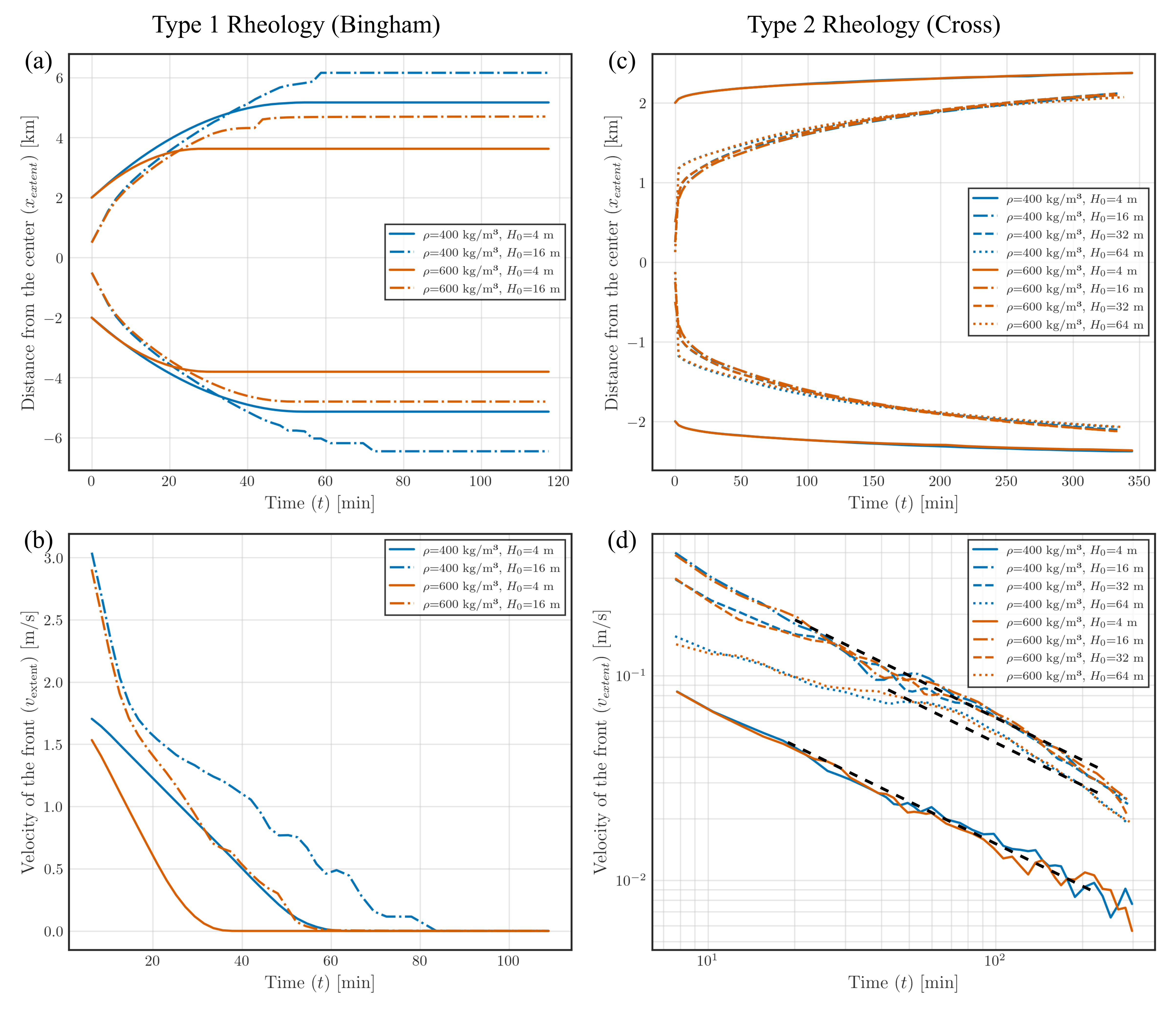} 
    \caption
    [Simulation results for scenario (SA)]
    {Horizontal extension \(x_{\text{extend}}\) and flow extension velocity of cryolava on Europa's surface, as part of scenario (SA) (column collapse). Only simulations where the fluid is not constrained by the simulation domain are plotted. Two rheologies are considered: Type 1 -- Bingham \citep{nishimura1996viscosity} and Type 2 -- Cross \citep{kern2004rheology}, each with two density values, \(400\) and \(600\) kg/m\(^3\). The power-law behavior of the Type 2 rheology is illustrated by the black dashed lines.}
    \label{fig:velocity_profiles}
\end{figure*}

For Type 2 flows, fitting Eq.~\eqref{eq:extension_velocity} to the simulation results yields $n_\text{ext} \approx -0.9$ with $\alpha_\text{ext}$ ranging from 3.7 to 5.5. This relationship allows estimation of flow durations for different distances:

\begin{equation}
    t = \left(\frac{x}{\alpha_\text{ext}}\right)^{1/n_\text{ext}}
    \label{eq:flow_duration}
\end{equation}

At distances of 1 km and 2 km, the calculated durations are so exceedingly long that the flow can effectively be considered stationary on geological or observational timescales.


The collapse of the cryolava column in Scenario A is driven solely by
gravitational potential energy, without any additional kinetic energy input.
The primary resistive forces acting on the flow are viscous dissipation and wall friction, the latter being handled through the dynamic boundary conditions \citep{cabrera2007boundary}, in which fixed boundary particles interact viscously with the fluid. For Type 2 flows, the higher apparent
viscosity plays a dominant role, effectively limiting the extent of flow
extension.

The calculated flow durations for Type 2 flows exceed the estimated surface
age of Europa, indicating that such flows would cease relatively quickly.
This analysis allows us to see that, in this scenario, it is possible to
recreate the smooth plains of Europa. Additionally, the power-law relationship
for extension velocity (\(v_\text{ext} \propto t^{-0.9}\)) provides a
quantitative means to relate flow morphology to emplacement timescales. These
results motivate further investigation into constant-flow eruption scenarios
and the role of shear stress in reducing the apparent viscosity
of the cryolava.


\subsection{Scenario B: Constant-Flow Eruption Dynamics}
\label{subsec:scenario_sb_results}

In Scenario~B, the cryolava is injected through a volcanic vent of half-width $R$ with a
parabolic velocity profile, providing a continuous supply of kinetic energy,
where $r$ denotes the radial distance from the vent center:
\begin{equation}
v_\text{erupt}(r) = v_\text{max}\left(1 - \frac{r^2}{R^2}\right),
\end{equation}
and the volumetric flow rate is given by:
\begin{equation}
Q_\mathcal{S} = \frac{4}{3} R v_\text{max}.
\end{equation}
The total kinetic energy of the system during the entire eruption $t_e$ is expressed as:
\begin{equation}
E_c = \frac{12}{35}\rho \mathcal{V^*} v_\text{max}^2,
\end{equation}
where \(\mathcal{V^*}\) is the cross-sectional volume per unit width
(\(16 \times 10^3~\text{m}^3/\text{m}\)). Setting $E_c = E_p = \frac{1}{2}\rho g \mathcal{V^*} H_0$ yields the equivalent maximum eruption velocity:
\begin{equation}
    v_\text{max} = \sqrt{\frac{35}{24} g H_0}.
\label{eq:kinetic_energy}
\end{equation}
This scenario contrasts with Scenario~A by introducing a sustained
energy input, which fundamentally alters the flow dynamics and morphological
evolution of the cryolava snow.

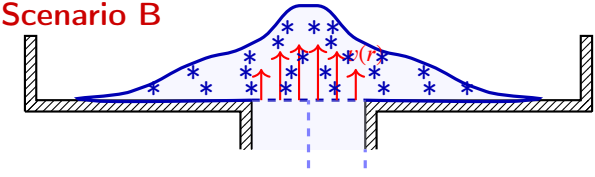
\begin{figure}
    \centering
    \begin{tikzpicture}[x=1cm,y=1cm, scale=0.5]
        \def\Xoff{1.5}
        \def\L{15}
        \def\R{1.5}
        \def\Hwall{2.0}
        \def\Hbase{0.3}
        \def\Hpeak{2.5}
        \def\ybase{\Hbase}
        \def\Hrect{3.5}
        \def\xrect{5}
        \def\rectw{2}
        \draw[thick] (0,0) -- (\L*0.5-\R-0.3,0) -- (\L*0.5-\R-0.3,-1.0);
        \draw[thick] (0,0) -- (0.0,\Hwall) -- (0.3,\Hwall) -- (0.3,0.3) -- (\L*0.5-\R,0.3) -- (\L*0.5-\R, -1.0);
        \draw[thick] (\L,0) -- (\L*0.5+\R+0.3,0) -- (\L*0.5+\R+0.3,-1.0);
        \draw[thick] (0\L,0) -- (0\L.0,\Hwall) -- (\L-0.3,\Hwall) -- (\L-0.3,0.3) -- (\L*0.5+\R,0.3) -- (\L*0.5+\R, -1.0);
        \begin{scope}
            \clip (0,-1.0) rectangle (\L,\Hwall);
            \fill[pattern=north east lines] (0,0) rectangle (0.3,\Hwall);
            \fill[pattern=north east lines] (\L-0.3,0) rectangle (\L,\Hwall);
            \fill[pattern=north east lines] (0,0) rectangle (\L*0.5-\R,0.3);
            \fill[pattern=north east lines] (\L*0.5-\R,-1.0) rectangle (\L*0.5-\R-0.3,0.0);
            \fill[pattern=north east lines] (\L*0.5+\R,0) rectangle (\L,0.3);
            \fill[pattern=north east lines] (\L*0.5+\R,-1.0) rectangle (\L*0.5+\R+0.3,0.0);
        \end{scope}
        \draw[very thick,blue!70!black,rounded corners=3pt,fill=blue!10,fill opacity=0.3]
            plot[smooth,tension=0.8] coordinates {
                (\L*0.5-\R, 0.3)
                (0.2+\Xoff, \ybase)
                (1.5+\Xoff, \ybase+0.3)
                (3+\Xoff, \ybase+1.0)
                (4.5+\Xoff, \ybase+1.5)
                (6+\Xoff, \ybase+\Hpeak)
                (7.5+\Xoff, \ybase+1.5)
                (9+\Xoff, \ybase+1.0)
                (10.5+\Xoff, \ybase+0.3)
                (11.8+\Xoff, \ybase)
                (\L*0.5+\R, 0.3)
            };
        \draw[very thick,blue!70!black, dashed] (\L*0.5-\R, 0.3) -- (\L*0.5+\R, 0.3);
        \fill[blue!10, opacity=0.3] (\L*0.5-\R,-1) rectangle (\L*0.5+\R,0.3);
        \draw[very thick,blue!50, dashed] (\L*0.5, -1.5) -- (\L*0.5, 0.3);
        \draw[very thick,blue!50, dashed] (\L*0.5+\R, -1.5) -- (\L*0.5+\R, -1.0);
        \foreach \x in {-1.25,-0.75,-0.25,0.25,0.75,1.25} {
            \pgfmathsetmacro{\yheight}{sqrt(1.5^2-((\x)^2))}
            \draw[red, thick, ->] (\x+\L*0.5,0+0.3) -- (\x+\L*0.5,\yheight+0.3);
        }
        \node[red] at (\L*0.5+\R, 1.5) {$v(r)$};
        \foreach \i in {0.7,2.1,3.3,4.2,5.5,6.8,8.0,9.}
            \foreach \j in {0.2}
                \node[blue!70!black] at (\i+\Xoff*1.8,\ybase+\j) {\large *};
        \foreach \i in {0.7,2.1,3.3,4.2,5.5,6.8}
            \foreach \j in {0.6}
                \node[blue!70!black] at (\i+\Xoff*2.5,\ybase+\j) {\large *};
        \foreach \i in {0.7,2.1,3.3,4.2}
            \foreach \j in {1.0}
                \node[blue!70!black] at (\i+\Xoff*3.5,\ybase+\j) {\large *};
        \foreach \i in {0.7,2.1}
            \foreach \j in {1.4}
                \node[blue!70!black] at (\i+\Xoff*4,\ybase+\j) {\large *};
        \foreach \i in {2.3,3.0,3.5}
            \foreach \j in {1.8}
                \node[blue!70!black] at (\i+\Xoff*3.1,\ybase+\j) {\large *};
        \node[red!80!black,font=\bfseries\large] at (1.5,\Hwall+0.6) {Scenario B};
    \end{tikzpicture}
    \caption{Scenario B: Constant-flow eruption with a parabolic
    velocity profile $v(r)$. The red arrows illustrate the
    velocity distribution at the vent, following the law
    $v(r) = v_\text{max}(1 - r^2/R^2)$.}
    \label{fig:scenario_b}
\end{figure}

\subsection*{Simulation Parameters}

The vent half-width (\(R\)) is set to 20 meters, that is in the estimated range of
the wideness done by \citet{craft2016fracturing}. Maximum velocities
(\(v_\text{max}\)) range from 2.77 to 11.1 m/s, as detailed in
Table~\ref{tab:conditions_initiale}. Eruption durations span from
\(10^3\) to \(10^5\) seconds to attain a total emitted volume
$\mathcal V^*$. The SPH resolution is defined at 0.5~meters
per particle. This value has been found to be a reasonable compromise between spatial resolution high enougth and not too long computation time.

\begin{table}
    \centering
    \begin{tabular}{c c}
        $H_0$ (m) & $v_{\text{max}}$ (m/s) \\
        \hline
        4  & 2.77 \\
        16 & 5.54 \\
        32 & 7.83 \\
        64 & 11.1 \\
    \end{tabular}
    \caption{Maximum eruption velocity $v_\text{max}$ for Scenario B,
    derived from the energy equivalence with Scenario A
    (Eq.~\ref{eq:kinetic_energy}). Note that $v_\text{max}$
    depends only on $H_0$ and not on density $\rho$.}
    \label{tab:conditions_scenario_b}
\end{table}

\subsubsection*{Flow Extension and Velocity Regimes}

The dynamics of cryolava flows on Europa reveal distinct behaviors
depending on their rheological type. For Type~1 flows, the
eruptive regime is marked by a nearly constant extension velocity,
enabling rapid and extensive spreading. The apparent viscosity
\(\mu_\text{app}\) decreases significantly under shear stress,
resulting in thin,
elongated flows whose propagation is limited more by Europa's terrain
than by internal resistance. As a demonstration of this result, even simulations with the
lowest kinetic energy quickly exit the domain boundaries, highlighting
the efficiency of shear-thinning in facilitating flow mobility.

\begin{figure*}
    \centering
    \includegraphics[width=0.95\linewidth]{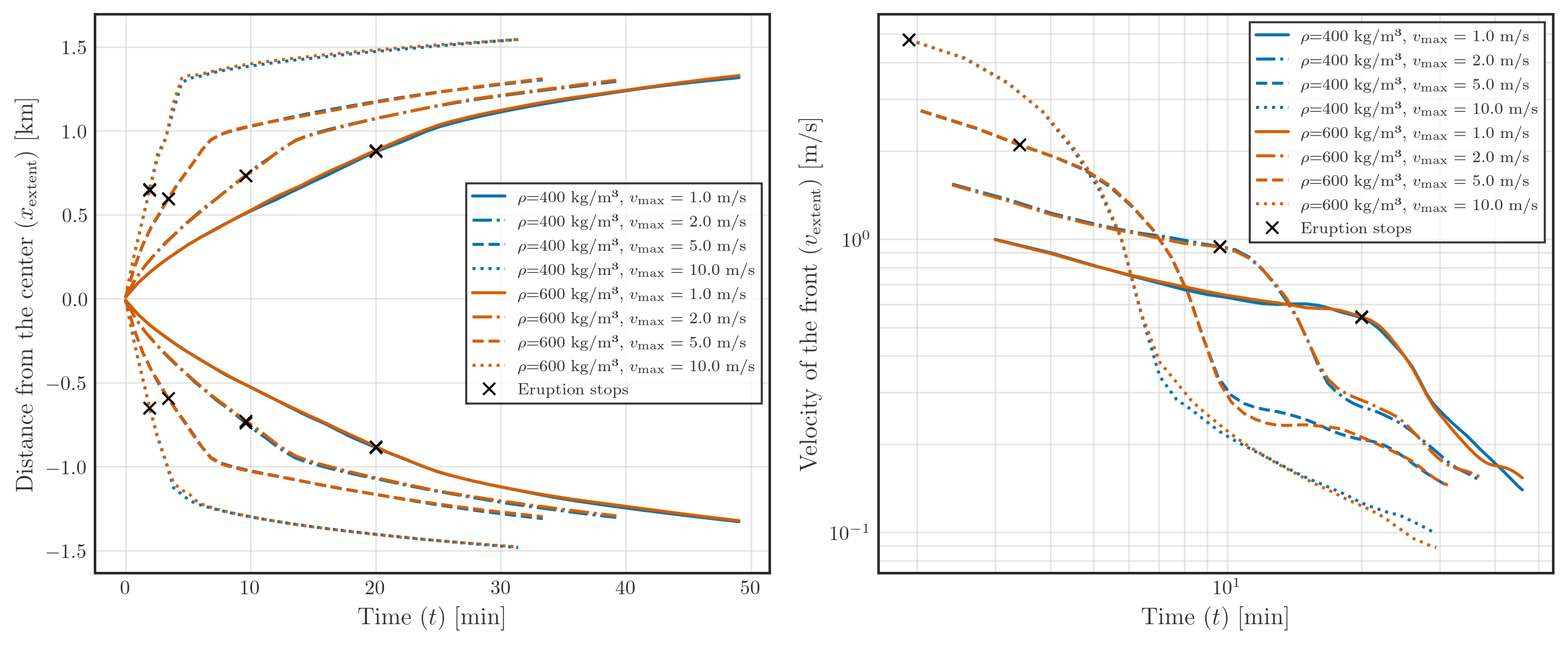}
    \caption{Extent (left) and spreading velocity (right) of Type~2 cryolava flow for
    maximum eruption velocities between 1~m/s (strait lines) and 10~m/s (dotted lines).}
    \label{fig:eruption_t2_extent_velocity}
\end{figure*}

Type~2 flows, however, exhibit two distinct velocity regimes.
During the eruptive phase, the flow extension velocity follows a
power-law relationship with time:
\begin{equation}
v_\text{extent}(t) \propto t^{\alpha},
\end{equation}
where \(\alpha\) varies between \(-1/3\) and \(-1/2\) depending on the
rheology and eruption conditions. The resulting flow thickness ranges
from 1 to 5~m. This behavior is illustrated in
Figure~\ref{fig:eruption_t2_extent_velocity}, which captures the abrupt
transition from the eruptive phase to relaxation.

In sum, the shear sensitivity
of the cryolava plays a crucial role in shaping flow morphology.
Type~1 rheology produces extensive, thin deposits constrained only by
terrain, while Type~2 rheology can independently replicate the
morphologies observed by Galileo. These findings underscore the need
for additional constraints and experimental data to refine our
understanding of cryolava composition and behavior.

The results of Scenario~B suggest that one single sustained eruption could
explain the formation of Europa's extensive smooth plains. The power-law
exponent \(\alpha\) during the eruptive regime emphasizes the influence
of eruption conditions, particularly the maximum velocity \(v_\text{max}\)
and vent radius \(R\). However, the current simulations assume a flat,
two-dimensional domain and homogeneous material properties. To improve
accuracy, future work should incorporate realistic 3D topography and
thermal effects, which are likely to refine predictions of flow extent
and morphology.

Ultimately, Scenario~B demonstrates that sustained kinetic
energy input is essential for modelling realistic cryolava flows
inferred from Europa's surface features.
These simulations demonstrated the need of new data and experiments
for constraining cryolava candidates, as show by 
the different behaviors of Type~1 and Type~2 rheologies.



\section{Determining the thermal profile of cryolava for
thermal anomaly detection with E-THEMIS}
\label{sec:thermal_profile_cryolava}

ESA's JUICE and NASA's Europa Clipper missions, launched in 2023 and
2024 respectively, will probably transform our understanding of
Europa's geology \citep{daubar2024planned}. While JUICE's Sub-millimeter Wave
Instrument (SWI) operates at 230--550~\textmu m \citep{esa_2011_juice_yellowbook}, Europa
Clipper's E-THEMIS \citep{christensen2024europa} covers 7--80~\textmu m, a range uniquely
suited to detect thermal anomalies from cryolava -- mixtures
of water, solutes, and ice grains constrained to
96--273~K~\cite{ashkenazy2019surface}. This study models the
thermal profile of cryolava to evaluate its detectability
using E-THEMIS, leveraging the instrument's Noise
Equivalent Delta Temperature (NEDT) and adaptive
operational modes.

\begin{figure}
    \centering
    \includegraphics[width=0.45\linewidth]{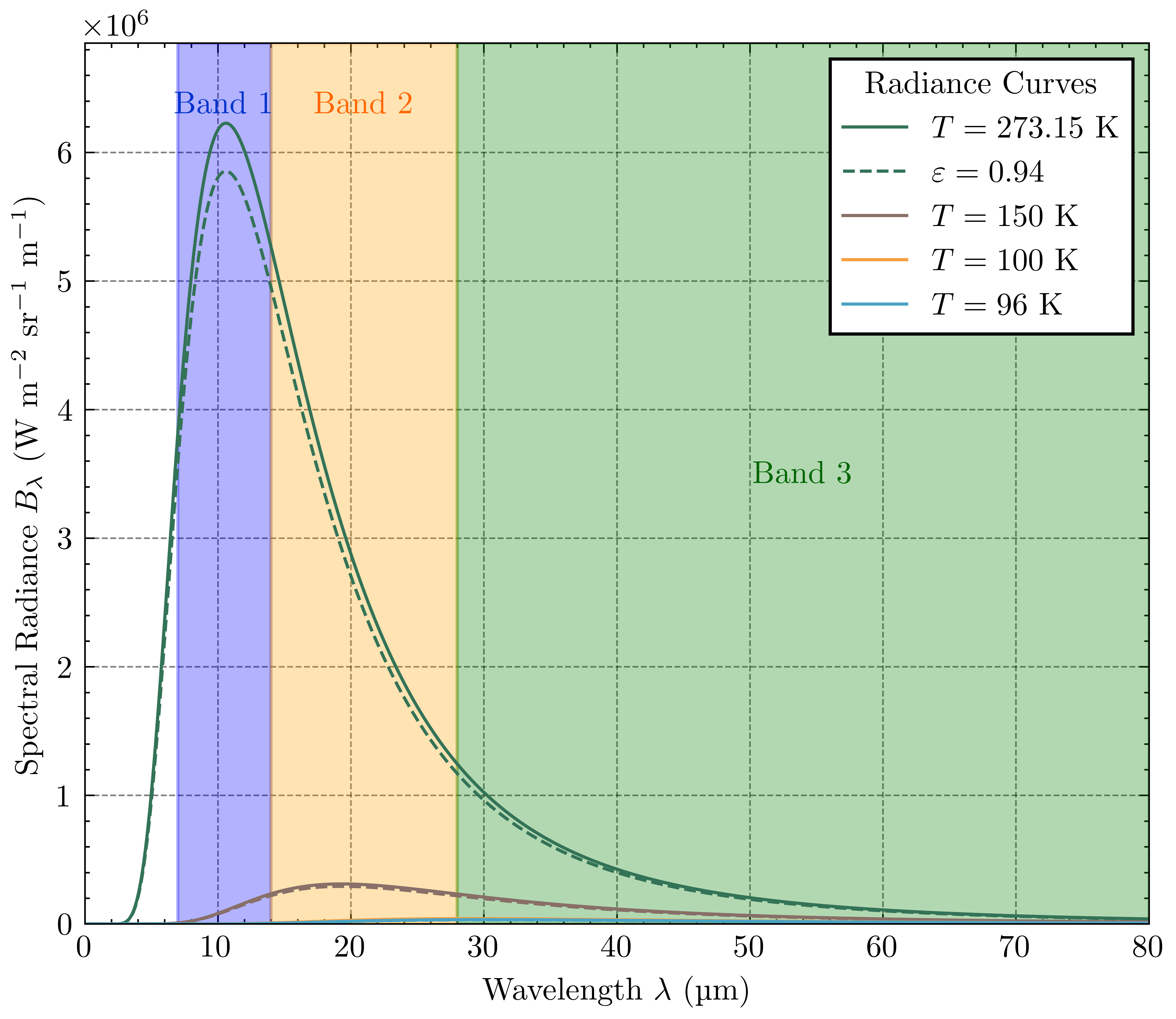}
    \caption{Spectral radiance curves for cryolava at 273~K
    and 96~K, compared to Europa's surface emissivity
    ($\varepsilon = 0.94$). E-THEMIS's bands (7--80~\textmu m)
    are highlighted, demonstrating its suitability for
    thermal anomaly detection.}
    \label{fig:radiance_spectrum}
\end{figure}

Given the wide range of snow emissivities, we adopt an emissivity for cryolava $\varepsilon \approx 0.94$ taken to be comparable to that inferred for Europa’s surface while allowing for variations associated with density, liquid water content, and contaminant load \citep{dozier1982effect, zhang2005influence}. Using Wien's law, peak emissions are
calculated at 10.61~\textmu m (273~K) and 30.19~\textmu m
(96~K), both within E-THEMIS's spectral range but beyond
SWI's sensitivity (Figure~\ref{fig:radiance_spectrum}).
This spectral alignment makes E-THEMIS the primary tool for
thermal anomaly detection during Europa Clipper's 53 planned flybys of Europa.

E-THEMIS's temperature resolution is defined by its NEDT,
derived as:
\begin{equation}
    \text{NEDT} = \frac{\text{NEP}}
    {\frac{\partial P}{\partial T}
    \sqrt{N_\text{TDI} N_\text{frames} N_\text{bining}}},
\end{equation}
where $\text{NEP} = 4.95 \times 10^{-12}$~W, and $P$ depends on
detector area ($1.85 \times 10^{-5}$~m$^2$), telescope
solid angle (0.66~sr), and operational
parameters~\cite{christensen2024europa}. Three modes optimize
resolution:
\begin{itemize}
    \item \textbf{Full-frame:} 920$\times$140 pixels at 60~fps,
    ideal for low-speed passes.
    \item \textbf{TDI (1--64 lines):} Line-by-line accumulation
    for high-speed flybys, improving signal-to-noise ratio
    (SNR) at the cost of temporal resolution.
    \item \textbf{Spatial aggregation:} Pixel binning
    (2$\times$2 to 5$\times$5) to balance data rate and
    radiometric precision.
\end{itemize}

\begin{figure*}
    \centering
    \begin{subfigure}[t]{0.33\linewidth}
        \centering
        \includegraphics[width=0.99\linewidth]{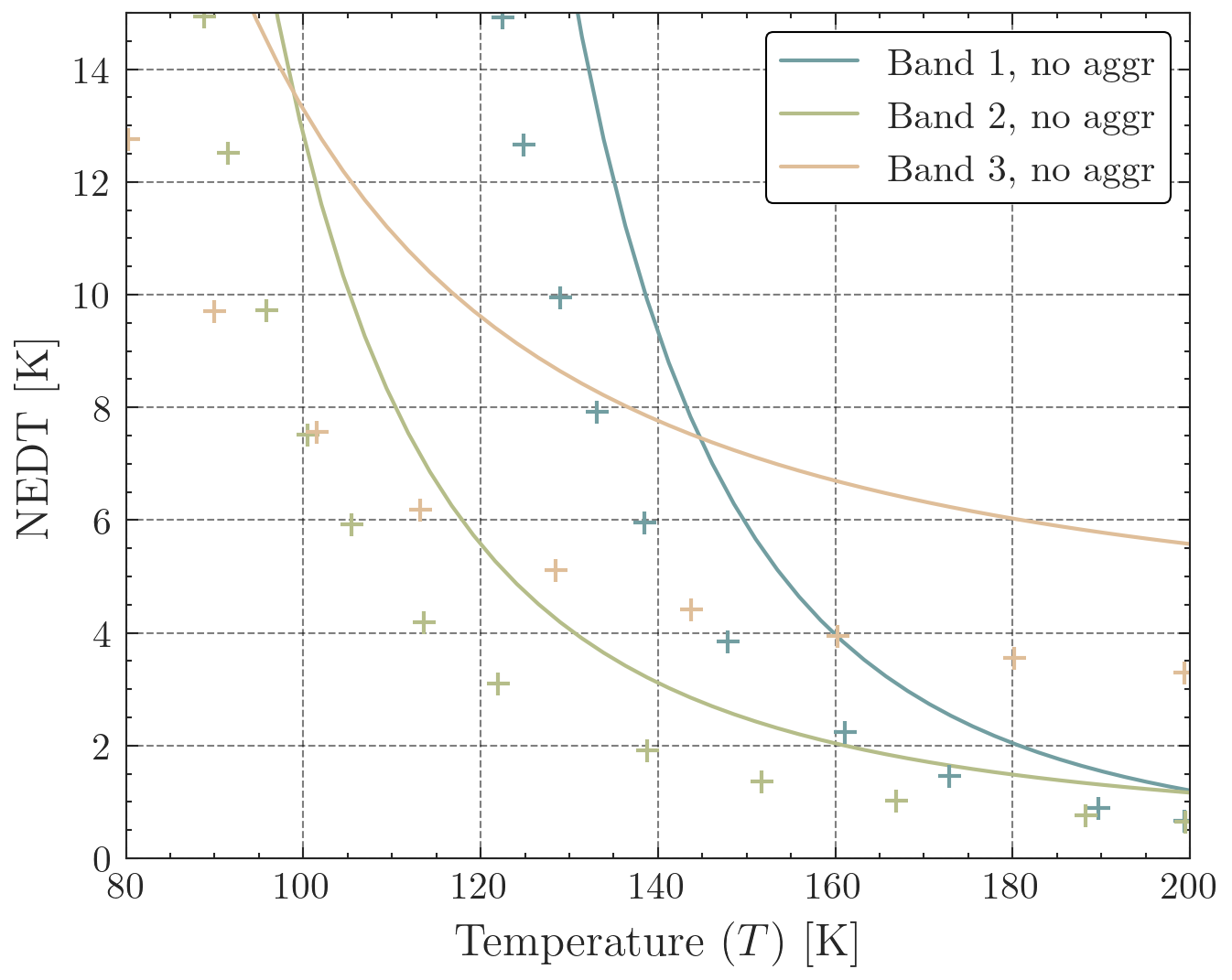}
        \caption{TDI=1, no spatial aggregation.}
    \end{subfigure}\hfill
    \begin{subfigure}[t]{0.33\linewidth}
        \centering
        \includegraphics[width=0.99\linewidth]{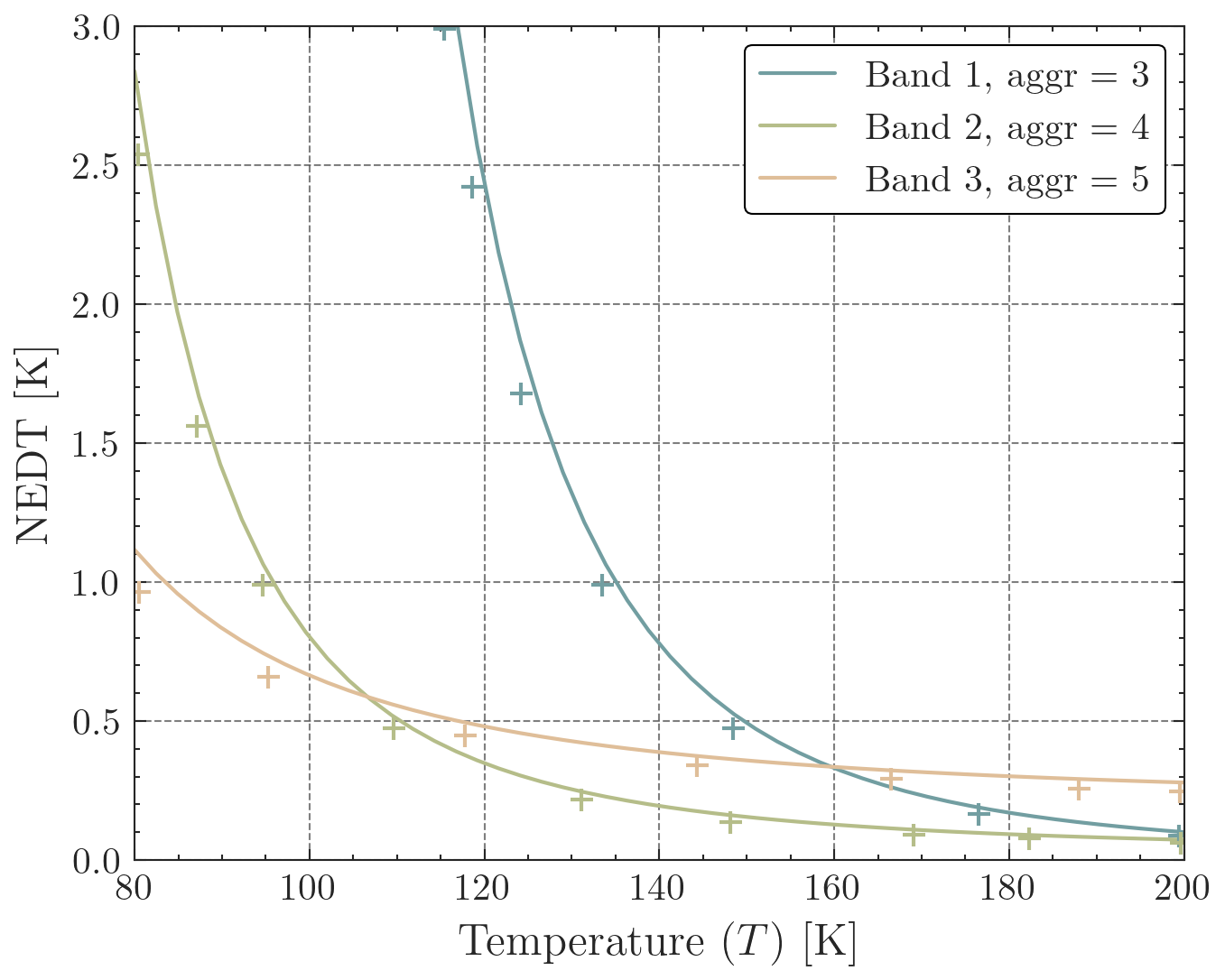}
        \caption{TDI=16, with aggregation.}
    \end{subfigure}\hfill
    \begin{subfigure}[t]{0.33\linewidth}
        \centering
        \includegraphics[width=0.99\linewidth]{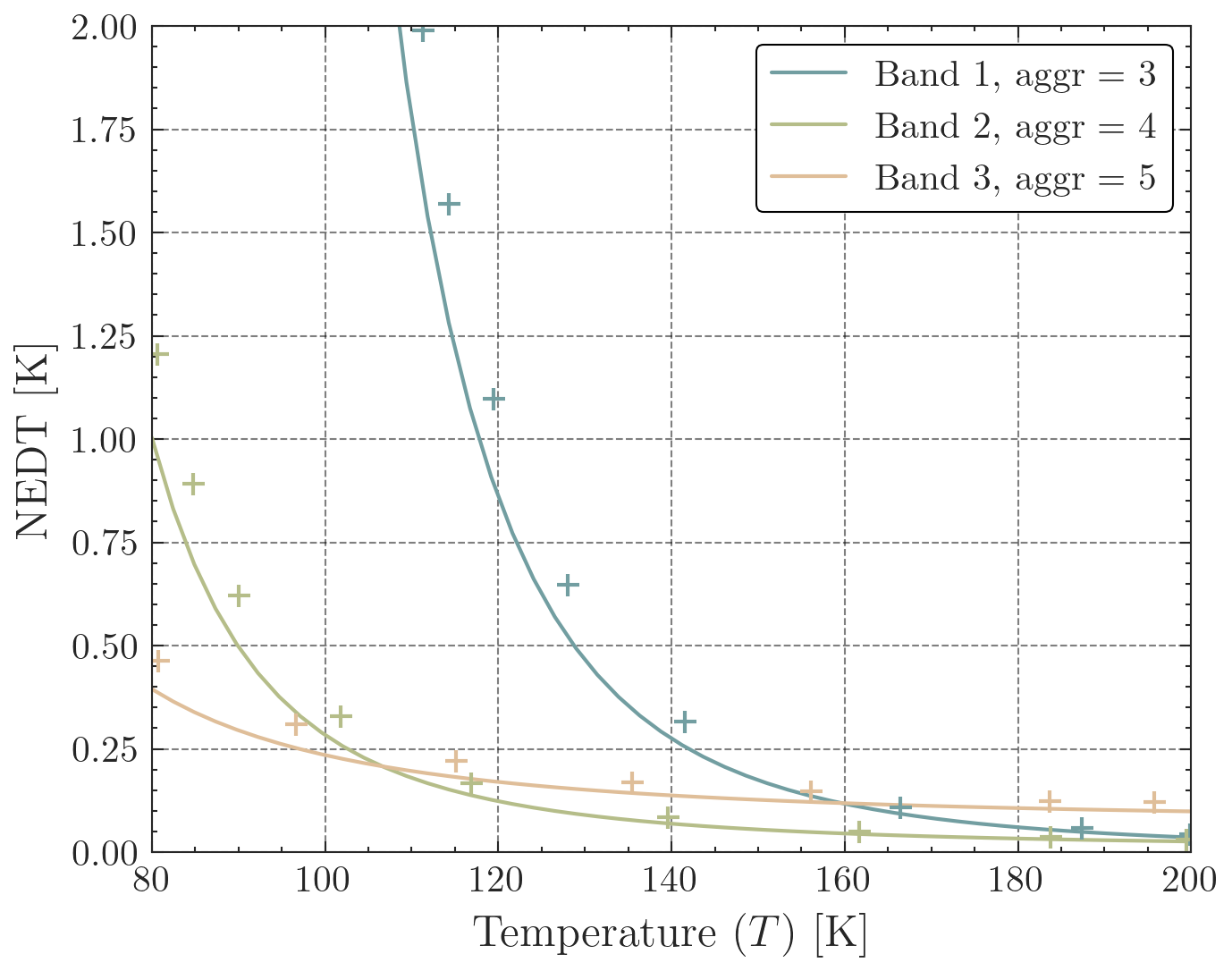}
        \caption{TDI=128, with aggregation.}
    \end{subfigure}
    \caption{Modeled NEDT (lines) vs. instrument
    specifications (crosses) for (a) TDI=1, (b) TDI=16, and
    (c) TDI=128. Agreement confirms the ability to simulate
    E-THEMIS's observational conditions.}
    \label{fig:nedt_band}
\end{figure*}

Our NEDT modeling (Figure~\ref{fig:nedt_band}) replicates
instrument specifications, validating detection thresholds
for simulated cryolava scenarios.

A detectable anomaly requires temperature deviations
exceeding the NEDT. For cryolava cooling from 273~K to
96~K, reducing this noise is achievable in TDI modes, particularly
with pixel aggregation. High-speed flybys may necessitate
TDI=64 or up to 5$\times$5 binning to mitigate motion blur while
preserving SNR.

Then E-THEMIS's adaptive modes enable robust detection of
cryolava thermal anomalies, with sensitivity dependent on
flyby parameters.

\subsection{Thermal Detection of a Cryomagmatic Reservoir}
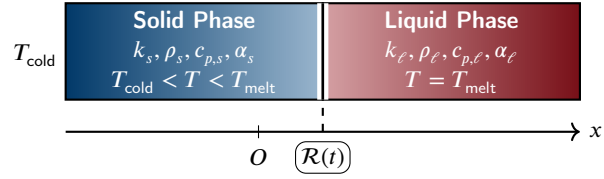
\begin{figure}
    \centering
    \begin{tikzpicture}[scale=0.85]
        \definecolor{solidColor}{RGB}{0, 70, 130}
        \definecolor{liquidColor}{RGB}{150, 20, 30}
        \draw[->,thick] (-4,0) -- (4,0) node[right] {$x$};
        \draw (-1,-0.1) -- (-1,0.1);
        \node at (-1,-0.4) {$O$};
        \fill[left color=solidColor!80!black, right
        color=solidColor!40!white] (-4,0.5) rectangle (0,2);
        \draw[thick] (-4,0.5) rectangle (0,2);
        \node[text=white] at (-2,1.25) {\begin{tabular}{c}
            \textbf{Solid Phase} \\
            $k_s, \rho_s, c_{p,s}, \alpha_s$ \\
            $T_\text{cold} < T < T_\text{melt}$
        \end{tabular}};
        \fill[left color=liquidColor!40!white, right
        color=liquidColor!80!black] (0,0.5) rectangle (4,2);
        \draw[thick] (0,0.5) rectangle (4,2);
        \node[text=white] at (2,1.25) {\begin{tabular}{c}
            \textbf{Liquid Phase} \\
            $k_\ell, \rho_\ell, c_{p,\ell}, \alpha_\ell$ \\
            $T = T_\text{melt}$
        \end{tabular}};
        \draw[ultra thick, white, double=black, double
        distance=1pt] (0,0.5) -- (0,2);
        \draw[dashed, thick] (0,0) -- (0,0.5);
        \node[draw, fill=white, rounded corners, inner
        sep=2pt] at (0,-0.4) {$\mathcal{R}(t)$};
        \node[below left] at (-4,1.5) {$T_\text{cold}$};
    \end{tikzpicture}
    \caption{Modeling the freezing of a cryomagmatic
    reservoir using the Stefan problem. The interface
    \(\mathcal{R}(t)\) separates solid and liquid phases.}
    \label{fig:stefan_reservoir}
\end{figure}

To identify potential sources of cryoclastic flows, we
investigate the detectability of cryomagmatic reservoirs -- 
subsurface bodies of cryomagma composed of water and salts,
with temperatures ranging from 268~K to 273~K
\citep{kargel1994cryovolcanism}. Following the scenarios advocated by \citet{lesage2020cryomagma, lesage2022simulation}, we focus on a pure water
reservoir with a melting point of 273.15~K, located at a
depth \(H\) below Europa's surface. Over time, the reservoir
cools and solidifies, forming an ice crust of thickness
\(\mathcal{R}\) (see Fig.~\ref{fig:stefan_reservoir}) before triggering an eruption
\cite{lesage2020cryomagma}. For a reservoir of radius
\(R = 600\)~m at a depth of \(H = 2\)~km, solidification
(\(\mathcal{R} \approx 10\)~m) occurs over approximately
2~years.

The Fourier number, defined as \(\text{Fo} = \alpha t / L^2\),
characterizes the regime of thermal conduction, where
\(\alpha\) is the thermal diffusivity, \(t\) is time, and \(L\)
is the characteristic length scale. For the 2~km-depth
reservoir, \(\text{Fo} = 1.98 \times 10^{-5} \ll 1\),
indicating that heat remains localized near the reservoir and not reaching the moon's surface.
Even for a deeper reservoir at \(H = 10\)~km with a radius
of \(R = 1.3\)~km and a solidification time of 300~years,
the Fourier number remains small (\(\text{Fo} \approx 1.19
\times 10^{-4} \ll 1\)). This suggests that thermal diffusion
is limited, and temperature changes are confined to the
vicinity of the reservoir. Shallower reservoirs are expected
to erupt more rapidly, further restricting the spatial
extent of any thermal anomaly.

To quantify this behavior, we model the solidification
process using the Stefan problem
(Figure~\ref{fig:stefan_reservoir}). The system is idealized
as an infinite domain with a solid phase (\(x < 0\))
initially at temperature \(T_\text{cold}\) and a liquid
phase (\(x > 0\)) at the melting temperature \(T_\text{melt}\).
The thermal properties of each phase -- thermal conductivity
\(k\), density \(\rho\), specific heat capacity \(c_\text{p}\),
and thermal diffusivity \(\alpha\) -- govern the evolution of
the solid-liquid interface \(\mathcal{R}(t)\). The position
of this interface is given by:
\begin{equation}
    \mathcal{R}(t) = 2\lambda \sqrt{\alpha_s t},
\end{equation}
where \(\lambda\) is determined by the transcendental equation:
\begin{equation}
    \lambda \exp(\lambda^2) (1 + \text{erf}\,\lambda) =
    \frac{c_{\text{p},s}(T_\text{melt} - T_\text{cold})}
    {\Delta H_\text{melt} \sqrt{\pi}},
\end{equation}
and \(\Delta H_\text{melt}\) is the enthalpy of fusion. The
temperature profile in the solid phase is described by:
\begin{equation}
    T(x < \mathcal{R}, t) = T_\text{cold} + \frac{
        T_\text{melt} - T_\text{cold}
    }{
        1 + \text{erf}\,\lambda
    } \left[1 + \text{erf}\left(\frac{x}{2\sqrt{\alpha_s t}}
    \right)\right].
\end{equation}

\begin{figure}
    \centering
    \begin{subfigure}{0.45\linewidth}
        \includegraphics[width=0.95\linewidth]
        {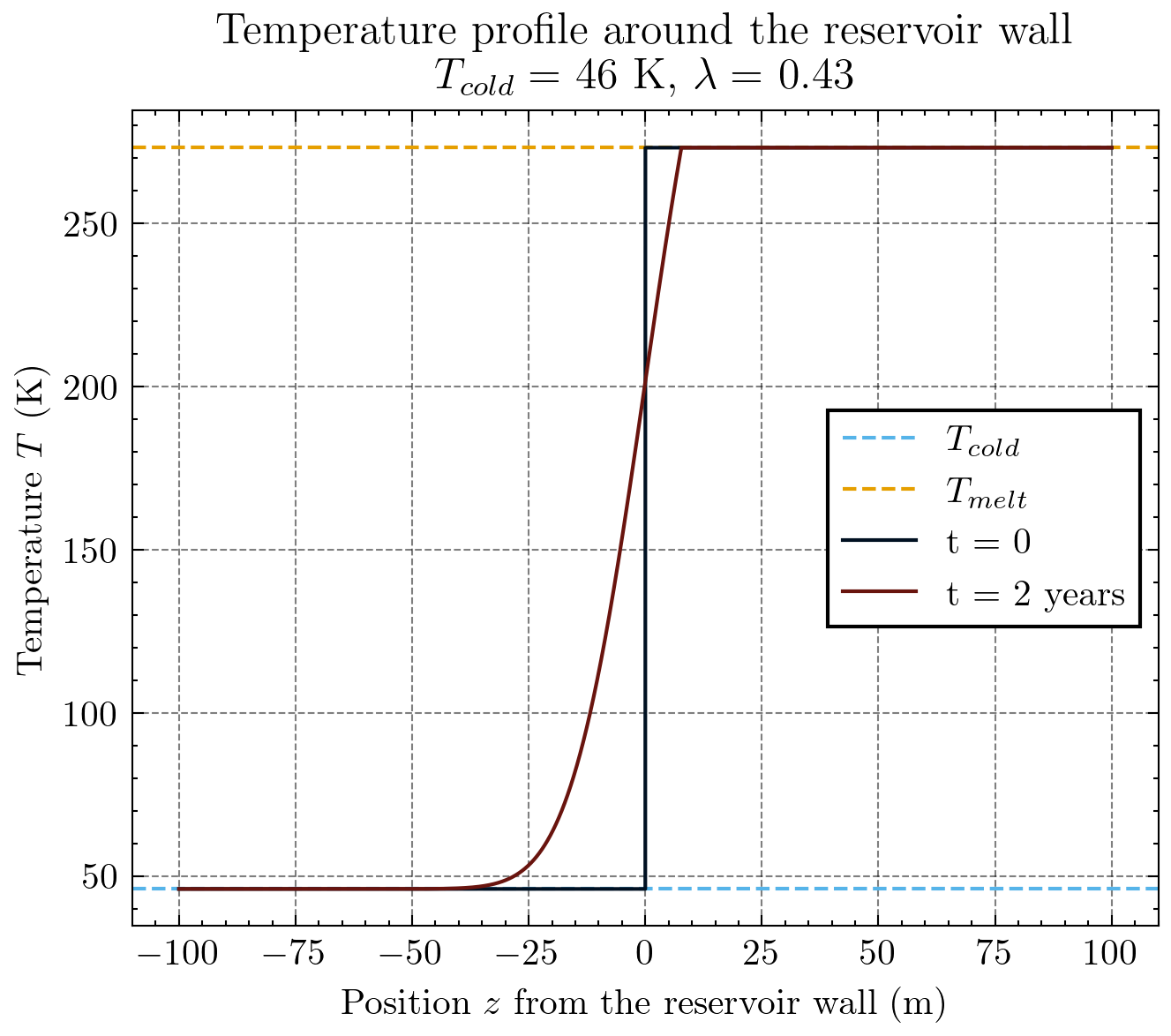}
        \caption{Profile for \(T_{\text{cold}} = 46\)~K
        (\(\lambda=0.43\)).}
        \label{fig:profil_46K}
    \end{subfigure}
    \hfil
    \begin{subfigure}{0.45\linewidth}
        \includegraphics[width=0.95\linewidth]
        {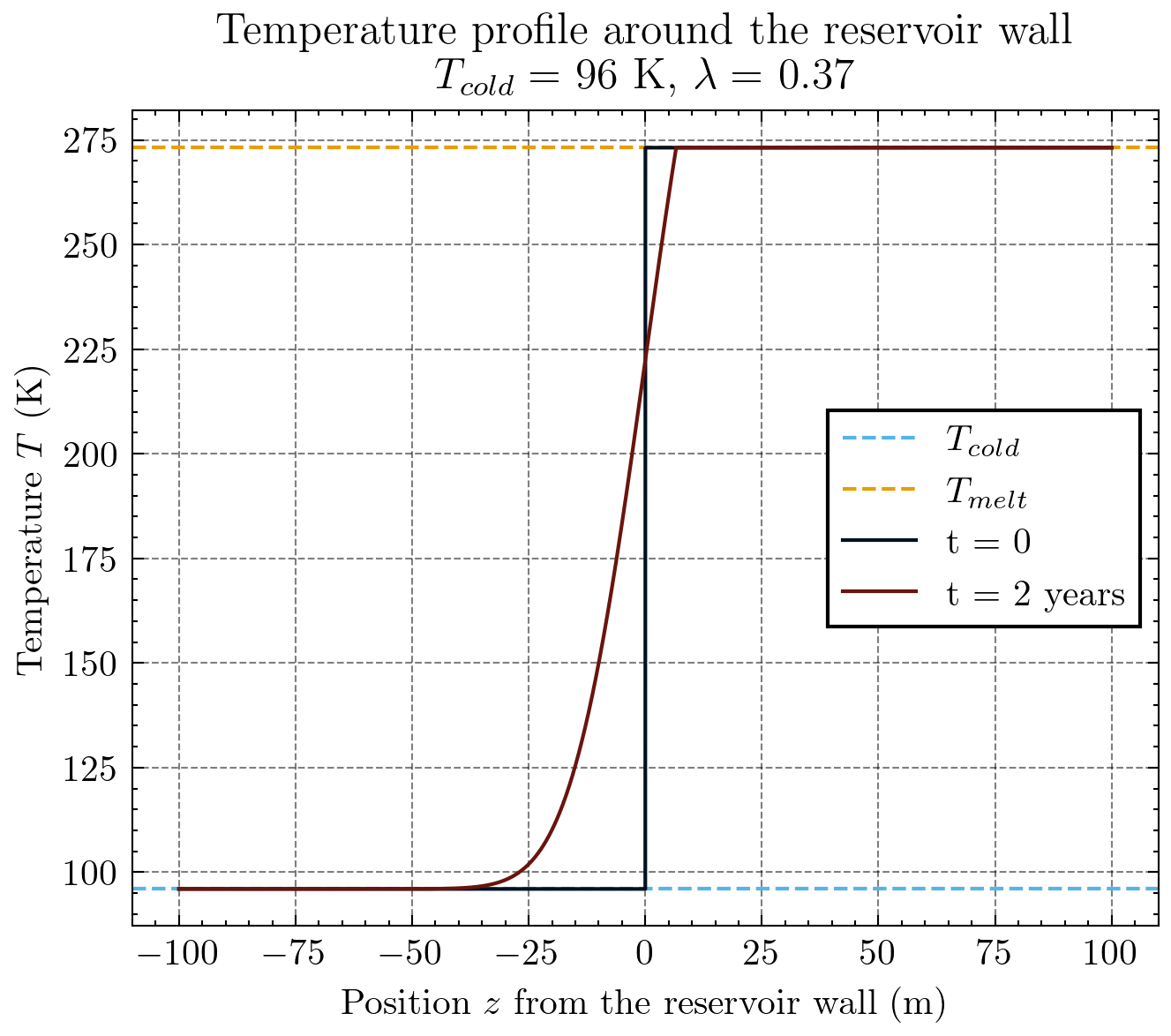}
        \caption{Profile for \(T_{\text{cold}} = 96\)~K
        (\(\lambda=0.38\)).}
        \label{fig:profil_96K}
    \end{subfigure}
    \caption{Temperature profiles around a reservoir wall
    (\(R=600\)~m, \(H=2\)~km) at \(t=0\) and \(t=2\)~years.}
    \label{fig:profils_temperature}
\end{figure}

Temperature profiles for Europa's polar (46~K) and
equatorial (96~K) surface conditions are shown in
Figure~\ref{fig:profils_temperature}. These profiles
demonstrate that temperature variations remain significant
only near the reservoir wall, diminishing rapidly with
distance. After two years, the thermal anomaly is still
confined to within a few tens of meters of the reservoir,
even for deeper systems. For shallower reservoirs, the
eruption timescale is much shorter, as the ice crust
required to trigger an eruption is thinner. Consequently,
any thermal anomaly is unlikely to reach the surface before
an eruption occurs.

These results confirm that thermal anomalies associated with
subsurface cryomagmatic reservoirs -- even those with
eruption timescales of hundreds or thousands of years -- 
will not produce a detectable thermal signature at Europa's
surface. Thus, such reservoirs are unlikely to be
identifiable by the instrumentation aboard the Europa
Clipper mission.

\subsection{Thermal Behavior of Cryolava Flows}
\begin{figure}
    \centering
    \includegraphics[width=0.45\linewidth]
    {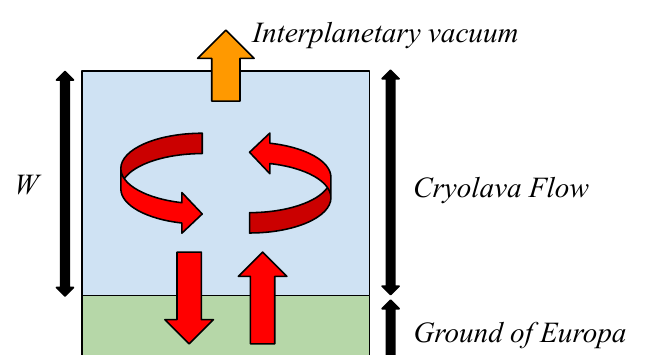}
    \caption{Illustration of the heat fluxes occurring within
    the cryolava (in blue). It cools by radiation (orange)
    and is subjected to conductive fluxes (red) within
    itself and through the surface (the \enquote{ground},
    in green).}
    \label{fig:flux_cryolave}
\end{figure}

To model the thermal evolution of a cryoclastic flow
cooling on Europa's surface, we account for all relevant
energy fluxes affecting the flow and its surroundings.
Europa's surface is subject to solar radiative flux,
radiative flux from Jupiter's magnetosphere, geophysical
flux due to tidal forces, and variations caused by eclipses
when Europa passes into Jupiter's shadow
\citep{ashkenazy2019surface,ruiz2005heat}. Since most of
cryovolcanic candidates are located near the equator, we
adopt an equilibrium surface temperature of
\(T_\text{env} = 96\)~K \citep{ashkenazy2019surface}. We
assume that cooling occurs over tens of Earth days or more,
allowing us to neglect daily temperature variations. These
combined fluxes are integrated into a single radiative
flux, described by Stefan-Boltzmann's law:
\begin{equation}
    F_\text{rad} = \epsilon \sigma \left(
    T_\text{surf}^4 - T_\text{env}^4\right).
\end{equation}
Given that cryolava is initially warmer than its
surroundings, heat is conducted from the flow into the icy
substrate, warming the surface in contact with the flow.
This conductive flux is expressed as:
\begin{equation}
    \vec{F}_\text{cond} = -k\vec{\nabla} T.
\end{equation}
Within the cryolava itself, heat conduction transports
energy from the interior to the surface. Due to Europa's
tenuous atmosphere, convective cooling above the flow is negligible
\citep{pappalardo2009europa}. Thus, the primary heat losses
for the cryolava are radiative cooling to the environment
and conductive transfer to the substrate
(Figure~\ref{fig:flux_cryolave}).

We implemented this physics in our \textit{CryoPy}
framework to simulate the thermal evolution of snowy
cryolava over time. The scenario considers a stationary
cryolava layer of thickness \(W\), cooling under the
influence of the described fluxes. We examine two cases: a
1~m thickness to model the flow's extremities and a 5~m
thickness to represent the flow's core, based on observed
thicknesses of up to 4--5~m \citep{lesage2021constraints}.
The simulations use periodic boundary conditions parallel
to the surface and assume a semi-infinite substrate, given
the conductive ice crust's thickness of approximately
10~km \citep{lesage2020cryomagma}. The substrate depth \(L\)
is determined such that the conductive flux at this depth
is negligible, using the Fourier number:
\begin{equation}
    L = \sqrt{10\, \alpha\, t_\text{sim}},
\end{equation}
where \(t_\text{sim}\) is the simulation's physical time.
The simulations focus on the surface temperature and heat
fluxes through the cryolava layer. For example, simulating
the cooling of a 5~m thick snow flow with a density of
\(600\)~kg/m\(^3\) over 650~days required 116,310 particles
and approximately 55~hours of computation on 
16 cores. 

\begin{figure}
    \centering
    \includegraphics[width=0.45\linewidth]
    {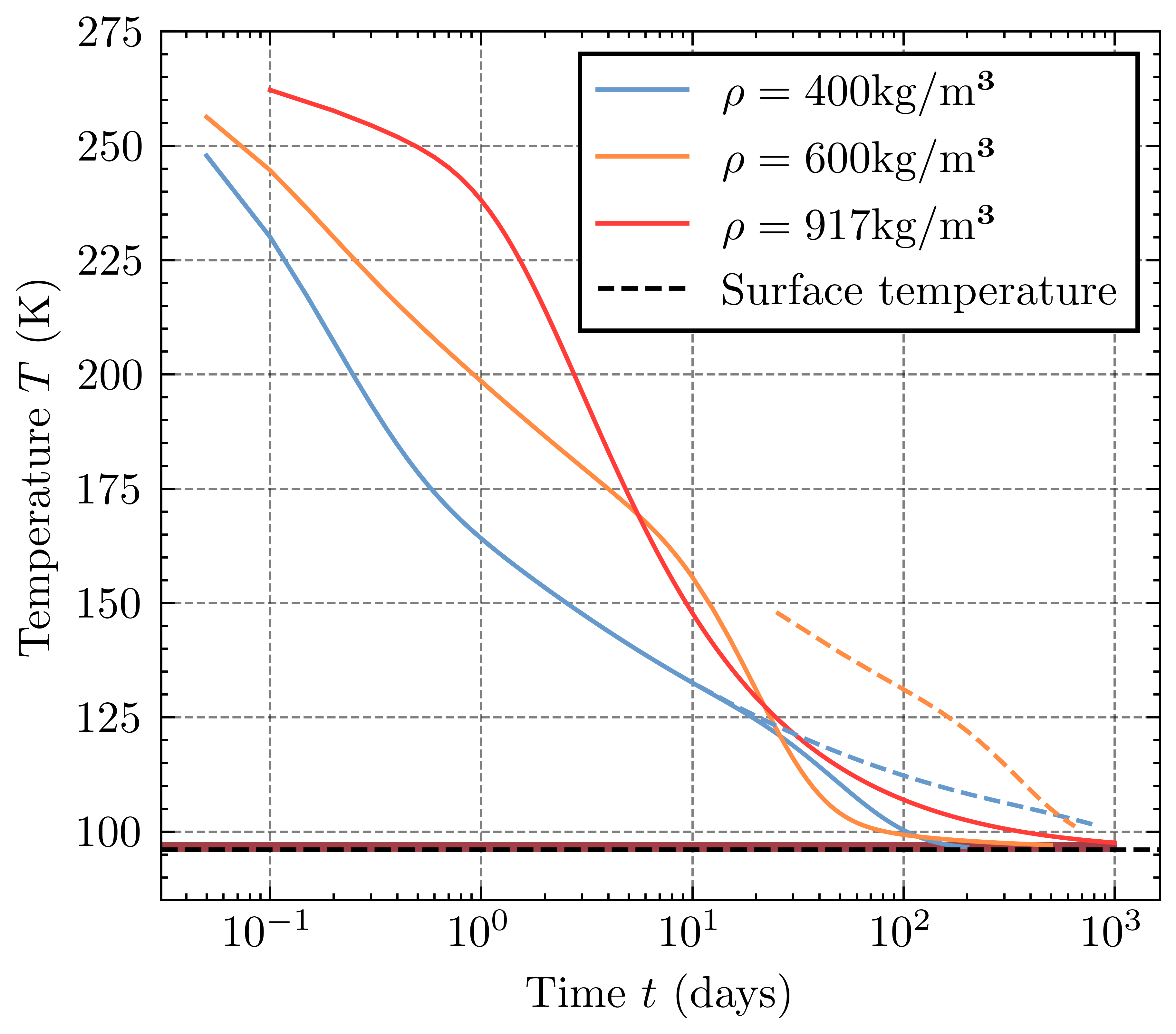}
    \caption{Temporal evolution (in Earth days) of the
    temperature of the surface of the cryolava flow for
    different densities of a one-meter layer of snow (solid
    lines) and 5~m (dashed lines) on Europa at the equator.
    The detectability limit considered for this analysis is
    set at 1.2~K (thick burgundy band near the surface
    temperature), i.e., the NEDT of Europa Clipper in band 2
    for a spatial aggregation of \(4 \times 4\) pixels and
    a TDI of 16.}
    \label{fig:snow_surface_temp}
\end{figure}

The cryolava is assumed to be at the melting temperature of
pure water at the simulation's start. The maximum detection
time is determined by the interval during which the
cryolava's surface temperature exceeds the detection limit
of E-THEMIS, set at 97.2~K for the instrument's regional
operating conditions (TDI of 16, \(4 \times 4\) pixel
aggregation). The temporal evolution of the surface
temperature is summarized in
Figure~\ref{fig:snow_surface_temp}, which shows that
detection is possible for several hundred days, depending
on the snow's density. However, for thin layers, the
thermal anomaly is only a few degrees above the detection
threshold after about fifty days.

\begin{figure*}
    \centering
    \begin{subfigure}{0.49\linewidth}
        \centering
        \includegraphics[width=0.99\linewidth]
        {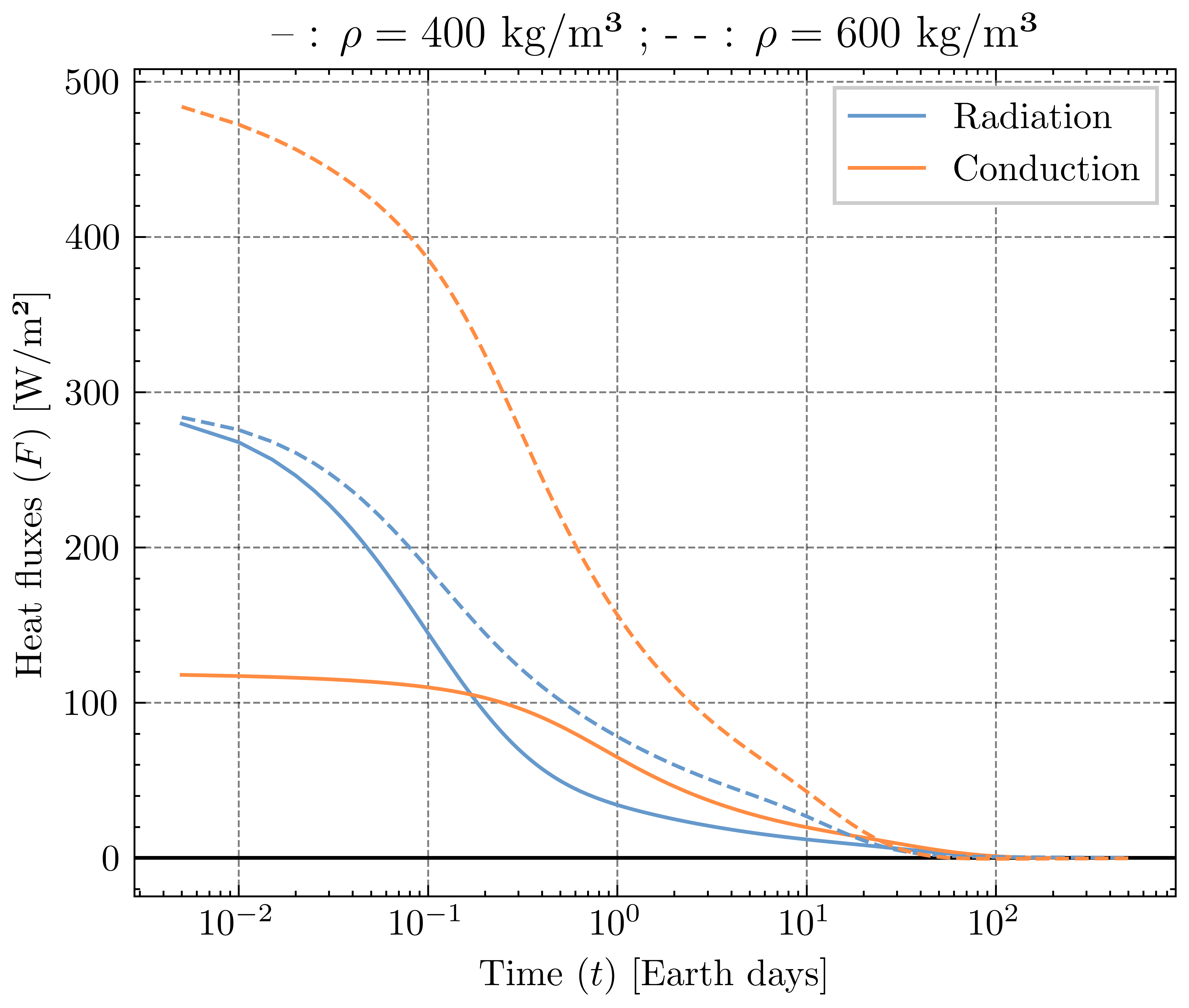}
        \caption{Heat losses at the interfaces for a 1~m
        thick snow.}
    \end{subfigure}
    \hfill
    \begin{subfigure}{0.49\linewidth}
        \centering
        \includegraphics[width=0.99\linewidth]
        {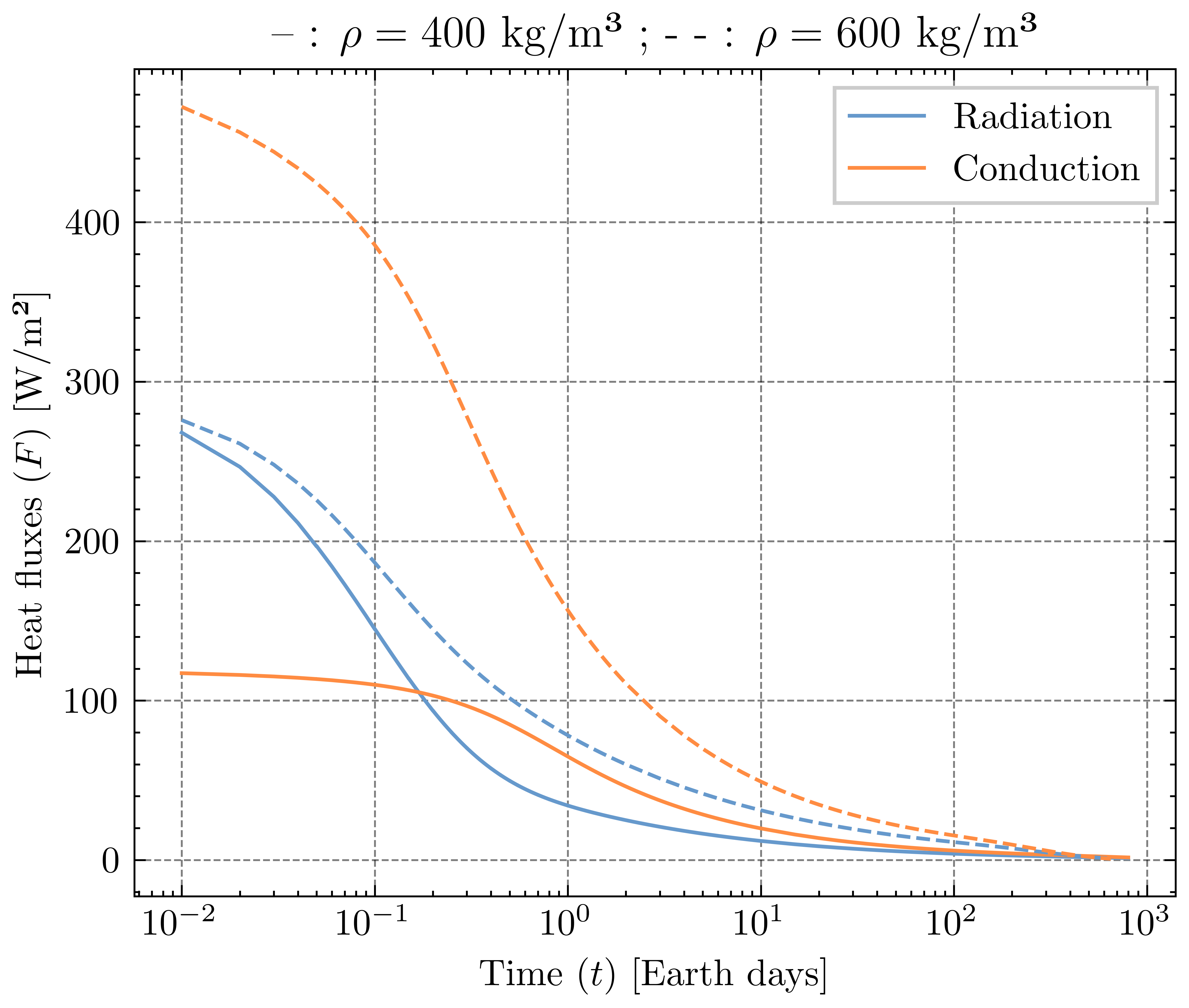}
        \caption{Heat losses at the interfaces for a 5~m
        thick snow.}
    \end{subfigure}
    \caption{Heat losses by radiation at the surface of the
    cryolava and by conduction with the icy substrate for
    different thicknesses of snow cooling on Europa's
    surface.}
    \label{fig:heat_losses_snow}
\end{figure*}

\begin{figure}
    \centering
    \includegraphics[width=0.45\linewidth]
    {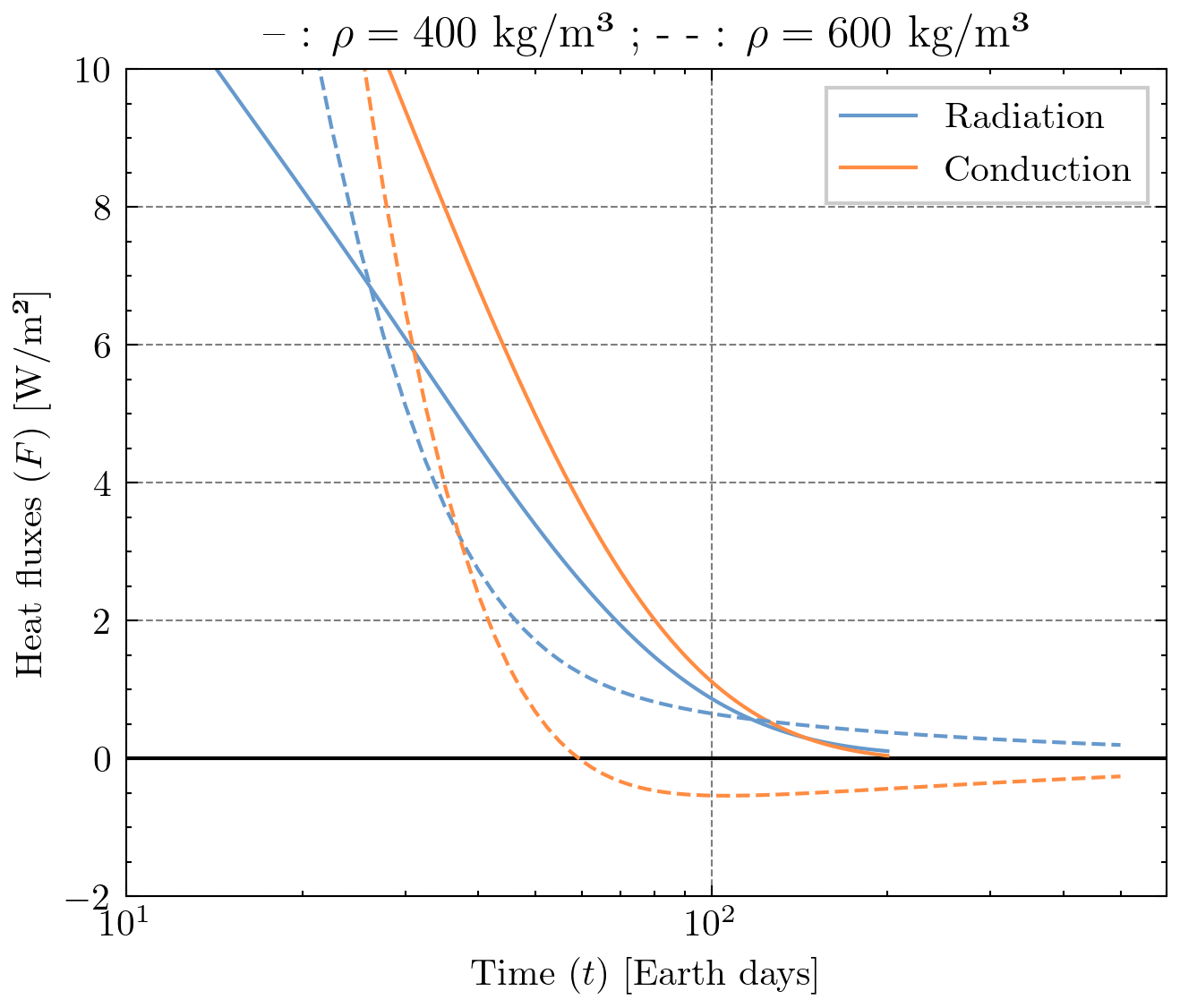}
    \caption{Magnification of the low heat flux area for
    1~m thick flows. We see a warming of the high-density
    snow by the substrate after 15~days of exposure to
    Europa's surface conditions. The same behavior is
    observed on the 5-meter flows.}
    \label{fig:zoom_heat_fluxes}
\end{figure}

Detection durations for different snow densities and
thicknesses are provided in Table~\ref{tab:duree_detection}.
To understand these thermal behaviors, we analyzed the
heat fluxes at the cryolava interfaces, particularly
conduction and radiation losses
(Figure~\ref{fig:heat_losses_snow}). For thin flows,
conduction fluxes become negative for the
\(600\)~kg/m\(^3\) case, indicating that the substrate
warms the flow (Figure~\ref{fig:zoom_heat_fluxes}). This
behavior arises from the higher thermal inertia of denser
snow (\(\sim 1,200\)~SI for \(600\)~kg/m\(^3\) vs.
\(\sim 600\)~SI for \(400\)~kg/m\(^3\)), which resists
temperature changes more effectively. Consequently, the
high-density flow loses less heat to the substrate and is
ultimately warmed by it.

\begin{table}[h]
    \centering
    \begin{tabular}{ccc}
    \hline
    \textbf{Layer Thickness} & \textbf{Density} &
    \textbf{Detection Duration} \\
    \textbf{[m]} & \textbf{[kg/m\(^3\)]} &
    \textbf{[Earth days]} \\
    \hline
    \multirow{2}{*}{1} & 400 & \(\sim\) 100 \\
    \cline{2-3}
     & 600 & \(\sim\) 130 \\
    \hline
    \multirow{2}{*}{5} & 400 & \(\sim\) 1,650 \\
    \cline{2-3}
     & 600 & \(\sim\) 1,550 \\
    \hline
    \end{tabular}
    \caption{Detection durations estimated from SPH thermal
    simulations performed for snow of \(400\)~kg/m\(^3\) and
    \(600\)~kg/m\(^3\) and for thicknesses of 1~m and 5~m.}
    \label{tab:duree_detection}
\end{table}

These results demonstrate that the thermal behavior of
cryoclastic flows is strongly dependent on snow density.
Detection durations can extend up to several years for the
thickest flows, highlighting the potential for long-term
identification of cryovolcanic activity on Europa.





\section{Discussions}
\label{sec:discussions}

Although effusive cryovolcanism has been proposed as a plausible
resurfacing mechanism on Europa, the limited spatial occurence
of smooth plains provides an important observational constraint
on the frequency and scale of such eruptions. Early geological
mapping from the Galileo mission 
\citep{greeley2000geologic, figueredo2004resurfacing}
identified smooth low-albedo patches that are morphologically
consistent with localized emplacement of fluid material,
potentially cryolava. However, these units remain relatively rare,
typically occurring as small, isolated features of only a few
kilometres in extent \citep{shirley2010europa}. Their scarcity
contrasts with the widespread distribution of tectonic structures
(bands, ridges, chaos terrains) suggesting that effusive processes,
if occurring, contribute only marginally to global resurfacing in
comparison to tectonic and endogenic fracturing mechanisms
\citep{fagents2003considerations}.

Effusive cryovolcanism is not expected to be a process unique to
Europa. The physical principles underlying effusive emplacement
(melt generation within or beneath an ice shell, transport through
fractures, and extrusion of a low-viscosity cryomagma at the surface)
are broadly governed by parameters common to many icy satellites:
thermal gradients, brine chemistry, ice-shell thickness, stress
field, and volatile content \citep{fagents2003considerations,lesage2021constraints}.
In this sense, a model developed for Europa provides a useful
baseline for studying similar processes on other ocean worlds.

However, the transferability of such a model strongly depends on local environmental
conditions, which differ substantially across bodies. For example, Enceladus
exhibits active cryovolcanism dominated by explosive or jet-driven venting from
the south polar fractures
\citep{porco2006cassini,spencer2009enceladus}, where plume
dynamics, vapor-dominated conduits, and tidal pumping dominate over sustained
effusive flows. An effusive model can still be applied, but only to hypothesized
surface flows expected from brine overflows or melt extrusion along tiger-striped
ridges—processes that remain unproven and likely subordinate to explosive venting.
Conversely, Ganymede possesses a thicker ice shell and lower tidal stresses, which
would inhibit frequent melt ascent but might still allow episodic effusive emplacement
if deep brines are mobilized during tectonic reorganizations. The smooth dark
terrains on Ganymede have been tentatively compared to cryolava-like resurfacing
in the past, though evidence remains ambiguous
\citep{solomonidou2021candidate}.

For Triton, the nitrogen--methane--CO$_2$ ice mixture produces
rheologies and volatile behaviors very different from Europa's
sulfate-rich brines. Although Triton shows clear signs of cryovolcanic
resurfacing \citep{smith1989voyager}, most inferred eruptions are
likely volatile-driven or explosive, rather than brine-fed effusive
flows. A Europan effusive model might still capture aspects of melt
migration or thermal gradients in Triton's crust but would require
substantial adaptation to account for lower melting temperatures,
volatile exsolution, and porous regolith flow regimes.

Satellites such as Titan, Dione, or Ariel represent intermediate cases.
Titan has long been proposed as a candidate for effusive cryovolcanic
resurfacing \citep{lopes2013cryovolcanism}, particularly where
ammonia-rich melts could reduce viscosity to Europan-like values.
Dione and Ariel host tectonically modified smooth terrains that could
hypothetically result from cryolava emplacement, though no definitive
morphological or compositional evidence exists. For these objects,
Europan effusive models are useful as first-order frameworks, especially
for estimating melt volumes, flow thicknesses, or thermal decay times, but
they cannot be applied without tailoring assumptions to each satellite's
composition, energy budget, and stress regime.

\section{Conclusion}
\label{sec:conclusion}

This study provides a comprehensive framework for
understanding the dynamics, thermal evolution, and
detectability of cryolava flows on Europa, with direct
relevance to the scientific objectives of the
\textit{Europa Clipper} mission. Through high-resolution
SPH simulations using the \textit{CryoPy} framework, we
demonstrate that the rheological behavior of cryolava -- 
distinguished as either fluidized (Type~1) or compact
(Type~2) snow -- fundamentally shapes the morphology and
extent of Europa's smooth plains. In Scenario~SA,
where gravitational collapse drives flow propagation,
Type~2 cryolava produces localized deposits due to its high
apparent viscosity, while Type~1 cryolava, characterized by
shear-thinning properties, enables rapid and extensive
spreading.
Scenario~SB, which simulates constant-flow
eruptions, further reveals that sustained kinetic energy
input is necessary to replicate the observed geomorphology,
with flow dynamics highly sensitive to eruption velocity
and vent geometry.

Thermal modeling complements these hydrodynamic insights by
quantifying the detectability of cryolava using
Europa Clipper's E-Themis instrument. Our results
indicate that surface temperature anomalies from
cryoclastic flows can remain detectable for up to
4.5~years for thicker deposits (5~m), well
within E-Themis's Noise Equivalent Delta Temperature
(NEDT) thresholds. However, subsurface cryomagmatic
reservoirs, even those with eruption timescales spanning
centuries, are unlikely to produce observable thermal
signatures at the surface due to the confined nature of
heat diffusion. This underscores the importance of focusing
observational efforts on freshly emplaced surface
flows, where thermal contrasts are most pronounced.

Furthermore, instruments synergy is critical for characterization.
The Mapping Imaging Spectrometer for Europa (MISE) \citep{blaney2024mapping},
spanning $0.8$ to $5~\text{\textmu m}$ at high spectral resolution, is
capable of detecting the warmest areas of newly emplaced cryolava and
snow deposits, provided the thermal anomaly exceeds $\sim190$~K and fills
at least 10\% of a MISE pixel \citep{blaney2024mapping}.
Beyond thermal detection, MISE's spectral range also enables retrieval
of grain size and ice crystallinity \citep{blaney2024mapping}, which
could reveal differences in snow texture between cryolava deposits
and the surrounding smooth plains. Such combined constraints, when
paired with E-THEMIS data, would help characterize both the temperature
distribution and the physical properties of the flow.

The integration of hydrodynamic and thermal analyses not
only advances our understanding of Europan cryovolcanism
but also provides actionable guidance for mission planning.
E-Themis's adaptive operational modes, such as Time-Delay
Integration (TDI) and spatial aggregation, will be
essential for maximizing detection sensitivity during
flybys, particularly in equatorial regions where thermal
anomalies are expected to be most persistent. These
findings also highlight the need for future work to
incorporate three-dimensional topography into simulations,
as well as experimental
validation of cryolava rheologies under Europan conditions.

Ultimately, this study bridges theoretical modeling with
mission-specific instrumentation, offering a robust
foundation for identifying active cryovolcanism on Europa.
By constraining the physical and thermal properties of
cryolava, we pave the way for \textit{Europa Clipper} to
uncover the geological processes shaping Europa's surface
and, by extension, the potential habitability of its
subsurface ocean.

\section*{Acknowledgements}
The present research was supported by the Programme National de Plan\'{e}tologie (PNP)
of CNRS-INSU co-funded by CNES,
and the Université de Reims Champagne-Ardenne.
Part of this work was performed at the Jet Propulsion Laboratory,
California Institute of Technology, under a contract with the National
Aeronautics and Space Administration (80NM0018D0004).
We acknowledge the use of
computational resources provided by the ROMEO high-performance computing
facility. The authors are grateful to Dr. Mathieu Choukroun for his support
and for facilitating collaborative visits, and to our colleagues at
the Jet Propulsion Laboratory and elsewhere for insightful discussions on lightning
processes in icy environments. The authors would also like to thank the entire
\verb|PySPH| community for their continuous support while developing this framework.

\appendix
\section*{Appendix}
\section{Numerical Implementation and SPH Framework}

This section provides a detailed overview of the Smoothed Particle Hydrodynamics (SPH) method used in this study, specifically focusing on the treatment of boundary conditions, surface detection, and the implementation within the PySPH framework.

\subsection{Fundamentals of SPH}

The SPH method is a meshless, Lagrangian particle method where the fluid is discretized into a set of particles $i$. Any field variable $A(\vec{r})$ is interpolated using a kernel function $W(\vec{r}-\vec{r}', h)$:

\begin{equation}
\langle A_i \rangle = \sum_j \frac{m_j}{\rho_j} A_j W_{ij}(h)
\label{eq:sph_interpolation}
\end{equation}

where $m_j$ and $\rho_j$ are the mass and density of neighbor particle $j$, and $h$ is the smoothing length. The governing equations (continuity and momentum) are solved in their discretized SPH form. In our model, we use the Weakly Compressible SPH (WCSPH) formulation.

\subsection{Dynamic Boundary Conditions}

To model the interaction between the cryolava and the icy substrate or vent walls, we implement the Dynamic Boundary Conditions (DBC) proposed by \citet{cabrera2007boundary}. 

\begin{figure}
\centering
\begin{tikzpicture}[
    fluid/.style={
        circle, draw=blue!80, fill=blue!20, 
        thick, inner sep=2pt
    },
    boundary/.style={
        circle, draw=black!80, fill=green!20, 
        thick, inner sep=2pt
    },
    vector_style/.style={
        -latex, red, thick
    },
    scale=1.5
]

    \newcommand{\deltap}{1.0} 

    \node[fluid] (P_f) at (2*\deltap, 1.5*\deltap) {};
    \node[anchor=west, blue] at (P_f.east) {~Fluid Particle};

    \foreach \i in {0,1,2,3,4} {
        \node[boundary] (T\i) at (\i*\deltap, 0) {};
    }

    \foreach \i in {0,1,2,3} {
        \node[boundary] (B\i) at (\i*\deltap + 0.5*\deltap, -0.5*\deltap) {};
    }

    \node[anchor=west] at (4.5*\deltap, -0.25*\deltap) {
        Two Staggered DBC Layers
    };

    \coordinate (Origin) at (2*\deltap, 0);
    
    \draw[vector_style] (Origin) -- ++(0, -0.5*\deltap) 
        node[midway, right, scale=0.8] {$\Delta p / 2$};
    
    \draw[vector_style] (Origin) -- ++(-0.5*\deltap, 0) 
        node[midway, below, scale=0.8] {$\Delta p / 2$};

\end{tikzpicture}
\caption{Dynamic Boundary Conditions (DBC) in a two-layer staggered arrangement, as show in \cite{cabrera2007boundary}.
The $\Delta p/2$ offset between the two rows prevents fluid particles from \enquote{leaking} through the gaps and provides a more isotropic density contribution near the solid interface.}
\label{fig:dbc_tikz}
\end{figure}
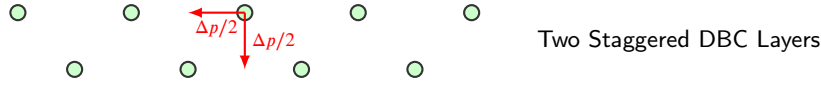

In this approach, boundary particles are treated similarly to fluid particles in that they satisfy the continuity equation and have a pressure derived from an Equation of State (EOS). As illustrated in Figure~\ref{fig:dbc_tikz}, boundary particles contribute to the density summation of nearby fluid particles, preventing them from penetrating the walls by exerting a pressure-based repulsive force. This ensures a consistent treatment of the pressure field at the interface without the need for ghost particles.

\subsection{Interface and Surface Detection}

To apply radiative cooling $q_{rad}$ and sublimation fluxes at the free surface, we employ the surface detection algorithm described by \citet{fraser2016adaptive}. This robust geometric method relies on two criteria: the number of neighbors and the position of the center of mass of the neighbors within the support domain.

\begin{figure}
\centering
\begin{tikzpicture}[scale=0.8]
    \begin{scope}[shift={(0,0)}]
        \draw[dashed, fill=blue!5] (0,0) circle (1.5); 
        \foreach \i in {1,...,15} {
            \fill[blue!60] ({rand*1.1},{rand*1.1}) circle (0.07);
        }
        \fill[red] (0,0) circle (0.1) node[below=4pt] {$\vec{r}_i$};
        \fill[black] (0.05, -0.1) circle (0.05) node[right=2pt] {$\vec{r}_{cm}$};
        \node at (0, 1.8) {$|\vec{r}_{cm} - \vec{r}_i| \to 0$};
    \end{scope}

    \begin{scope}[shift={(5,0)}]
        \begin{scope}
            \clip (-1.5,-1.5) rectangle (1.5,0);
            \draw[dashed, fill=blue!20] (0,0) circle (1.5);
            \foreach \i in {1,...,9} {
                \fill[blue!60] ({rand*1.1},{-abs(rand)*1.1 - 0.1}) circle (0.07);
            }
        \end{scope}
        \draw[thick] (-1.5,0) -- (1.5,0) node[right] {Surface};
        \draw[dashed] (0,0) ++(1.5,0) arc (0:180:1.5);
        
        \fill[red] (0,0) circle (0.1) node[above=4pt, right=4pt] {$\vec{r}_i$};
        \fill[black] (0, 0.7) circle (0.08) node[right=4pt] {$\vec{r}_{cm}$};
        
        \draw[-{Stealth}, thick] (0,0) -- (0,0.7) node[midway, left] {$> \ell h$};
        
        \node at (0, 1.8) {$|\vec{r}_{cm} - \vec{r}_i| \gg 0$};
    \end{scope}
\end{tikzpicture}
\caption{Free surface detection logic based on Fraser (2016). In the bulk (left), neighbors are balanced and $\vec{r}_{cm}$ is near the particle. At the surface (right), the missing neighbors shift $\vec{r}_{cm}$ inwards, triggering the surface flag.}
\label{fig:fraser_tikz}
\end{figure}
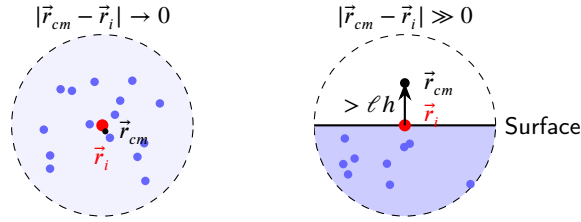

First, the geometric center of mass (centroid) of the neighbors of particle $i$, denoted $\vec{r}_{i,cm}$, is calculated, for each coordinate $\alpha$, as:
\begin{equation}
\vec{r}_{i,cm}^\alpha = \frac{1}{M} \sum_{j=1}^{N_i} \vec{r}_{ij}^\alpha
\label{eq:centroid}
\end{equation}
where $M$ is the total mass of the cluster of particles, $N_i$ is the number of neighbors of particle $i$, and $\vec{r}_{ij} = \vec{r}_i - \vec{r_j}$ . A particle is identified as being on the surface if it satisfies either of the following conditions:
\begin{enumerate}
    \item The number of neighbors is below a threshold: $N_i < N_{min}$.
    \item The distance to the centroid exceeds a fraction of the smoothing length: $|\vec r_{i,cm}| > \ell h$.
\end{enumerate}
Typical values used in this study are $N_{min} \approx 12-13$ and $\ell \approx 0.25$ to $0.4$, depending on the kernel type. As shown in Fig. \ref{fig:fraser_tikz}, for a bulk particle, the neighbors are distributed symmetrically around it, keeping the centroid close to the particle's position. For a surface particle, the truncated support shifts the centroid toward the interior of the fluid.

\subsection{Custom Implementation in PySPH}

The simulations were performed using \textbf{PySPH}, an open-source framework for SPH. Below are examples of how custom equations and applications are structured in our model.

\subsubsection{Custom Equation for Heat Transfer}

The following Python snippet demonstrates the implementation of a custom cooling law using the \texttt{Equation} class in PySPH.

\begin{lstlisting}[language=Python, caption={Custom Heat Transfer Equation for cryolava cooling.}]
from pysph.sph.equation import Equation

class CryolavaCooling(Equation):
    def __init__(self, dest, sources, Stefan_Boltzmann, epsilon):
        self.sb = Stefan_Boltzmann
        self.eps = epsilon
        super(CryolavaCooling, self).__init__(dest, sources)

    def loop(self, d_idx, d_temp, d_rho, d_m, d_u, d_at_surface):
        # Simplified surface radiative loss calculation (Fraser detection)
        # d_u represents the internal energy
        if d_at_surface[d_idx] > 0:
            # Compute radiative flux: sigma * epsilon * T^4
            q_rad = self.sb * self.eps * (d_temp[d_idx]**4)
            d_u[d_idx] -= q_rad / d_rho[d_idx]
\end{lstlisting}

\subsubsection{Custom Application Setup}

The code below illustrates the setup of the main application loop including the WCSPH scheme.

\begin{lstlisting}[language=Python, caption={PySPH Application setup for cryovolcanic flows.}]
from pysph.solver.application import Application
from pysph.sph.scheme import WCSPHScheme

class CryoPyApp(Application):
    def add_user_options(self, group):
        group.add_argument("--viscosity", type=float, default=0.1)

    def consume_user_options(self):
        self.visc = self.options.viscosity

    def create_particles(self):
        # Define fluid and boundary particles (DBC)
        # Initialization logic for m_fluid, m_boundary, etc.
        pass

    def create_scheme(self):
        # Initialize WCSPH scheme with Cabrera-Crespo boundary conditions
        s = WCSPHScheme(
            fluids=['lava'], solids=['ground'],
            dim=2, rho0=1000.0, c0=10.0, h0=0.1
        )
        return s

    def create_equations(self):
        # Append custom equations to the standard SPH suite
        eqs = self.scheme.get_equations()
        # Custom surface detection logic can be inserted here
        return eqs

if __name__ == '__main__':
    app = CryoPyApp()
    app.run()
\end{lstlisting}

\bibliographystyle{apalike}
\bibliography{biblio_abbreviated}

\begin{thebibliography}{}

\bibitem[Anderson, 1976]{anderson_1976}
Anderson, E.~A. (1976).
\newblock A point energy and mass balance model of a snow cover.
\newblock Technical report, US Department of Commerce, National Oceanic and
  Atmospheric Administration, National Weather Service.

\bibitem[Andersson and Inaba, 2005]{andersson2005thermal}
Andersson, O. and Inaba, A. (2005).
\newblock Thermal conductivity of crystalline and amorphous ices and its
  implications on amorphization and glassy water.
\newblock {\em Phys. Chem. Chem. Phys.}, 7:1441--1449.

\bibitem[Arenson et~al., 2021]{arenson2021physical}
Arenson, L., Colgan, W., and Marshall, H.~P. (2021).
\newblock Physical, thermal, and mechanical properties of snow, ice, and
  permafrost.
\newblock In {\em Snow and ice-related hazards, risks, and disasters}, pages
  35--71. Elsevier.

\bibitem[Ashkenazy, 2019]{ashkenazy2019surface}
Ashkenazy, Y. (2019).
\newblock {The surface temperature of Europa}.
\newblock {\em Heliyon}, 5(6):e01908.

\bibitem[Bartak, 1990]{bartak1990study}
Bartak, J. (1990).
\newblock A study of the rapid depressurization of hot water and the dynamics
  of vapour bubble generation in superheated water.
\newblock {\em Int. J. of Multiphase Flow}, 16(5):789--798.

\bibitem[{Becker} et~al., 2023]{becker2023SRUeuropa}
{Becker}, H.~N., {Lunine}, J.~I., {Schenk}, P.~M., {Florence}, M.~M.,
  {Brennan}, M.~J., {Hansen}, C.~J., {Martos}, Y.~M., {Bolton}, S.~J., and
  {Alexander}, J.~W. (2023).
\newblock {A Complex Region of Europa's Surface With Hints of Recent Activity
  Revealed by Juno's Stellar Reference Unit}.
\newblock {\em Journal of Geophysical Research (Planets)},
  128(12):e2023JE008105.

\bibitem[Bernoulli, 1738]{bernoulli_hydrodynamica_1738}
Bernoulli, D. (1738).
\newblock {\em Hydrodynamica, sive de viribus et motibus fluidorum commentarii
  : opus academicum ab auctore, dum Petropoli ageret, congestum}.
\newblock sumptibus Johannis Reinholdi Dulseckeri : Typis Joh. Deckeri,
  typographi Basiliensis, ETH-Bibliothek Zürich, Rar 5503.
\newblock Public Domain Mark.

\bibitem[Bilotta et~al., 2016]{bilotta2016gpusph}
Bilotta, G., H{\'e}rault, A., Cappello, A., Ganci, G., and Del~Negro, C.
  (2016).
\newblock Gpusph: a smoothed particle hydrodynamics model for the thermal and
  rheological evolution of lava flows.
\newblock {\em Geological Soc. London, Special Publications}, 426(1):387--408.

\bibitem[Bird et~al., 2007]{bird2007transport}
Bird, R.~B., Stewart, W.~E., and Lightfoot, E.~N. (2007).
\newblock {\em Transport Phenomena}.
\newblock John Wiley \& Sons, revised 2nd edition edition.

\bibitem[Blaney et~al., 2024]{blaney2024mapping}
Blaney, D.~L., Hibbitts, K., Diniega, S., Davies, A.~G., Clark, R.~N., Green,
  R.~O., Hedman, M., Langevin, Y., Lunine, J., McCord, T.~B., et~al. (2024).
\newblock The mapping imaging spectrometer for europa (mise).
\newblock {\em Space Sci. Rev.}, 220(7):80.

\bibitem[Boinovich and Emelyanenko, 2014]{boinovich2014experimental}
Boinovich, L. and Emelyanenko, A. (2014).
\newblock Experimental determination of the surface energy of polycrystalline
  ice.
\newblock In {\em Doklady Physical Chemistry}, volume 459, pages 198--202.
  Springer.

\bibitem[{Brennen}, 1995]{brennen_1995}
{Brennen}, C.~E. (1995).
\newblock {\em {Cavitation and Bubble Dynamics}}.
\newblock Oxford University Press, New York, Oxford.

\bibitem[{Bro\v{z}} et~al., 2025]{broz_etal_2025}
{Bro\v{z}}, P., {Pato\v{c}ka}, V., {Butcher}, F., {Sylvest}, M., and {Patel},
  M. (2025).
\newblock {The complexity of water freezing under reduced atmospheric
  pressure}.
\newblock {\em Earth Planet. Sci. Lett.}, 668:119531.

\bibitem[Cabrera~Crespo et~al., 2007]{cabrera2007boundary}
Cabrera~Crespo, A.~J., G{\'o}mez~Gesteira, R., Dalrymple, R.~A., et~al. (2007).
\newblock Boundary conditions generated by dynamic particles in sph methods.
\newblock {\em Computers, Mater. \& Continua}.

\bibitem[Carslaw and Jaeger, 1959]{Carslaw1959}
Carslaw, H.~S. and Jaeger, J.~C. (1959).
\newblock {\em Conduction of Heat in Solids}.
\newblock Oxford University Press, Oxford, 2 edition.

\bibitem[Chen and Millero, 1977]{chen1977speed}
Chen, C.-T. and Millero, F.~J. (1977).
\newblock Speed of sound in seawater at high pressures.
\newblock {\em the J. of the Acoust. Soc. of America}, 62(5):1129--1135.

\bibitem[{Chivers} et~al., 2021]{chivers2021JGRE}
{Chivers}, C.~J., {Buffo}, J.~J., and {Schmidt}, B.~E. (2021).
\newblock {Thermal and Chemical Evolution of Small, Shallow Water Bodies in
  Europa's Ice Shell}.
\newblock {\em Journal of Geophysical Research (Planets)}, 126(5):e06692.

\bibitem[Christensen et~al., 2024]{christensen2024europa}
Christensen, P.~R., Spencer, J.~R., Mehall, G.~L., Patel, M., Anwar, S., Brick,
  M., Bowles, H., Farkas, Z., Fisher, T., Gjellum, D., et~al. (2024).
\newblock The europa thermal emission imaging system (e-themis) investigation
  for the europa clipper mission.
\newblock {\em Space Sci. Rev.}, 220(4):38.

\bibitem[Colebrook, 1939]{colebrook1939}
Colebrook, C.~F. (1939).
\newblock Turbulent flow in pipes, with particular reference to the transition
  region between the smooth and the rough pipe laws.
\newblock {\em J. of the Institution of Civ. Engineers}, 11(4):133--156.

\bibitem[{Cordier} et~al., 2024]{cordier_etal_2024}
{Cordier}, D., {Liger-Belair}, G., {Bonhommeau}, D.~A., {S\'{e}on}, T.,
  {App\'{e}r\'{e}}, T., and {Carrasco}, N. (2024).
\newblock {Capillary Processes in Extraterrestrial Contexts}.
\newblock {\em J. Geophys. Res. Planets}, 129(5):e2023JE008248.

\bibitem[Courant et~al., 1928]{courant1928partiellen}
Courant, R., Friedrichs, K., and Lewy, H. (1928).
\newblock {\"U}ber die partiellen differenzengleichungen der mathematischen
  physik.
\newblock {\em Mathematische annalen}, 100(1):32--74.

\bibitem[Craft et~al., 2016]{craft2016fracturing}
Craft, K.~L., Patterson, G.~W., Lowell, R.~P., and Germanovich, L. (2016).
\newblock Fracturing and flow: Investigations on the formation of shallow water
  sills on europa.
\newblock {\em Icarus}, 274:297--313.

\bibitem[Crooks and Bouard, 1997]{CROOKS1997123}
Crooks, M. and Bouard, R. (1997).
\newblock D\'etermination des coefficients de pertes de pression dans des
  changements de direction en \'ecoulements diphasiques gaz-solide.
\newblock {\em Powder Technology}, 91(2):123--130.

\bibitem[Darcy, 1856]{darcy1856fontaines}
Darcy, H. (1856).
\newblock {\em Les fontaines publiques de la ville de Dijon: exposition et
  application des principes à suivre et des formules à employer dans les
  questions de distribution d'eau}.
\newblock Victor Dalmont, Paris.

\bibitem[Daubar et~al., 2024]{daubar2024planned}
Daubar, I.~J., Hayes, A.~G., Collins, G., Craft, K.~L., Rathbun, J., Spencer,
  J.~R., Wyrick, D.~Y., Bland, M.~T., Davies, A.~G., Ernst, C.~M., et~al.
  (2024).
\newblock Planned geological investigations of the europa clipper mission.
\newblock {\em Space Sci. Rev.}, 220(1):18.

\bibitem[Dean and Timmerhaus, 1963]{Dean1963}
Dean, J. and Timmerhaus, K. (1963).
\newblock Thermal conductivity of solid {H}2{O} and {D}2{O} at low
  temperatures.
\newblock In Timmerhaus, K., editor, {\em Advances in Cryogenic Engineering},
  volume~8, pages 263--267. Springer, Boston, MA.

\bibitem[{Debenedetti}, 1997]{debenedetti_1997a}
{Debenedetti}, P.~G. (1997).
\newblock {\em {1. Introduction: Metastable Liquids in Nature and Technology}},
  pages 1--62.
\newblock Princeton University Press, Princeton.

\bibitem[Dempsey et~al., 1999]{dempsey1999scale}
Dempsey, J., Adamson, R., and Mulmule, S. (1999).
\newblock Scale effects on the in-situ tensile strength and fracture of ice.
  part ii: First-year sea ice at resolute, nwt.
\newblock {\em Int. J. of fracture}, 95(1):347--366.

\bibitem[Dillard and Timmerhaus, 1969]{DillardTimmerhaus1969}
Dillard, D.~S. and Timmerhaus, K.~D. (1969).
\newblock Thermal conductivity of solid h2o and d2o at low temperatures.
\newblock In Ho, C.~Y. and Taylor, B.~E., editors, {\em Proceedings of the
  Eighth Conference on Thermal Conductivity}, page 949, New York. Plenum.

\bibitem[{Dougherty} et~al., 2006]{dougherty2006SciEncPlume}
{Dougherty}, M.~K., {Khurana}, K.~K., {Neubauer}, F.~M., {Russell}, C.~T.,
  {Saur}, J., {Leisner}, J.~S., and {Burton}, M.~E. (2006).
\newblock {Identification of a Dynamic Atmosphere at Enceladus with the Cassini
  Magnetometer}.
\newblock {\em Science}, 311(5766):1406--1409.

\bibitem[Dozier and Warren, 1982]{dozier1982effect}
Dozier, J. and Warren, S.~G. (1982).
\newblock Effect of viewing angle on the infrared brightness temperature of
  snow.
\newblock {\em Water Resources Res.}, 18(5):1424--1434.

\bibitem[Eames et~al., 1997]{eames_etal_1997}
Eames, I., Marr, N., and Sabir, H. (1997).
\newblock The evaporation coefficient of water: a review.
\newblock {\em Int. J. of Heat and Mass Transfer}, 40(12):2963--2973.

\bibitem[Elias and Chambré, 1993]{elias1993flashing}
Elias, E. and Chambré, P.~L. (1993).
\newblock Flashing inception in water during rapid decompression.
\newblock {\em J. of Heat Transfer}, 115(1):231--238.

\bibitem[{ESA}, 2011]{esa_2011_juice_yellowbook}
{ESA} (2011).
\newblock Juice: Exploring the emergence of habitable worlds around gas giants
  (yellow book).
\newblock Assessment study report.
\newblock Accessed on 2025-04-01.

\bibitem[Eucken, 1911]{eucken1911uber}
Eucken, A. (1911).
\newblock Über die temperaturabhängigkeit der wärmeleitfähigkeit fester
  nichtmetalle.
\newblock {\em Ann. Phys.}, 34(2):185--219.

\bibitem[Fagents, 2003]{fagents2003considerations}
Fagents, S.~A. (2003).
\newblock Considerations for effusive cryovolcanism on europa: The post-galileo
  perspective.
\newblock {\em J. of Geophys. Res. Planets}, 108(E12).

\bibitem[{Fagents} et~al., 2000]{Fagents2000IcarusEurVolc}
{Fagents}, S.~A., {Greeley}, R., {Sullivan}, R.~J., {Pappalardo}, R.~T.,
  {Prockter}, L.~M., and {Galileo SSI Team} (2000).
\newblock {Cryomagmatic Mechanisms for the Formation of Rhadamanthys Linea,
  Triple Band Margins, and Other Low-Albedo Features on Europa}.
\newblock {\em Icarus}, 144(1):54--88.

\bibitem[Figueredo and Greeley, 2004]{figueredo2004resurfacing}
Figueredo, P.~H. and Greeley, R. (2004).
\newblock Resurfacing history of europa from pole-to-pole geological mapping.
\newblock {\em Icarus}, 167(2):287--312.

\bibitem[{Filacchione} et~al., 2019]{filacchione2019europa}
{Filacchione}, G., {Adriani}, A., {Mura}, A., {Tosi}, F., {Lunine}, J.~I.,
  {Raponi}, A., {Ciarniello}, M., {Grassi}, D., {Piccioni}, G., {Moriconi},
  M.~L., {Altieri}, F., {Plainaki}, C., {Sindoni}, G., {Noschese}, R.,
  {Cicchetti}, A., {Bolton}, S.~J., and {Brooks}, S. (2019).
\newblock {Serendipitous infrared observations of Europa by Juno/JIRAM}.
\newblock {\em Icarus}, 328:1--13.

\bibitem[Fraser et~al., 2016]{fraser2016adaptive}
Fraser, K., St-Georges, L., and Kiss, L. (2016).
\newblock Adaptive thermal boundary conditions for smoothed particle
  hydrodynamics.
\newblock In {\em Proceedings of the 14th Internation LS-DYNA Conference,
  Detroit, MI, USA}, pages 12--14.

\bibitem[Fukusako, 1990]{fukusako_1990}
Fukusako, S. (1990).
\newblock Thermophysical properties of ice, snow, and sea ice.
\newblock {\em Int. J. of Thermophysics}, 11(2):353--372.

\bibitem[Galilei, 1610]{galileo1610sidereus}
Galilei, G. (1610).
\newblock Sidereus nuncius.

\bibitem[Gingold and Monaghan, 1977]{gingold1977smoothed}
Gingold, R.~A. and Monaghan, J.~J. (1977).
\newblock Smoothed particle hydrodynamics: theory and application to
  non-spherical stars.
\newblock {\em Monthly notices of the R. Astron. Soc.}, 181(3):375--389.

\bibitem[Greeley et~al., 2000]{greeley2000geologic}
Greeley, R., Figueredo, P.~H., Williams, D.~A., Chuang, F.~C., Klemaszewski,
  J.~E., Kadel, S.~D., Prockter, L.~M., Pappalardo, R.~T., Head~III, J.~W.,
  Collins, G.~C., Spaun, N.~A., Sullivan, R.~J., Moore, J.~M., Senske, D.~A.,
  Tufts, B.~R., Johnson, T.~V., Belton, M. J.~S., and Tanaka, K.~L. (2000).
\newblock Geologic mapping of europa.
\newblock {\em J. of Geophys. Res. Planets}, 105(E9):22559--22578.

\bibitem[Guerra et~al., 2012]{guerra2012understanding}
Guerra, C., Scheibert, J., Bonamy, D., and Dalmas, D. (2012).
\newblock Understanding fast macroscale fracture from microcrack post mortem
  patterns.
\newblock {\em Proc. of the Natl. Acad. of Sci.}, 109(2):390--394.

\bibitem[Guildner et~al., 1976]{guildner1976vapor}
Guildner, L., Johnson, D., and Jones, F. (1976).
\newblock Vapor pressure of water at its triple point.
\newblock {\em Journal of research of the National Bureau of Standards. Section
  A, Physics and chemistry}, 80(3):505.

\bibitem[Hall et~al., 1995]{hall1995detection}
Hall, D.~T., Strobel, D., Feldman, P., McGrath, M., and Weaver, H. (1995).
\newblock Detection of an oxygen atmosphere on jupiter's moon europa.
\newblock {\em Nature}, 373(6516):677--679.

\bibitem[Hansen et~al., 2006]{hansen2006enceladus}
Hansen, C.~J., Esposito, L., Stewart, A., Colwell, J., Hendrix, A., Pryor, W.,
  Shemansky, D., and West, R. (2006).
\newblock Enceladus' water vapor plume.
\newblock {\em Sci.}, 311(5766):1422--1425.

\bibitem[{Hansen} et~al., 2024]{hansen2024juno}
{Hansen}, C.~J., {Ravine}, M.~A., {Schenk}, P.~M., {Collins}, G.~C., {Leonard},
  E.~J., {Phillips}, C.~B., {Caplinger}, M.~A., {Tosi}, F., {Bolton}, S.~J.,
  and {J{\'o}nsson}, B. (2024).
\newblock {Juno's JunoCam Images of Europa}.
\newblock {\em Planet. Sci. J.}, 5(3):76.

\bibitem[Head et~al., 1998]{head1998cryovolcanism}
Head, J., Sherman, N., Pappalardo, R., Thomas, C., Greeley, R., Team, G.~S.,
  et~al. (1998).
\newblock Cryovolcanism on {E}uropa: {E}vidence for the {E}mplacement of
  {F}lows and {R}elated {D}eposits in the {E}4 {R}egion (5{N}, 305{W}) and
  {I}nterpreted {E}ruption {C}onditions.
\newblock In {\em 29th Lunar and Planetary Science Conference}.
\newblock Abstract \#1491.

\bibitem[Hobbs, 2010]{hobbs2010ice}
Hobbs, P.~V. (2010).
\newblock {\em Ice Physics}.
\newblock Oxford Classic Texts in the Physical Sciences. OUP Oxford, Oxford.
\newblock Originally published 1974.

\bibitem[Ito, 1960]{Ito1960}
Ito, H. (1960).
\newblock Pressure losses in smooth pipe bends.
\newblock {\em J. of Basic Engineering}, 82(1):131--140.

\bibitem[Johnson et~al., 1992]{johnson1992space}
Johnson, T.~V., Yeates, C., and Young, R. (1992).
\newblock Space science reviews volume on galileo mission overview.
\newblock {\em Space Sci. Rev.}, 60(1):3--21.

\bibitem[Kargel, 1995]{kargel1994cryovolcanism}
Kargel, J. (1995).
\newblock Cryovolcanism on the icy satellites.
\newblock {\em Earth Moon and Planets}, 67(1):101--113.

\bibitem[Kargel et~al., 1991]{kargel1991rheological}
Kargel, J., Croft, S., Lunine, J., and Lewis, J. (1991).
\newblock Rheological properties of ammonia-water liquids and crystal-liquid
  slurries: Planetological applications.
\newblock {\em Icarus}, 89(1):93--112.

\bibitem[{Kashchiev}, 2000]{kashchiev_2000}
{Kashchiev}, D. (2000).
\newblock {\em {Nucleation: Basic Theory with Applications}}.
\newblock Butterworth–Heinemann, Oxford/Boston.

\bibitem[{Kattenhorn} and {Prockter}, 2014]{kattenhorn2014NatGeo}
{Kattenhorn}, S.~A. and {Prockter}, L.~M. (2014).
\newblock {Evidence for subduction in the ice shell of Europa}.
\newblock {\em Nature Geoscience}, 7(10):762--767.

\bibitem[Kern et~al., 2004]{kern2004rheology}
Kern, M., Tiefenbacher, F., and McElwaine, J. (2004).
\newblock The rheology of snow in large chute flows.
\newblock {\em Cold Regions Sci. and Technology}, 39(2-3):181--192.

\bibitem[{Ketcham} and {Hobbs}, 1969]{ketcham_Hobbs_1969}
{Ketcham}, W.~M. and {Hobbs}, S.~G. (1969).
\newblock {An experimental determination of the surface energies of ice}.
\newblock {\em Philos. Mag.}, 19:1161--1173.

\bibitem[{Knight}, 1971]{knight_1971}
{Knight}, C.~A. (1971).
\newblock {Experiments on the contact angle of water on ice}.
\newblock {\em Philos. Mag}, 23.

\bibitem[Landauer, 1952]{Landauer1952}
Landauer, R. (1952).
\newblock The electrical resistance of binary metallic mixtures.
\newblock {\em J. of Appl. Phys.}, 23(7):779--784.

\bibitem[{Leonard} et~al., 2023]{leonard2023Europageomap}
{Leonard}, E.~J., {Patthoff}, D.~A., and {Senske}, D. (2023).
\newblock {Global geologic map of Europa: U.S. Geological Survey Scientific
  Investigations Map 3513, scale 1:15,000,000}.
\newblock {\em USGS pamphlet}, page~18.

\bibitem[Lesage et~al., 2022]{lesage2022simulation}
Lesage, E., Massol, H., Howell, S.~M., and Schmidt, F. (2022).
\newblock Simulation of freezing cryomagma reservoirs in viscoelastic ice
  shells.
\newblock {\em the Planet. Sci. J.}, 3(7):170.

\bibitem[Lesage et~al., 2020]{lesage2020cryomagma}
Lesage, E., Massol, H., and Schmidt, F. (2020).
\newblock Cryomagma ascent on europa.
\newblock {\em Icarus}, 335:113369.

\bibitem[Lesage et~al., 2021]{lesage2021constraints}
Lesage, E., Schmidt, F., Andrieu, F., and Massol, H. (2021).
\newblock Constraints on effusive cryovolcanic eruptions on europa using
  topography obtained from galileo images.
\newblock {\em Icarus}, 361:114373.

\bibitem[Li et~al., 2022]{li2022influence}
Li, T., Kuang, S., Hu, L., Nie, P., Ramaswamy, H.~S., and Yu, Y. (2022).
\newblock Influence of the pressure shift freezing and thawing on the
  microstructure of largemouth bass.
\newblock {\em Innovative Food Sci. \& Emerging Technologies}, 82:103176.

\bibitem[Lopes et~al., 2013]{lopes2013cryovolcanism}
Lopes, R.~M., Kirk, R.~L., Mitchell, K.~L., LeGall, A., Barnes, J.~W., Hayes,
  A., Kargel, J., Wye, L., Radebaugh, J., Stofan, E., et~al. (2013).
\newblock Cryovolcanism on titan: New results from cassini radar and vims.
\newblock {\em J. of Geophys. Res. Planets}, 118(3):416--435.

\bibitem[Manga and Michaut, 2017]{manga2017formation}
Manga, M. and Michaut, C. (2017).
\newblock Formation of lenticulae on europa by saucer-shaped sills.
\newblock {\em Icarus}, 286:261--269.

\bibitem[Menzies and van~der Meer, 2018]{menzies2018past}
Menzies, J. and van~der Meer, J.~J., editors (2018).
\newblock {\em Past Glacial Environments}.
\newblock Elsevier, 2nd edition.

\bibitem[{Mitchell} et~al., 2024]{mitchell_etal_2024}
{Mitchell}, K.~L., {Rabinovitch}, J., {Scamardella}, J.~C., and {Cable}, M.~L.
  (2024).
\newblock {A Proposed Model for Cryovolcanic Activity on Enceladus Driven by
  Volatile Exsolution}.
\newblock {\em Journal of Geophysical Research: Planets}, 129(7):e2023JE007977.
\newblock e2023JE007977 2023JE007977.

\bibitem[{Moore} et~al., 2001]{moore2001IcarusEurGEM}
{Moore}, J.~M., {Asphaug}, E., {Belton}, M. J.~S., {Bierhaus}, B., {Breneman},
  H.~H., {Brooks}, S.~M., {Chapman}, C.~R., {Chuang}, F.~C., {Collins}, G.~C.,
  {Giese}, B., {Greeley}, R., {Head}, J.~W., {Kadel}, S., {Klaasen}, K.~P.,
  {Klemaszewski}, J.~E., {Magee}, K.~P., {Moreau}, J., {Morrison}, D.,
  {Neukum}, G., {Pappalardo}, R.~T., {Phillips}, C.~B., {Schenk}, P.~M.,
  {Senske}, D.~A., {Sullivan}, R.~J., {Turtle}, E.~P., and {Williams}, K.~K.
  (2001).
\newblock {Impact Features on Europa: Results of the Galileo Europa Mission
  (GEM)}.
\newblock {\em Icarus}, 151(1):93--111.

\bibitem[Morrison et~al., 2023]{morrison2023reevaluation}
Morrison, A.~A., Whittington, A.~G., and Mitchell, K.~L. (2023).
\newblock A reevaluation of cryolava flow evolution: Assumptions, physical
  properties, and conceptualization.
\newblock {\em J. of Geophys. Res. Planets}, 128(1):e2022JE007383.

\bibitem[Murphy and Koop, 2005]{murphy2005review}
Murphy, D.~M. and Koop, T. (2005).
\newblock Review of the vapour pressures of ice and supercooled water for
  atmospheric applications.
\newblock {\em Quarterly Journal of the Royal Meteorological Society: A journal
  of the atmospheric sciences, applied meteorology and physical oceanography},
  131(608):1539--1565.

\bibitem[Nishimura, 1996]{nishimura1996viscosity}
Nishimura, K. (1996).
\newblock Viscosity of fluidized snow.
\newblock {\em Cold regions Sci. and technology}, 24(2):117--127.

\bibitem[Papanastasiou, 1987]{papanastasiou1987flows}
Papanastasiou, T.~C. (1987).
\newblock Flows of materials with yield.
\newblock {\em J. of rheology}, 31(5):385--404.

\bibitem[Pappalardo et~al., 1998]{pappalardo1998geological}
Pappalardo, R., Head, J., Greeley, R., Sullivan, R., Pilcher, C., Schubert, G.,
  Moore, W., Carr, M., Moore, J., Belton, M., et~al. (1998).
\newblock Geological evidence for solid-state convection in europa's ice shell.
\newblock {\em Nature}, 391(6665):365--368.

\bibitem[Pappalardo et~al., 2009]{pappalardo2009europa}
Pappalardo, R., McKinnon, W., and Khurana, K. (2009).
\newblock {\em Europa}.
\newblock The University of Arizona Space Science Series. University of Arizona
  Press.

\bibitem[{Pappalardo} et~al., 2024]{pappalardo2024SSR}
{Pappalardo}, R.~T., {Buratti}, B.~J., {Korth}, H., {Senske}, D.~A., {Blaney},
  D.~L., {Blankenship}, D.~D., {Burch}, J.~L., {Christensen}, P.~R., {Kempf},
  S., {Kivelson}, M.~G., {Mazarico}, E., {Retherford}, K.~D., {Turtle}, E.~P.,
  {Westlake}, J.~H., {Paczkowski}, B.~G., {Ray}, T.~L., {Kampmeier}, J.,
  {Craft}, K.~L., {Howell}, S.~M., {Klima}, R.~L., {Leonard}, E.~J., {Matiella
  Novak}, A., {Phillips}, C.~B., {Daubar}, I.~J., {Blacksberg}, J., {Brooks},
  S.~M., {Choukroun}, M.~N., {Cochrane}, C.~J., {Diniega}, S., {Elder}, C.~M.,
  {Ernst}, C.~M., {Gudipati}, M.~S., {Luspay-Kuti}, A., {Piqueux}, S., {Rymer},
  A.~M., {Roberts}, J.~H., {Steinbr{\"u}gge}, G., {Cable}, M.~L., {Scully}, J.
  E.~C., {Castillo-Rogez}, J.~C., {Hay}, H. C.~F.~C., {Persaud}, D.~M.,
  {Glein}, C.~R., {McKinnon}, W.~B., {Moore}, J.~M., {Raymond}, C.~A.,
  {Schroeder}, D.~M., {Vance}, S.~D., {Wyrick}, D.~Y., {Zolotov}, M.~Y.,
  {Hand}, K.~P., {Nimmo}, F., {McGrath}, M.~A., {Spencer}, J.~R., {Lunine},
  J.~I., {Paty}, C.~S., {Soderblom}, J.~M., {Collins}, G.~C., {Schmidt}, B.~E.,
  {Rathbun}, J.~A., {Shock}, E.~L., {Becker}, T.~C., {Hayes}, A.~G.,
  {Prockter}, L.~M., {Weiss}, B.~P., {Hibbitts}, C.~A., {Moussessian}, A.,
  {Brockwell}, T.~G., {Hsu}, H.-W., {Jia}, X., {Gladstone}, G.~R., {McEwen},
  A.~S., {Patterson}, G.~W., {McNutt}, R.~L., {Evans}, J.~P., {Larson}, T.~W.,
  {Cangahuala}, L.~A., {Havens}, G.~G., {Buffington}, B.~B., {Bradley}, B.,
  {Campagnola}, S., {Hardman}, S.~H., {Srinivasan}, J.~M., {Short}, K.~L.,
  {Jedrey}, T.~C., {St. Vaughn}, J.~A., {Clark}, K.~P., {Vertesi}, J., and
  {Niebur}, C. (2024).
\newblock {Science Overview of the Europa Clipper Mission}.
\newblock {\em Space Sci. Rev.}, 220(4):40.

\bibitem[Parsons et~al., 1987]{parsons1987preliminary}
Parsons, B.~L., Snellen, J.~B., and Hill, B. (1987).
\newblock Preliminary measurements of terminal crack velocity in ice.
\newblock {\em Cold Regions Sci. and Technology}, 13(3):233--238.

\bibitem[Petrenko and Whitworth, 1999]{petrenko1999physics}
Petrenko, V.~F. and Whitworth, R.~W. (1999).
\newblock {\em Physics of ice}.
\newblock OUP Oxford.

\bibitem[Porco et~al., 2006]{porco2006cassini}
Porco, C.~C., Helfenstein, P., Thomas, P., Ingersoll, A., Wisdom, J., West, R.,
  Neukum, G., Denk, T., Wagner, R., Roatsch, T., et~al. (2006).
\newblock {Cassini observes the active south pole of Enceladus}.
\newblock {\em Sci.}, 311(5766):1393--1401.

\bibitem[Prakash and Cleary, 2011]{prakash2011three}
Prakash, M. and Cleary, P.~W. (2011).
\newblock Three dimensional modelling of lava flow using smoothed particle
  hydrodynamics.
\newblock {\em Appl. Math. Model.}, 35(6):3021--3035.

\bibitem[{Prockter} and {Schenk}, 2005]{prockter2005macula}
{Prockter}, L.~M. and {Schenk}, P.~M. (2005).
\newblock {Origin and evolution of Castalia Macula: an anomalous young
  depression on Europa}.
\newblock {\em Icarus}, 177:305--326.

\bibitem[Prohaska et~al., 2022]{prohaska_2022}
Prohaska, T., Irrgeher, J., Benefield, J., B{\"o}hlke, J.~K., Chesson, L.~A.,
  Coplen, T.~B., Ding, T., Dunn, P. J.~H., Gr{\"o}ning, M., Holden, N.~E.,
  Meijer, H. A.~J., Moossen, H., Possolo, A., Takahashi, Y., Vogl, J., Walczyk,
  T., Wang, J., Wieser, M.~E., Yoneda, S., Zhu, X.-K., and Meija, J. (2022).
\newblock Standard atomic weights of the elements 2021.
\newblock {\em Pure and Appl. Chem.}, 94(5):573--600.
\newblock Received June 23, 2019; accepted January 9, 2022.

\bibitem[Quick et~al., 2013]{quick2013constraints}
Quick, L.~C., Barnouin, O.~S., Prockter, L.~M., and Patterson, G.~W. (2013).
\newblock {Constraints on the detection of cryovolcanic plumes on Europa}.
\newblock {\em Planet. and Space Sci.}, 86:1--9.

\bibitem[{Quick} et~al., 2022]{quick2022Icarus}
{Quick}, L.~C., {Fagents}, S.~A., {N{\'u}{\~n}ez}, K.~A., {Wilk}, K.~A.,
  {Beyer}, R.~A., {Beddingfield}, C.~B., {Martin}, E.~S., {Prockter}, L.~M.,
  and {Hurford}, T.~A. (2022).
\newblock {Cryolava dome growth resulting from active eruptions on Jupiter's
  moon Europa}.
\newblock {\em Icarus}, 387:115185.

\bibitem[Quick et~al., 2017]{quick2017cryovolcanic}
Quick, L.~C., Glaze, L.~S., and Baloga, S.~M. (2017).
\newblock {Cryovolcanic emplacement of domes on Europa}.
\newblock {\em Icarus}, 284:477--488.

\bibitem[Quick and Hedman, 2020]{quick2020characterizing}
Quick, L.~C. and Hedman, M.~M. (2020).
\newblock {Characterizing deposits emplaced by cryovolcanic plumes on Europa}.
\newblock {\em Icarus}, 343:113667.

\bibitem[Ramachandran et~al., 2021]{ramachandran2021pysph}
Ramachandran, P., Bhosale, A., Puri, K., Negi, P., Muta, A., Dinesh, A., Menon,
  D., Govind, R., Sanka, S., Sebastian, A.~S., Sen, A., Kaushik, R., Kumar, A.,
  Kurapati, V., Patil, M., Tavker, D., Pandey, P., Kaushik, C., Dutt, A., and
  Agarwal, A. (2021).
\newblock Pysph: A python-based framework for smoothed particle hydrodynamics.
\newblock {\em ACM Trans. Math. Softw.}, 47(4).

\bibitem[Riche and Schneebeli, 2013]{riche2013thermal}
Riche, F. and Schneebeli, M. (2013).
\newblock Thermal conductivity of snow measured by three independent methods
  and anisotropy considerations.
\newblock {\em the Cryosphere}, 7(1):217--227.

\bibitem[Roth, 2021]{roth2021stable}
Roth, L. (2021).
\newblock A stable h2o atmosphere on europa’s trailing hemisphere from hst
  images.
\newblock {\em Geophys. Res. Lett.}, 48(20):e2021GL094289.

\bibitem[Roth et~al., 2014]{roth2014vapor}
Roth, L., Saur, J., Retherford, K.~D., Strobel, D.~F., Feldman, P.~D., McGrath,
  M.~A., and Nimmo, F. (2014).
\newblock Transient water vapor at europa’s south pole.
\newblock {\em Sci.}, 343(6167):171--174.

\bibitem[Ruiz, 2005]{ruiz2005heat}
Ruiz, J. (2005).
\newblock The heat flow of europa.
\newblock {\em Icarus}, 177(2):438--446.
\newblock Europa Icy Shell.

\bibitem[Shirley et~al., 2010]{shirley2010europa}
Shirley, J.~H., Dalton~III, J.~B., Prockter, L.~M., and Kamp, L.~W. (2010).
\newblock Europa’s ridged plains and smooth low albedo plains: Distinctive
  compositions and compositional gradients at the leading side--trailing side
  boundary.
\newblock {\em Icarus}, 210(1):358--384.

\bibitem[Smith et~al., 1989]{smith1989voyager}
Smith, B.~A., Soderblom, L.~A., Banfield, D., c.~Barnet, Basilevsky, A.~T.,
  Beebe, R.~F., Bollinger, K., Boyce, J.~M., Brahic, A., Briggs, G.~A., Brown,
  R.~H., c.~Chyba, s.~A.~Collins, Colvin, T., Cook, A.~F., Crisp, D., Croft,
  S.~K., Cruikshank, D., Cuzzi, J.~N., Danielson, G.~E., Davies, M.~E., Jong,
  E.~D., Dones, L., Godfrey, D., Goguen, J., Grenier, I., Haemmerle, V.~R.,
  Hammel, H., c.~J.~Hansen, c.~P.~Helfenstein, Howell, C., Hunt, G.~E.,
  Ingersoll, A.~P., Johnson, T.~V., Kargel, J., Kirk, R., Kuehn, D.~I., Limaye,
  S., Masursky, H., McEwen, A., Morrison, D., Owen, T., Owen, W., Pollack,
  J.~B., c.~c. Porco, Rages, K., Rogers, P., Rudy, D., Sagan, C., Schwartz, J.,
  Shoemaker, E.~M., Showalter, M., Sicardy, B., Simonelli, D., Spencer, J.,
  Sromovsky, L.~A., Stoker, C., Strom, R.~G., Suomi, V.~E., Synott, S.~P.,
  Terrile, R.~J., Thomas, P., Thompson, W.~R., Verbiscer, A., and Veverka, J.
  (1989).
\newblock Voyager 2 at neptune: Imaging science results.
\newblock {\em Sci.}, 246(4936):1422--1449.

\bibitem[Smith et~al., 1979]{smith1979jupiter}
Smith, B.~A., Soderblom, L.~A., Johnson, T.~V., Ingersoll, A.~P., Collins,
  S.~A., Shoemaker, E.~M., Hunt, G.~E., Masursky, H., Carr, M.~H., Davies,
  M.~E., Cook, A.~F., Boyce, J., Danielson, G.~E., Owen, T., Sagan, C., Beebe,
  R.~F., Veverka, J., Strom, R.~G., McCauley, J.~F., Morrison, D., Briggs,
  G.~A., and Suomi, V.~E. (1979).
\newblock The jupiter system through the eyes of voyager 1.
\newblock {\em Sci.}, 204(4396):951--972.

\bibitem[Solomonidou et~al., 2021]{solomonidou2021candidate}
Solomonidou, A., Stephan, K., Kalousova, K., and Soderlund, K. (2021).
\newblock Candidate cryovolcanic regions on ganymede: a target priority for
  juice.
\newblock Technical report, Copernicus Meetings.

\bibitem[Sotin et~al., 2002]{sotin2002europa}
Sotin, C., Head~III, J.~W., and Tobie, G. (2002).
\newblock Europa: Tidal heating of upwelling thermal plumes and the origin of
  lenticulae and chaos melting.
\newblock {\em Geophys. Res. Lett.}, 29(8):74--1--74--4.

\bibitem[Spahn et~al., 2006]{spahn2006cassini}
Spahn, F., Schmidt, J., Albers, N., H{\"{o}}rning, M., Makuch, M., Sei{\ss},
  M., Kempf, S., Srama, R., Dikarev, V., Helfert, S., et~al. (2006).
\newblock Cassini dust measurements at enceladus and implications for the
  origin of the e ring.
\newblock {\em Sci.}, 311(5766):1416--1418.

\bibitem[Spencer et~al., 2009]{spencer2009enceladus}
Spencer, J.~R., Barr, A.~C., Esposito, L.~W., Helfenstein, P., Ingersoll,
  A.~P., Jaumann, R., McKay, C.~P., Nimmo, F., and Waite, J.~H. (2009).
\newblock Enceladus: An active cryovolcanic satellite.
\newblock In {\em Saturn from Cassini-Huygens}, pages 683--724. Springer.

\bibitem[Stockli and Rixen, 2000]{stockli2000characteristics}
Stockli, V. and Rixen, C. (2000).
\newblock Characteristics of artificial snow and its effect on vegetation.
\newblock In {\em Proceedings of the 2000 International Snow Science Workshop,
  October 1-6, Big Sky, Montana}, pages 468--471. Montana State University
  Library.

\bibitem[Sturm et~al., 1997]{sturm1997thermal}
Sturm, M., Holmgren, J., König, M., and Morris, K. (1997).
\newblock The thermal conductivity of seasonal snow.
\newblock {\em J. of Glaciol.}, 43(143):26--41.

\bibitem[{Turnbull}, 1950]{turnbull_1950}
{Turnbull}, D. (1950).
\newblock {Kinetics of Heterogeneous Nucleation}.
\newblock {\em J. Chem. Phys.}, 18(2):198--203.

\bibitem[Valovirta and Vinha, 2004]{valovirta2004water}
Valovirta, I. and Vinha, J. (2004).
\newblock Water vapor permeability and thermal conductivity as a function of
  temperature and relative humidity.
\newblock In {\em Performance of Exterior Envelopes of Whole Buildings, IX
  International Conference, Florida, USA, December 5-10, 2004}, page~16.

\bibitem[Vance et~al., 2023]{vance2023investigating}
Vance, S.~D., Craft, K.~L., Shock, E., Schmidt, B.~E., Lunine, J., Hand, K.~P.,
  McKinnon, W.~B., Spiers, E.~M., Chivers, C., Lawrence, J.~D., et~al. (2023).
\newblock Investigating europa’s habitability with the europa clipper.
\newblock {\em Space Sci. Rev.}, 219(8):81.

\bibitem[{Villanueva} et~al., 2023]{villanueva_etal_2023}
{Villanueva}, G.~L., {Hammel}, H.~B., {Milam}, S.~N., {Faggi}, S., {Kofman},
  V., {Roth}, L., {Hand}, K.~P., {Paganini}, L., {Stansberry}, J., {Spencer},
  J., {Protopapa}, S., {Strazzulla}, G., {Cruz-Mermy}, G., {Glein}, C.~R.,
  {Cartwright}, R., and {Liuzzi}, G. (2023).
\newblock {Endogenous CO$_{2}$ ice mixture on the surface of Europa and no
  detection of plume activity}.
\newblock {\em Science}, 381(6664):1305--1308.

\bibitem[Vogt et~al., 2008]{vogt2008speed}
Vogt, C., Laihem, K., and Wiebusch, C. (2008).
\newblock Speed of sound in bubble-free ice.
\newblock {\em the J. of the Acoust. Soc. of America}, 124(6):3613--3618.

\bibitem[{Waite} et~al., 2006]{waite2006SciEncPlume}
{Waite}, J.~H., {Combi}, M.~R., {Ip}, W.-H., {Cravens}, T.~E., {McNutt}, R.~L.,
  {Kasprzak}, W., {Yelle}, R., {Luhmann}, J., {Niemann}, H., {Gell}, D.,
  {Magee}, B., {Fletcher}, G., {Lunine}, J., and {Tseng}, W.-L. (2006).
\newblock {Cassini Ion and Neutral Mass Spectrometer: Enceladus Plume
  Composition and Structure}.
\newblock {\em Science}, 311(5766):1419--1422.

\bibitem[{Wilson} et~al., 1997]{wilson1997JGRflowsEuropa}
{Wilson}, L., {Head}, J.~W., and {Pappalardo}, R.~T. (1997).
\newblock {Eruption of lava flows on Europa: Theory and application to Thrace
  Macula}.
\newblock {\em J. Geophys. Res.}, 102(E4):9263--9272.

\bibitem[Wolfenbarger et~al., 2021]{wolfenbarger_etal_2021}
Wolfenbarger, N.~S., Carnahan, E., Jordan, J.~S., and Hesse, M.~A. (2021).
\newblock A comprehensive dataset for the thermal conductivity of ice ih for
  application to planetary ice shells.
\newblock {\em Data in Brief}, 36:107091.

\bibitem[Yen, 1981]{yen_1981}
Yen, Y.-C. (1981).
\newblock Review of thermal properties of snow, ice and sea ice.
\newblock Technical Report CRREL Report 81-10, United States Army Corps of
  Engineers, Cold Regions Research and Engineering Laboratory, Hanover, New
  Hampshire, U.S.A.

\bibitem[Zahnle et~al., 2003]{zahnle2003cratering}
Zahnle, K., Schenk, P., Levison, H., and Dones, L. (2003).
\newblock Cratering rates in the outer solar system.
\newblock {\em Icarus}, 163(2):263--289.

\bibitem[Zhang et~al., 2022]{zhang2022smoothed}
Zhang, C., Zhu, Y.-j., Wu, D., Adams, N.~A., and Hu, X. (2022).
\newblock Smoothed particle hydrodynamics: Methodology development and recent
  achievement.
\newblock {\em J. of Hydrodyn.}, 34(5):767--805.

\bibitem[Zhang, 2005]{zhang2005influence}
Zhang, T. (2005).
\newblock Influence of the seasonal snow cover on the ground thermal regime: An
  overview.
\newblock {\em Rev. of Geophysics}, 43(4).

\end{thebibliography}

\end{document}